\documentclass[superscriptaddress,showpacs,twocolumn,prd]{revtex4}
\usepackage{graphicx}

\newcommand{\be}{\begin{equation}}
\newcommand{\ee}{\end{equation}}
\newcommand{\ba}{\begin{eqnarray}}
\newcommand{\ea}{\end{eqnarray}}
\newcommand\eqref[1]{(#1)}

\newcommand{\Tr}{\mathrm{Tr}}
\newcommand{\Str}{\mathrm{Str}}

\newcommand{\nn}{\nonumber\\}
\newcommand{\lgl}{\langle}
\newcommand{\rgl}{\rangle}
\begin{document}

\onecolumngrid
\thispagestyle{empty}
\begin{flushright}
{\large 
LU  TP  06-07 \\
HISKP - TH - 05-26 \\
\large hep-lat/0602003\\[0.1cm]
\large revised February 2006}
\end{flushright}
\vskip2cm
\begin{center}
{\Large\bf
Three-Flavor Partially Quenched Chiral Perturbation Theory\\
at NNLO for Meson Masses and Decay Constants}

\vfill

{\large \bf Johan Bijnens$^a$, Niclas Danielsson$^{a,b}$ and 
Timo A. L\"ahde$^c$}\\[1cm]
{$^a$Department of Theoretical Physics, Lund University, \\
S\"olvegatan 14A, SE - 223 62 Lund, Sweden\\[1cm]
$^b$Division of Mathematical Physics, 
Lund Institute of Technology, Lund University, \\ 
Box 118, SE - 221 00 Lund, Sweden\\[1cm]
$^c$Helmholtz-Institut f\"ur Strahlen- und Kernphysik (HISKP),
           Bonn University, \\ Nu\ss allee 14-16, D - 53115 Bonn, 
           Germany
}

\vfill

{\large\bf Abstract}

\vskip1cm

\parbox{14cm}{\large

We discuss Partially Quenched Chiral Perturbation Theory (PQ$\chi$PT) 
and possible fitting strategies to Lattice QCD data at 
next-to-next-to-leading order (NNLO) in the mesonic sector. We also 
present a complete calculation of the masses of the charged 
pseudoscalar mesons, in the supersymmetric formulation of PQ$\chi$PT. 
Explicit analytical results are given for up to three nondegenerate sea 
quark flavors, along with the previously unpublished expression for the 
pseudoscalar meson decay constant for three nondegenerate sea quark 
flavors. The numerical analysis in this paper demonstrates that the 
corrections at NNLO are sizable, as expected from earlier work.}

\vskip2cm

\end{center}
\vskip2cm
\twocolumngrid
\setcounter{page}{0}

\title{Three-Flavor Partially Quenched Chiral Perturbation Theory
at NNLO \\ for Meson Masses and Decay Constants}

\author{Johan Bijnens}
\affiliation{Department of Theoretical Physics, Lund University,\\
S\"olvegatan 14A, SE - 223 62 Lund, Sweden}
\author{Niclas Danielsson}
\affiliation{Department of Theoretical Physics, Lund University,\\
S\"olvegatan 14A, SE - 223 62 Lund, Sweden}
\affiliation{Division of Mathematical Physics, Lund Institute of 
Technology, Lund University,\\ 
Box 118, SE - 221 00 Lund, Sweden}
\author{Timo A. L\"ahde}
\affiliation{Helmholtz-Institut f\"ur Strahlen- und Kernphysik (HISKP),
           Bonn University, \\ Nu\ss allee 14-16, D - 53115 Bonn, 
Germany}


\begin{abstract} 
We discuss Partially Quenched Chiral Perturbation Theory (PQ$\chi$PT) 
and possible fitting strategies to Lattice QCD data at 
next-to-next-to-leading order (NNLO) in the mesonic sector. We also 
present a complete calculation of the masses of the charged 
pseudoscalar mesons, in the supersymmetric formulation of PQ$\chi$PT. 
Explicit analytical results are given for up to three nondegenerate sea 
quark flavors, along with the previously unpublished expression for the 
pseudoscalar meson decay constant for three nondegenerate sea quark 
flavors. The numerical analysis in this paper demonstrates that the 
corrections at NNLO are sizable, as expected from earlier work.
\end{abstract}

\maketitle

\section{Introduction}
\label{intro}

The generally accepted theory of the strong interaction, Quantum Chromo 
Dynamics (QCD), provides in principle a way to derive various 
low-energy hadronic observables, such as masses and decay constants. 
However, an analytic \mbox{\it ab initio} derivation of such properties 
has not yet been possible. An alternative approach called Lattice QCD, 
where the functional integrals are evaluated using numerical Monte 
Carlo techniques on a discretized spacetime lattice, suggests a way 
around this problem, but is nevertheless not free of difficulties of 
its own. In particular, the properties of low-mass particles are 
especially difficult to calculate as they can propagate over large 
distances on the lattice, giving rise to large nonlocal correlations 
which limit the precision obtainable with given computer resources. As 
a consequence, most simulations have been performed with heavier quark 
masses than those of the physical world. The quark masses 
used in present Lattice QCD simulations typically fulfill $m_{u,d}\ge 
m_s/8$. The results then have to be extrapolated down to the physical 
masses of $\sim m_s/25$.

The preferable way to perform the extrapolation to physical quark 
masses is by means of Chiral Perturbation Theory ($\chi$PT) 
\cite{Weinberg,GL}, which provides a theoretically correct description 
of the low-energy properties of QCD within the range of quark masses 
where $\chi$PT calculations can be considered accurate enough. In 
particular, the appearance of so-called chiral logarithms renders 
simple polynomial extrapolation insufficient. If the Lattice QCD 
simulations are performed with quark masses in the chiral regime, i.e. 
with quark masses for which $\chi$PT calculations are reliable, then 
they can be used to e.g. determine the effective low-energy constants 
(LEC:s) of $\chi$PT. Nevertheless, in practice it is still difficult to 
reach the chiral regime and therefore many large scale lattice 
simulations of so-called quenched QCD have been performed, in which the 
effects of closed (sea) quark loops are neglected. Consideration of 
such loop effects requires repeated evaluation of fermion determinants, 
which is computationally extremely expensive. It should be noted that 
first Lattice QCD calculations with sea quark masses significantly 
below $m_s$ are now becoming available.

From a computational point of view, it is however much cheaper to vary 
the valence quark masses $m_{\mathrm{val}}$ rather than those of the 
sea quarks, $m_{\mathrm{sea}}$. Therefore it is not uncommon to perform 
simulations where the sea quark loops are not neglected, but rather 
suppressed by choosing $m_{\mathrm{sea}} > m_{\mathrm{val}}$, or even 
$n_{\mathrm{sea}} \ne n_{\mathrm{val}}$, where $n_{\mathrm{val}}$ and 
$n_{\mathrm{sea}}$ denote the number of each quark species in the 
theory. Such procedures are referred to as partial quenching and lead 
into a space of unphysical theories.

\subsection{Partially Quenched Theories}

At first, one might expect that partially quenched (PQ) theories could 
yield only qualitative information about QCD. However, since unquenched 
QCD may be recovered from partially quenched QCD (PQQCD) in the limit 
of equal sea and valence quark masses, it follows that QCD and PQQCD 
are continuously connected by variation of sea quark masses. In 
contrast, this is not true for fully quenched QCD. Nevertheless, 
$\chi$PT can be extended to the case of quenched QCD, which was 
done by Bernard, Golterman and Sharpe \cite{BG1,SharpeA} after earlier 
work by Morel \cite{Morel}. The extension to the partially quenched 
case was done in Ref.~\cite{BG2}. A more extensive discussion can be 
found in the work of Sharpe and Shoresh \cite{Sharpe1,Sharpe2}.

The formulation of Partially Quenched Chiral Perturbation Theory 
(PQ$\chi$PT) is such that the dependence on the sea quark masses is 
explicit, and thus the limit of equal sea and valence quark masses can 
also be considered for PQ$\chi$PT. It follows that $\chi$PT is 
recovered as a continuous limit of PQ$\chi$PT just as QCD is from 
PQQCD. In particular, the LEC:s of $\chi$PT, which are of physical 
significance, can be obtained directly from those of PQ$\chi$PT. In 
addition, the ability to vary valence and sea quark masses separately 
allows more information to be extracted at fewer values of the sea 
quark masses. This is another reason to consider PQQCD in lattice 
simulations as discussed in detail in Ref.~\cite{Sharpe1}.

This paper is devoted to the extension of PQ$\chi$PT to 
next-to-next-to-leading-order (NNLO). The discussions already given in 
Refs.~\cite{BDL,BL1,BL2} are completed and the calculations of the 
masses and decay constants of the charged, or off-diagonal, 
pseudoscalar mesons are completed for the case of $n_{\mathrm{sea}} = 
3$. It should be noted that indications already exist, by the [qq+q] 
collaboration~\cite{Latt1} and the MILC collaboration \cite{Latt2}, 
that the ${\cal O}(p^6)$ contributions to these quantities are sizable, 
and that the inclusion of such effects can have a significant impact on 
the chiral extrapolations down to the physical quark masses. More 
examples can be found in the proceedings of the Lattice 2005 
conference.

\subsection{Partially Quenched $\chi$PT at NNLO}

The meson masses and decay constants of the pseudoscalar mesons in 
PQ$\chi$PT for $n_{\mathrm{sea}} = 3$ to NLO, one loop, or ${\cal 
O}(p^4)$ in the momentum expansion, were calculated in 
Refs.~\cite{BG2,Sharpe1,Sharpe2}. First results for these quantities at 
NNLO, two loops, or ${\cal O}(p^6)$, namely the mass of a charged 
pseudoscalar meson for degenerate sea and degenerate valence quark 
masses, can be found in Ref.~\cite{BDL}. The NNLO expression for the 
decay constant of a charged pseudoscalar meson with two nondegenerate 
sea quarks has been calculated in Ref.~\cite{BL1}. The full results for 
pseudoscalar meson masses and decay constants for $n_{\mathrm{sea}} = 
2$ were given in Ref.~\cite{BL2}. This paper presents the full 
nondegenerate calculations of the charged pseudoscalar meson mass to 
NNLO for $n_{\mathrm{sea}} = 3$, along with the calculation of the 
decay constant for three nondegenerate sea quarks.

In general, the NNLO expressions are very long, but it is possible to 
shorten them considerably by introducing a specialized notation which 
can accommodate the complications inherent in PQ$\chi$PT. Furthermore, 
this notation satisfies many different algebraic relations, which 
allows for a systematic simplification to be carried out. In the 
next sections, the technical background for the NNLO calculations, 
together with the above-mentioned notation is presented, followed by 
the results for the NNLO masses and decay constants of the charged 
pseudoscalar mesons, and a numerical analysis. All analytical formulas 
are given explicitly, but they can also be downloaded from the 
website~\cite{website}.

This paper is organized in the following manner: Sect.~\ref{PQCHPT} 
introduces PQ$\chi$PT and discusses the various aspects relevant for 
calculations at NNLO. In particular, the relations between the various 
sets of LEC:s are highlighted. This section also includes an overview 
of all notation used for loop integrals and combinations of quark 
masses. Sects.~\ref{masses} and~\ref{decayconstant} contain the 
analytical NNLO expressions for the masses and decay constants of the 
charged pseudoscalar mesons. Sect.~\ref{discussion} presents a 
discussion of the checks performed on the analytical and numerical 
calculations, a numerical analysis of the results, and an elaboration 
on the extraction of the various LEC:s from Lattice QCD
calculations is given in Sect.~\ref{LECdetermination}. 
Our conclusions are shortly discussed in Sect.~\ref{conclusions}.

\section{Technical Aspects of PQ$\chi$PT}
\label{PQCHPT}

The technical aspects of PQ$\chi$PT calculations to NLO have been 
thoroughly covered, for the supersymmetric formulation of PQ$\chi$PT, 
in Refs.~\cite{Sharpe1,Sharpe2}. The new parts specific to the NNLO 
calculations have been covered briefly in Refs.~\cite{BDL,BL1,BL2}. The 
aim of this section is to collect all this information in one place and 
describe the NNLO aspects in somewhat greater detail.

\subsection{Supersymmetric PQ$\chi$PT}

Let us first recall a few aspects of $\chi$PT, and note in passing that 
introductions to the subject can be found e.g. in 
Ref.~\cite{CHPTlectures}. For $n_f$ massless quark flavors, QCD has a 
chiral symmetry
\begin{equation}
G=SU(n_f)_L\times SU(n_f)_R\,,  
\label{chsym}
\end{equation}
which is spontaneously broken by the vacuum to its diagonal subgroup 
$H=SU(n_f)_{V}$, where $V = L+R$. The Goldstone bosons produced by this 
spontaneous breakdown live on the coset $G/H$ which is also an 
$SU(n_f)$ manifold and can be parameterized in terms of a special 
unitary matrix $U$. This matrix is conventionally written in terms of 
a traceless Hermitian matrix according to
\begin{equation}
U \equiv \exp\left(i\sqrt{2}\,\phi/F\right)\,  
\end{equation}
where the constant $F$ with dimension of energy can be shown to be 
related to the decay constant of the pion. For three quark flavors 
($u,d,s$), the matrix $\phi$ has a flavor structure
\begin{equation}
\label{phiQCD}
\phi =  \left(\begin{array}{ccc}u\bar u & u\bar d & u \bar s\\
d\bar u & d \bar d & d\bar s\\
s\bar u & s \bar d & s\bar s\end{array}\right)\,.
\end{equation}

The above considerations do not distinguish between valence and sea 
quarks whereas the construction of a PQ theory requires a mechanism 
which gives different masses to sea quarks and valence quarks. This 
effect may be produced in a systematical way by adding to $\chi$PT 
explicit sea quarks as well as unphysical bosonic ghost quarks. 
The latter cancel exactly all effects of closed valence quark loops due 
to their different statistics, provided that their masses are identical 
to those of the valence quarks. Supersymmetric PQ$\chi$PT thus contains 
a set of fermionic as well as bosonic quarks. This leads to a
modification of the chiral symmetry group as given by 
Eq.~(\ref{chsym}), which in the PQ theory has the graded structure
\begin{equation}
G = SU(n_\mathrm{val}+n_\mathrm{sea} | n_\mathrm{val})_L
\times SU(n_\mathrm{val}+n_\mathrm{sea} | n_\mathrm{val})_R\,.
\end{equation}
The precise structure of $G$ is somewhat different as discussed
in~\cite{Sharpe1,Sharpe2}. This theory 
contains $n_\mathrm{val}$ valence and $n_\mathrm{val}$ ghost quarks,
as well as 
$n_\mathrm{sea}$ flavors of sea quarks. As the PQ theories include 
bosonic ghost quarks, they are not normal relativistic quantum field 
theories since they violate the spin-statistics theorem. However, under 
the assumption that the low-energy structure of such a theory can be 
described similarly to the case of normal QCD, one arrives at an 
effective low-energy theory in terms of a matrix $U$, according to
\begin{equation}
U \equiv \exp\left(i\sqrt{2}\,\Phi/\hat F\right)\,,
\end{equation}
where the matrix $\Phi$ now has a more complicated flavor structure
than $\phi$ in Eq.~(\ref{phiQCD}) because of the different types of 
quarks present. In terms of a sub-matrix notation for the flavor 
structure
\begin{equation}
\label{SubMatrix}
q_a\bar q_b =
\left(\begin{array}{ccc}
u_a \bar u_b & u_a \bar d_b & u_a \bar s_b \\ d_a \bar u_b &
d_a \bar d_b & d_a \bar s_b \\ s_a \bar u_b & s_a \bar d_b & s_a \bar s_b    
\end{array}\right)\,,
\end{equation}
the matrix $\Phi$ becomes
\begin{equation}
\label{SUSY_FieldMatrix}
\Phi =
\left(\begin{array}{ccc}
\Big[\;\;q_V\bar q_V\;\;\Big] & 
\Big[\;\;q_V\bar q_S\;\;\Big] &
\Big[\;\;q_V\bar q_B\;\;\Big] \\ \\ 
\Big[\;\;q_S\bar q_V\;\;\Big] &
\Big[\;\;q_S\bar q_S\;\;\Big] & 
\Big[\;\;q_S\bar q_B\;\;\Big]\\ \\
\Big[\;\;q_B\bar q_V\;\;\Big] & 
\Big[\;\;q_B\bar q_S\;\;\Big] &
\Big[\;\;q_B\bar q_B\;\;\Big]
\end{array}\right)\,,
\label{pqmatr}
\end{equation}
where the labels $V,S$ and $B$ stand for valence, sea and bosonic 
quarks, respectively. The size of each sub-matrix depends on the exact 
number of quark flavors used. 

The quarks $q_V$, $q_S$ and their respective antiquarks are fermions, 
while the quarks $q_B$ and their antiquarks are bosons. Thus a 
combination of a fermionic quark and a bosonic quark yield a fermionic 
(anticommuting) field, while a combination of two fermionic or two 
bosonic quarks result in a bosonic field. Each sub-matrix in 
Eq.~(\ref{pqmatr}) therefore consists of either fermionic of bosonic fields 
only. Although in principle arbitrary, the bosonic quarks are given the 
same masses as the corresponding valence quarks in order to cancel the 
contributions from closed valence quark loops. The above formalism is 
often referred to as supersymmetric PQ$\chi$PT, although this only 
refers to the graded group structure of the matrices used in the 
construction of the theory. Furthermore, the term 'bosonic' merely 
indicates that those quark terms are treated as commuting variables. 
They are still spin $1/2$ particles since the quark-antiquark pairs 
should build up particles with spin 0, representing the mesons of the 
theory. This violation of the spin-statistics theorem implies, as 
stated above, that in general the PQ theory is not a (fully causal) 
field theory.

In the version of PQ$\chi$PT used in this paper, the singlet $\Phi_0$ 
field has been dropped as discussed in Ref.~\cite{Sharpe2}. This 
PQ singlet is expected to be heavy due to the axial anomaly in the same 
way as the $\eta'$ which is not included in unquenched $\chi$PT. Having 
defined the supersymmetric field matrix $\Phi$ in Eq.~(\ref{pqmatr}), 
one then proceeds by constructing an effective Lagrangian for the PQ 
theory by requiring that this Lagrangian should be invariant under 
$SU(n_\mathrm{val}+n_\mathrm{sea} \vert n_\mathrm{val})_L\times
SU(n_\mathrm{val}+n_\mathrm{sea} \vert n_\mathrm{val})_R$ in the same 
way as the Lagrangian of $\chi$PT was required to be invariant under 
$SU(n_f)_L\times SU(n_f)_R$. This modification results in the same
Lagrangian structure as for $\chi$PT, provided that the traces of 
matrix products in those Lagrangians are replaced by supertraces. The 
supertraces are defined in terms of ordinary traces by
\begin{equation}
\Str \left(\begin{array}{cc} A & B \\ C & D \end{array}\right)
=\Tr\,A - \Tr\,D\,,
\end{equation}
where $A, B, C$ and $D$ denote block matrices. For example, the block 
$B$ corresponds to the $[ q_V \bar q_B ]$ and $[ q_S \bar q_B ]$ 
sectors of the field matrix in Eq.~(\ref{SUSY_FieldMatrix}) and 
contains anticommuting fields, while the block $D$ represents the $[ 
q_B \bar q_B ]$ sector of Eq.~(\ref{SUSY_FieldMatrix}). Note that the 
removal of $\Phi_0$ implies the assumption
\begin{equation}
\Str\,(\Phi) = 0\,.
\end{equation}
The entire external field formalism introduced by Gasser and 
Leutwyler~\cite{GL} for $\chi$PT can be generalized to include the 
extra degrees of freedom. In practice, external fields will be used in 
the valence sector only.

\subsection{Lagrangians and LECs}

We now proceed with the construction of the Lagrangians for the 
Goldstone bosons and organize them according to the Weinberg power 
counting scheme. The external fields included are the vector and 
axial-vector fields $v_\mu$ and $a_\mu$, as well as the scalar and 
pseudo-scalar external sources $s$ and $p$. These are the suitably 
generalized versions of those used in standard $\chi$PT~\cite{GL}.
Under a symmetry transformation $g_{L,R}\in SU(n_\mathrm{val}+
n_\mathrm{sea}|n_\mathrm{val})_{L,R}$ the fields transform as
\begin{eqnarray}
U &\to& g_R\,U\,g_L^\dagger,
\nonumber \\
\chi\equiv 2 \hat B \left(s+ip\right)&\to& g_R\,\chi\,g_L^\dagger,
\nonumber \\
l_\mu\equiv v_\mu-a_\mu&\to & g_L\,l_\mu\,g_L^\dagger
-i \partial_\mu\,g_L\,g_L^\dagger,
\nonumber \\
r_\mu\equiv v_\mu+a_\mu&\to & g_R\,r_\mu\,g_R^\dagger
-i \partial_\mu\,g_R\,g_R^\dagger.
\end{eqnarray}
However, one can also define another set of quantities which behave 
differently under chiral symmetry transformations, and which turn out 
to be more useful for the construction of the Lagrangians. If one 
considers 
\begin{equation}
u\equiv\exp\left(i\Phi/(\sqrt{2}\hat F)\right),
\end{equation}
then it is possible to find a matrix $h$ such that
\begin{equation}
u\to g_R\,u\,h^\dagger = h\,u\,g_L^\dagger\,.
\end{equation}
One can then proceed by constructing a set of quantities that 
transform under chiral symmetry as $X\to h\,X\,h^\dagger$. Such 
quantities are
\begin{eqnarray}
u_\mu &=& i\left\{
u^\dagger(\partial_\mu-i r_\mu)\,u -
u\,(\partial_\mu-i l_\mu)\,u^\dagger\right\},
\nonumber \\
\chi_\pm &=& u^\dagger\chi\,u^\dagger\pm u\,\chi^\dagger\,u,
\nonumber \\
f_\pm^{\mu\nu} &=& u\,F_L^{\mu\nu}\,u^\dagger\pm 
u^\dagger F_R^{\mu\nu}\,u,
\label{uquant}
\end{eqnarray}
where $F_L$ and $F_R$ denote the field strengths of the external fields 
$l$ and $r$, such that $F_L^{\mu\nu} = 
\partial^\mu l^\nu-\partial^\nu l^\mu-i\left[l^\mu,l^\nu\right]$. 
$F_R^{\mu\nu}$ is then defined analogously in terms of $r$. It should 
be noted that the Minkowski convention is used throughout this paper 
instead of the Euclidean one needed in lattice QCD, in order to 
maintain compatibility with the existing literature in NNLO $\chi$PT. 
In terms of the quantities defined in Eq.~(\ref{uquant}) the lowest 
order, or $\mathcal{O}(p^2)$ Lagrangian is given by
\begin{equation}
{\cal L}_2 = \frac{\hat F^2}{4} \langle u^\mu u_\mu + \chi_+\rangle,
\end{equation}
where the shorthand notation
\begin{equation}
\langle A\rangle = \Str\,A
\end{equation}
has been introduced. At this level there are two parameters $\hat F$ 
and $\hat B$, which depend on the number of sea quark flavors in the PQ 
theory. In the power counting scheme, each derivative or factor of 
$l_\mu,r_\mu$ counts as one, and each factor of $s,p$ as two powers of 
the momentum $p$. The order in $p$ is indicated by the subscript of the 
Lagrangian. The $\mathcal{O}(p^4)$ Lagrangian has the generic 
form~\cite{GL}
\begin{eqnarray}
\label{L4}
{\cal L}_4 &=& \sum_{i=0}^{12} {\hat L}_i X_i + \mbox{contact terms}
\nn  
&=& {\hat L}_0\,\lgl u^\mu u^\nu u_\mu u_\nu \rgl 
+{\hat L}_1\,\lgl  u^\mu u_\mu \rgl^2 
+{\hat L}_2\,\lgl u^\mu u^\nu \rgl \lgl u_\mu u_\nu \rgl
\nn
&+&{\hat L}_3\,\lgl (u^\mu u_\mu)^2 \rgl
+ {\hat L}_4\,\lgl u^\mu u_\mu \rgl \lgl \chi_+\rgl 
+ {\hat L}_5\,\lgl u^\mu u_\mu \chi_+ \rgl 
\nn
&+& {\hat L}_6\,\lgl \chi_+ \rgl^2 
+ {\hat L}_7\,\lgl \chi_- \rgl^2
+ \frac{{\hat L}_8}{2}\,\lgl \chi_+^2 + \chi_-^2 \rgl 
\nn
&-& i{\hat L}_9\,\lgl f_+^{\mu\nu} u_\mu u_\nu \rgl 
+ \frac{{\hat L}_{10}}{4}\,\lgl f_+^2 - f_-^2 \rgl 
\nn
&+& i{\hat L}_{11}\,\left\lgl \hat\chi_-\left( \nabla^\mu u_\mu - 
\frac{i}{2} \hat\chi_- \right) \right\rgl
\nn 
&+& {\hat L}_{12}\,\left\lgl \left( \nabla^\mu u_\mu - 
\frac{i}{2} \hat\chi_- \right)^2 \right\rgl 
\nn 
&+& \hat H_1\,\lgl F_L^2+F_R^2\rgl 
+ \hat H_2\,\lgl\chi\chi^\dagger\rgl,
\end{eqnarray}
where the definition $\hat\chi_- \equiv \chi_- -\lgl 
\chi_-\rgl/n_\mathrm{sea}$ has been applied. Furthermore, the lowest 
order equation of motion is given by
\begin{equation}
\nabla^\mu u_\mu - \frac{i}{2}\hat\chi_- = 0.
\end{equation}
The Lagrangian of Eq.~(\ref{L4}) contains three types of terms. Of 
these, the terms proportional to $\hat{H_i}$ are contact terms which 
contain external fields only. Thus they are not relevant for low-energy 
phenomenology, but they are necessary for the computation of 
operator expectation values. Their values are determined by the precise 
definition used for the QCD currents, and they are conventionally 
labeled $h_i^r$ and $H_i^r$ for unquenched $\chi$PT with $n_f = 2$ and 
$n_f = 3$ quark flavors, respectively. The terms that depend on $\hat 
L_{11}$ and $\hat L_{12}$ are proportional to the equations of motion, 
and as such they can always be reabsorbed in higher order LEC:s, see 
Ref.~\cite{BCE2} for a full proof. The Lagrangian at $\mathcal{O}(p^6)$ 
is also known~\cite{BCE1} and can be written in the form
\begin{equation}
{\cal L}_6 = \sum_{i=1}^{112} \hat K_i Y_i + \mbox{contact terms},
\end{equation}
where the form of the 
operators $Y_i$ can be found in Refs.~\cite{BCE1,BCE2}. The PQ$\chi$PT
operators may again be obtained from the results of Refs.~\cite{GL,BCE1,BCE2} 
if all traces are replaced by supertraces. All manipulations and 
identities used there to decrease the number of terms in the 
$\mathcal{O}(p^6)$ Lagrangian, such as integration by parts and usage 
of the equation of motion, remain valid under this modification. This 
is why it is possible to apply the $n_f$ flavor results of 
Refs.~\cite{GL,BCE1,BCE2} to PQ$\chi$PT by using the appropriate graded 
matrices and supertraces.

\begin{table}[b]
\begin{center}
\caption{The different sets of LEC:s for unquenched ($n_f$) and 
partially quenched ($n_\mathrm{sea}$) $\chi$PT. The number of 
physically relevant terms in each set is indicated by $n_\mathrm{ph}$,
and the number of contact terms by $n_\mathrm{ct}$. The relationships 
between the various LEC:s are discussed in the text.}
\vspace{.3cm}
\begin{tabular}{c||c|c|c|c|c}
 & \,\,$\chi$PT\,\, & \,\,$\chi$PT\,\, & $\chi$PT 
& \,\,PQ$\chi$PT\,\, & \,\,PQ$\chi$PT\,\, \\[1mm]
$n_f, n_\mathrm{sea}$ &  2 & 3 & $n$ & 2 & 3\\ 
 & & & & & \vspace{-.2cm} \\
\hline\hline 
 & & & & & \vspace{-.2cm} \\
LO & $F,B$ & $F_0,B_0$ & $F_0^{(n)},B_0^{(n)}$ & $F,B$ & $F_0,B_0$\\
 & & & & & \vspace{-.2cm} \\ \hline
 & & & & & \vspace{-.2cm} \\
NLO & $l_i^r$ & $L_i^r$ &$ L_i^{r(n)}$
 &$ L_i^{r(2pq)}$ &$ L_i^{r(3pq)}$\\
$n_\mathrm{ph}+n_\mathrm{ct}$ 
& 7\,+\,3 & 10\,+\,2 & 11\,+\,2 & 11\,+\,2 & 11\,+\,2 \\
 & & & & & \vspace{-.2cm} \\ \hline
 & & & & & \vspace{-.2cm} \\
\,NNLO\, & $c_i^r$ & $C_i^r$ & $K_i^{r(n)}$ & $K_i^{r(2pq)}$ & 
$K_i^{r(3pq)}$\\
$n_\mathrm{ph}+n_\mathrm{ct}$ 
& \,53\,+\,4\, & \,90\,+\,4\, & \,112\,+\,3\, & \,112\,+\,3\, 
& \,112\,+\,3\,
\end{tabular}
\label{tab:LEC}
\end{center}
\end{table}

The divergences of PQ$\chi$PT follow directly from those of $n_f$ 
flavor $\chi$PT calculated in Ref.~\cite{BCE2}, provided that $n_f$ is 
set equal to the number of sea-quark flavors. This is so since all the 
manipulations in Ref.~\cite{BCE2} for $n_f$ flavors remain valid when 
traces are replaced with supertraces. Alternatively, this derivation of 
the Lagrangians of PQ$\chi$PT can be argued for with the Replica method 
of Ref.~\cite{replica}. The renormalization procedure is thus identical 
with the one of $\chi$PT also in the partially quenched case. An extensive 
discussion of this procedure can be found in Ref.~\cite{BCE2} and 
references therein. The finite parts of the ${\hat L}_i$ and 
${\hat K}_i$ for the different theories considered are summarized in 
Table~\ref{tab:LEC}. For the case of $\chi$PT with two or three flavors 
the generic Lagrangians referred to above can be further simplified 
using the Cayley-Hamilton relations for $2\times2$ or $3\times3$ 
matrices. These allow for a further reduction of the number of 
operators, resulting in the different sets of constants shown in 
Table~\ref{tab:LEC}. The superscript $r$ for the NLO and NNLO LEC:s 
indicates that they are the renormalized versions as defined in 
Ref.~\cite{BCE2}. In particular, the NNLO coefficients are given by the 
finite parts of the $\hat K_i$, multiplied by $\hat F^2$ 
in order to obtain dimensionless quantities.

Consider next the LEC:s of PQ$\chi$PT with $n$ flavors of sea-quarks, 
which to lowest order are identical to those of unquenched $n$ flavor 
$\chi$PT. At NLO, they are $L^{r(npq)}_0$ through $L^{r(npq)}_{12}$, 
but as argued above the two terms $L_{11}^{r(npq)}$ and 
$L_{12}^{r(npq)}$ can be removed using field redefinitions or the 
equations of motion~\cite{BCE1,BCE2}. For three flavors of sea-quarks, 
the standard choice is
\begin{equation}
L_{11}^{r(3pq)} = L_{12}^{r(3pq)} = 0,
\end{equation}
whereas the unquenched two-flavor Lagrangian as defined in the first
paper of Ref.~\cite{GL} differs by an $L_{11}^r$ type term. This makes 
no difference at NLO since that term does not contribute, but in order 
to get the correct correspondence at NNLO between the $l_i^r,c_i^r$ and 
the $L_i^{r(2pq)},K_i^{r(2pq)}$ one should take
\begin{equation}
L^{r(2pq)}_{11} = -l^r_4 / 4, \quad L^{r(2pq)}_{12} = 0.
\end{equation}
For the case of three sea-quarks, the $L_i^r$ are simple linear 
combinations of the $L_i^{r(3pq)}$, which have been given in 
Refs.~\cite{BL1,BCE1}. Explicitly,
\begin{eqnarray}
L_1^r &=& L_1^{r(3pq)} + L_0^{r(3pq)}/2, \nonumber \\
L_2^r &=& L_2^{r(3pq)} + L_0^{r(3pq)}, \nonumber \\
L_3^r &=& L_3^{r(3pq)} - 2\,L_0^{r(3pq)},
\end{eqnarray}
and
\begin{eqnarray}
L_{4\ldots 12}^r &=& L_{4\ldots 12}^{r(3pq)},
\end{eqnarray}
which indicates, as also discussed in Ref.~\cite{Sharpevandewater}, 
that $L_0^{r(3pq)}$ is an independent LEC in partially quenched 
$\chi$PT with $n_\mathrm{sea} = 3$, but in unquenched $\chi$PT it can 
be absorbed into $L_1^r$, $L_2^r$ and $L_3^r$. One can therefore 
conclude that the numerical value of $L_0^{r(3pq)}$ cannot be 
determined by experiment, but is accessible via partially 
quenched Lattice QCD simulations. Similarly at NNLO, the 
$C_i^r$ of unquenched $\chi$PT are linear combinations of the 
$K_i^{r(3pq)}$, and the corresponding relations have been derived in 
Ref.~\cite{BCE1}. 

The expressions given in the following sections of this paper concern 
PQ$\chi$PT with $n_\mathrm{sea} = 3$ only, and therefore the 
superscripts $(3pq)$ of the $L_i^{r(3pq)}$ and the $K_i^{r(3pq)}$ have
been suppressed in most of the remaining equations. 

\subsection{Quark Masses and Propagators}

The version of PQ$\chi$PT considered in this paper has three flavors of 
valence quarks ($n_\mathrm{val} = 3$), three flavors of sea quarks 
($n_\mathrm{sea} = 3$) and consequently three flavors of bosonic 
'ghost' quarks. For simplicity, the different quark masses are 
identified in the following calculations by the flavor indices 
$i=1,\ldots,9$, rather than by the indices $u,d,s$ and $V,S,B$ of 
Eqs.~(\ref{SubMatrix}) and~(\ref{SUSY_FieldMatrix}). The results are 
expressed in terms of the quark masses $m_q$ via the quantities 
$\chi_i=2B_0\,m_{qi}$ such that $\chi_1,\chi_2,\chi_3$, belong to the 
valence sector, $\chi_4,\chi_5,\chi_6$ to the sea sector, and 
$\chi_7,\chi_8,\chi_9$ to the ghost sector. The latter ones do not 
appear in the results since the ghost quark masses are always set equal 
to the masses of the corresponding valence quarks, such that 
$\chi_7=\chi_1, \chi_8=\chi_2$ and $\chi_9=\chi_3$. As discussed above, 
this is necessary in order to cancel the disconnected valence quark 
loops, which are replaced by the shifted mass contributions that 
involve the sea quarks.

Furthermore, the quantities $d_{\mathrm{val}}$ and $d_{\mathrm{sea}}$ 
are used to indicate the number of nondegenerate quark masses in each 
sector. Thus $d_{\mathrm{val}}=1$ has all valence quark masses 
degenerate, such that $\chi_1 = \chi_2 = \chi_3$, while 
$d_{\mathrm{val}}=2$ indicates that $\chi_1 = \chi_2 \ne \chi_3$. 
Finally, in the case of $d_{\mathrm{val}}=3$, which is not needed in 
the present calculations, $\chi_1 \ne \chi_2 \ne \chi_3$. In a similar 
way, for the sea quark masses $d_{\mathrm{sea}}=1$ implies that $\chi_4 
= \chi_5 = \chi_6$, while for $d_{\mathrm{sea}}=2$ one has $\chi_4 = 
\chi_5 \ne \chi_6$, and  for $d_{\mathrm{sea}}=3$ all the sea 
quark masses are nondegenerate, such that $\chi_4 \ne \chi_5 \ne 
\chi_6$. The notation $(d_{\mathrm{val}}+d_{\mathrm{sea}})$ is often 
used in this paper to indicate what degree of degeneracy in the quark 
masses is being considered. For example, the expression for the 
pseudoscalar meson mass for $n_\mathrm{val}=2$ and $n_\mathrm{sea}=3$ 
in the (1+2)~case depends on one (degenerate) valence quark mass and 
two distinct (nondegenerate) sea quark masses.

The meson propagators for the supersymmetric formulation of PQ$\chi$PT 
have been calculated in Ref.~\cite{Sharpe2}, and they correspond to the 
limit $m_0\to\infty$ of the results in Ref.~\cite{BG1}, where $m_0$ is 
the mass parameter of the singlet field $\Phi_0$. Again, for 
calculational reasons, the results of Ref.~\cite{Sharpe2} have been 
translated from the Euclidean formalism into Minkowski space. In 
general, three distinct types of propagators are encountered in the 
calculations. The simplest one is the propagator of a charged, or 
flavor-off-diagonal meson, connecting the meson field $q_i \bar q_j$ 
with its antifield $q_j \bar q_i$. It is given by~\cite{Sharpe2}
\begin{eqnarray}
-i\,G_{ij}^c (k) &=& 
\frac{\epsilon_j}{k^2 - \chi_{ij} + i\varepsilon}\quad (i \neq j)\,.
\label{propc}
\end{eqnarray}   
where the combination of quark masses $\chi_{ij} \equiv (\chi_i + 
\chi_j) / 2$ corresponds to the lowest order meson masses, and the 
signature vector $\epsilon_j$ is defined as $+1$ for the flavor indices 
of the $n_\mathrm{val} + n_\mathrm{sea}$ fermionic quarks, and as $-1$ 
for the flavor indices of the $n_\mathrm{val}$ bosonic ghost quarks. In 
the present calculation, with the number of valence and sea quarks as 
given above, $\epsilon_j$ assumes the values
\begin{equation}
\epsilon_j=\left\{
  \begin{array}{cl}
    +1 &  \mathrm{for} \;\;\;j=1,\dots,6\\
    -1 & \mathrm{for}\;\;\;j=7,8,9\,.
  \end{array}
\right.
\end{equation}
The propagator of a neutral, or flavor-diagonal meson, has a more 
complicated structure, since it connects mesons with different flavor 
indices as well. A propagator which connects the meson fields $q_i \bar 
q_i$ and $q_j \bar q_j$ is written in 
the form~\cite{Sharpe2} 
\begin{eqnarray}
G_{ij}^n (k) &=& G_{ij}^c (k)\,\delta_{ij} 
- G_{ij}^q (k) / n_\mathrm{sea},
\label{propn}
\end{eqnarray}
where the second term $G_{ij}^q (k)$ is expressed in 
terms of a sum of single and double poles~\cite{Sharpe1,Sharpe2}. There 
are two distinct cases, as the double pole only appears if 
either $i=j$ or $\chi_i=\chi_j$. For $i\ne j$ and $\chi_i\ne \chi_j$, 
the propagator $G_{ij}^q$ is given by
\begin{eqnarray}
-i\,G_{ij}^q (k) &=& \frac{R^{i}_{j\pi\eta}}{k^2 - \chi_i + i\varepsilon}
+ \frac{R^{j}_{i\pi\eta}}{k^2 - \chi_j + i\varepsilon} 
\nonumber \\ 
&+& \frac{R_{\eta ij}^\pi} {k^2 - \chi_\pi + i\varepsilon} 
+ \frac{R_{\pi ij}^\eta}{k^2 - \chi_\eta + i\varepsilon},
\label{npropij} 
\end{eqnarray}
where the factors $R$ are referred to collectively as the single pole 
residues. For $i=j$ or $\chi_i=\chi_j$, the residues of the first two 
terms in eq.~(\ref{npropij}) become singular and the whole expression 
should be replaced by
\begin{eqnarray}
-i\,G_{ij}^q (k) &=& \frac{R^d_i}{(k^2 - \chi_i + i\varepsilon)^2} 
+ \frac{R^c_i}{k^2 - \chi_i + i\varepsilon} 
\nonumber \\
&+& \frac{R_{\eta ii}^\pi} {k^2 - \chi_\pi + i\varepsilon} 
+ \frac{R_{\pi ii}^\eta}{k^2 - \chi_\eta + i\varepsilon},
\label{npropij2} 
\end{eqnarray}
where the residue of the double pole is denoted $R_i^d$. Note also the 
appearance of an auxiliary residue $R_i^c$ in eq.~(\ref{npropij2}). All 
expressions for the propagator $G_{ij}^q$ depend on the lowest order 
neutral pion and eta meson masses in the sea quark sector, which are 
denoted by $\chi_\pi$ and $\chi_\eta$. For $d_{\mathrm{sea}}=3$ these 
are determined by the relations
\begin{eqnarray}
\label{neutral_sea_masses}
\chi_\pi+\chi_\eta &=& \frac{2}{3}\left(\chi_4+\chi_5+\chi_6\right),
\nonumber\\
\chi_\pi \chi_\eta &=& \frac{1}{3}
\left(\chi_4\chi_5+\chi_5\chi_6+\chi_4\chi_6\right),
\end{eqnarray}
which have no polynomial solution, but for $d_{\mathrm{sea}}=2$ one 
has $\chi_\pi=\chi_4$ and $\chi_\eta=1/3\chi_4+2/3\chi_6$. The 
propagators for $d_{\mathrm{sea}}=2$ can then be obtained from 
Eqs.~(\ref{npropij}) and~(\ref{npropij2}) by taking the 
appropriate limits. In particular, the terms with a pole in 
$\chi_\pi$ are no longer present. For $d_{\mathrm{sea}}=1$, the fact 
that $\chi_\pi=\chi_\eta=\chi_4$ gives rise to several further 
simplifications, and in that case the terms with a pole in $\chi_\eta$ 
vanish as well.

The residues $R$ of the neutral meson propagators in PQ$\chi$PT are in 
general rational functions of the sea and valence quark masses which 
can be expressed in terms of the more general quantities
$R^z_{a\ldots b}$ defined by  
\begin{eqnarray}
R^z_{ab} &=& \chi_a - \chi_b, \nonumber \\
R^z_{abc} &=& \frac{\chi_a - \chi_b}{\chi_a - \chi_c}, \nonumber \\ 
R^z_{abcd} &=& \frac{(\chi_a - \chi_b)(\chi_a - \chi_c)}
{\chi_a - \chi_d}, \nonumber \\
R^z_{abcdefg} &=& \frac{(\chi_a - \chi_b)(\chi_a - \chi_c)(\chi_a - \chi_d)}
{(\chi_a - \chi_e)(\chi_a - \chi_f)(\chi_a - \chi_g)},
\label{RSfunc}
\end{eqnarray}
and so on. Note that $R^z_{a\ldots b}$ has the same dimension as 
$\chi_i$ for an even number of indices and is dimensionless for an odd 
number of indices. For the case of $d_{\mathrm{sea}} = 3$, the residues 
generated by Eqs.~(\ref{npropij}) and~(\ref{npropij2}) are
\begin{eqnarray}
R_{jkl}^{i} &=& R^z_{i456jkl}, \nonumber \\
R_{i}^{d} &=& R^z_{i456\pi\eta}, \nonumber \\
R_{i}^{c} &=& R^i_{4\pi\eta} + R^i_{5\pi\eta} + R^i_{6\pi\eta}
          - R^i_{\pi\eta\eta} - R^i_{\pi\pi\eta}.
\end{eqnarray}
From these definitions, it is apparent that residues of the type 
$R_i^d$ or $R_{j\pi\eta}^{i}$ vanish if $i$ is a sea-quark index. 
Thus the propagators in the sea-quark sector of PQ$\chi$PT contain 
no double poles as expected, since the origin of the double poles lies 
in the quenching of the valence quark loops. For the case of 
$d_{\mathrm{sea}} = 2$, due to the cancellations in the sea-quark 
sector (as discussed above) the residues simplify to
\begin{eqnarray}
R_{jk}^{i} &=& R^z_{i46jk}, \nonumber \\
R_{i}^{d} &=& R^z_{i46\eta},\nonumber \\
R_{i}^{c} &=& R^i_{4\eta} + R^i_{6\eta} - R^i_{\eta\eta},
\end{eqnarray}
so that the index $\pi$ no longer appears. For the case of 
$d_{\mathrm{sea}} = 1$, all residues associated with the sea quark 
sector have reduced to numbers. Some nontrivial residues can still 
appear if the valence quarks are nondegenerate. These are
\begin{eqnarray}
R_{j}^{i} &=& R^z_{i4j}, \nonumber \\
R_{i}^{d} &=& R^z_{i4}\,,
\end{eqnarray}
where the double-pole residue has been retained mainly for notational 
consistency. Finally, it should be noted that if the sea quark masses 
are set equal to the valence quark masses, the propagator residues of 
PQ$\chi$PT reduce so that the $\pi^0$ and $\eta$ 
meson propagators of unquenched $\chi$PT are recovered.

Typically, a direct NNLO calculation with the propagators of 
Eqs.~(\ref{npropij}) and~(\ref{npropij2}) produces a large number of 
redundant residues in the output. This problem is especially 
troublesome for the larger values of $d_{\mathrm{sea}}$ and 
$d_{\mathrm{val}}$. It is thus useful to note that the various residues 
$R$ satisfy a large number of algebraic relations, which provide an 
efficient, albeit tedious, way to simplify and compress the end results 
of the NNLO calculations.

\subsection{Loop Integrals at NNLO}

The expressions for the NNLO masses and decay constants of the charged 
pseudoscalar mesons in PQ$\chi$PT depend on several one- and two-loop 
integrals. After regularization and renormalization has been carried 
out, the finite contributions from these integrals are written in terms 
of the functions $\bar{A},\bar{B}$ and $\bar{C}$, which are defined as
\begin{eqnarray}
\bar A(\chi) &=& -\pi_{16}\, \chi \log(\chi/\mu^2), \nonumber \\
\bar B(\chi_i,\chi_j;0) &=& -\pi_{16}\, \frac{\chi_i\log(\chi_i/\mu^2) 
- \chi_j\log(\chi_j/\mu^2)}{\chi_i - \chi_j},
\nonumber\\
\bar C(\chi,\chi,\chi;0) &=& -\pi_{16}/(2 \chi)\, ,
\end{eqnarray}  
where $\mu$ denotes the renormalization scale and $\pi_{16} = 1/(16 
\pi^2)$. These integrals are often referred to as chiral logarithms, 
although the integral $\bar C$ does not contain any logarithm when all 
three arguments are equal. In the often encountered limit 
$\chi_i=\chi_j$, the expression for $\bar B$ reduces to
\begin{eqnarray}
\bar B(\chi,\chi;0) &=& -\pi_{16}\left(1 + \log(\chi/\mu^2) \right).
\end{eqnarray}
The functions described above are in principle sufficient to express 
all one-loop integrals encountered in the NNLO calculations, but the 
introduction of further combinations of integrals is desirable in order 
to reduce the size and complexity of the results. For this purpose, the 
following three combinations of integrals have been introduced,
\begin{eqnarray}
\bar A(\chi;\varepsilon) &=& \bar A(\chi)^2 / (2\pi_{16}\,\chi)
\nonumber \\
&+& \pi_{16}\,\chi\,(\pi^2/12 + 1/2), \nonumber \\
\bar B(\chi,\chi;0,\varepsilon) &=& 
\bar A(\chi)\bar B(\chi,\chi;0) / (\pi_{16}\,\chi)
\nonumber \\
&-& \bar A(\chi)^2 / (2\pi_{16}\,\chi^2) 
\nonumber \\
&+& \pi_{16}\,(\pi^2/12 + 1/2),
\nonumber\\
\bar B(\chi_i,\chi_j;0,k) &=& \chi_i \bar B(\chi_i,\chi_j;0) 
+ \bar A(\chi_j),
\end{eqnarray}
of which the first two expressions are naturally generated by the 
dimensional regularization procedure, whereas the third one is useful 
since it is symmetric under the interchange of $\chi_i$ and $\chi_j$.

The NNLO calculation of pseudoscalar meson masses and decay constants 
also introduces a number of genuine, nonfactorizable two-loop 
integrals, which can be evaluated using a generalization of the methods 
described in Ref.~\cite{ABT1}. The two-loop integrals encountered have 
the following general structure,
\begin{eqnarray}
\langle\langle X \rangle\rangle&=&\frac{1}{i^2}\int \frac{d^d q}{(2\pi)^d}
\frac{d^d r}{(2\pi)^d}\times
\label{Hint} \\
&& \frac{X}{(q^2-\chi_1)^a(r^2-\chi_2)^b((q+r-p)^2-\chi_3)^c}\,,\nonumber
\end{eqnarray}   
where $a,b,c=1$ or $2$, and $X$ represents the different combinations 
of momentum factors $q$ and $r$ that can occur. The integrals thus 
generated by Eq.~(\ref{Hint}) are sometimes referred to as sunset 
integrals, and they can be expressed in terms of the $H$ functions 
according to the definitions in Ref.~\cite{ABT1}. The $H$ functions 
satisfy a number of integral relations such that only $H$, $H_1$ 
and $H_{21}$ are required to express all sunset integrals encountered. 
Furthermore, after regularization of the $H$ functions, only the finite 
parts denoted $H^{F}$, $H_1^{F}$ and $H_{21}^{F}$ appear in the end 
results. In the case of the decay constants, sunset integrals 
differentiated with respect to $p^2$ are also required, and are denoted 
by the primed quantities $H^{F'}$, $H_1^{F'}$ and $H_{21}^{F'}$.

\begin{table}[b]
\begin{center}
\caption{Overview of the notation for the possible configurations of 
double poles in the $H$ functions generated by Eq.~(\ref{Hint}) in 
PQ$\chi$PT. Redundant configurations are given in parentheses.}
\vspace{.3cm}
\begin{tabular}{c|c c c}
 & \quad $ a $ \quad & \quad $ b $ \quad & \quad $ c $ \quad \\ & 
\vspace{-.2cm} \\
\hline\hline
& \vspace{-.2cm} \\
$n=1$ \quad & \quad 1 \quad & \quad 1 & \quad 1\quad\\
\vspace{-.2cm} &&& \\ \hline \vspace{-.2cm} &&& \\
$n=2$ \quad & \quad 2 \quad & \quad 1 & \quad 1\quad\\
$n=3$ \quad & \quad 1 \quad & \quad 2 & \quad 1\quad\\ 
($n=4$) \quad & \quad 1 \quad & \quad 1 & \quad 2\quad\\
\vspace{-.2cm} &&& \\ \hline \vspace{-.2cm} &&& \\
$n=5$ \quad & \quad 2 \quad & \quad 2 & \quad 1\quad\\
($n=6$) \quad & \quad 2 \quad & \quad 1 & \quad 2\quad\\
$n=7$ \quad & \quad 1 \quad & \quad 2 & \quad 2\quad\\
\vspace{-.2cm} &&& \\ \hline \vspace{-.2cm} &&& \\
$n=8$ \quad & \quad 2 \quad & \quad 2 & \quad 2\quad
\end{tabular}
\label{tabint}
\end{center}
\end{table}

However, the appearance of double poles in the neutral meson 
propagators of Eqs.~(\ref{npropij}) and~(\ref{npropij2}) leads to 
a significant complication in Eq.~(\ref{Hint}), as cases with 
$a,b,c\ne1$ will then show up. This added layer of complexity is 
accounted for by an extra integer argument $n$, which indicates the 
pole configuration of Eq.~(\ref{Hint}) according to Table~\ref{tabint}. 
The finite parts of the $H$ functions are thus generalized as
\begin{equation}
H^F(\chi_i,\chi_j,\chi_k;p^2) \rightarrow 
H^F(n,\chi_i,\chi_j,\chi_k;p^2),
\end{equation} 
and similarly for the $H_1^F$ and $H_{21}^F$. In principle, eight 
different configurations can show up, and the value of $n$ for each one 
of them is given in Table~\ref{tabint}. However, this expanded set of  
$H$ functions obeys a generalization of the symmetries (under 
the exchange of mass arguments) discussed in Ref.~\cite{ABT1}, and thus 
some of the pole configurations turn out to be redundant, such that 
only $n=1,2,3,5$ and $7$ are required for a complete NNLO calculation 
of the pseudoscalar meson masses and decay constants. The case 
with $n=8$ would however appear for a NNLO calculation of the 
coefficient of the double pole in the neutral meson propagator. Such a 
calculation has not yet been performed, but it should be noted that a 
NLO calculation of that quantity has been published in 
Ref.~\cite{Sharpe1}.

The explicit expressions for the expanded set of $H$ functions with 
$n\ne 1$ can be obtained by differentiation with respect to the 
masses of the expressions for $n=1$ given in Ref.~\cite{ABT1}. It 
should also be noted that the finite contributions from the set of $H$ 
functions contain integrals which have to be evaluated by numerical 
integration, and these are again generalizations of the integrals for 
$n=1$ in Ref.~\cite{ABT1}.

\subsection{Notation for NNLO Results}

The length of the final NNLO expressions presents a problem. One major 
culprit has already been identified in the discussion of the PQ 
propagator residues, but even after that problem is dealt with, the 
size of the PQ expressions far exceeds that of the analogous ones in 
unquenched $\chi$PT. One reason for this is that the NNLO results in 
PQ$\chi$PT are highly symmetric under the interchange of quark masses, 
both in the sea and valence quark sectors, and thus contain a lot of 
unnecessary repetition. This observation suggests that the NNLO 
expressions could be efficiently compressed by summation over sea and 
valence quark indices.

The summation conventions in the sea-quark sector are implemented 
through two new indices $s$ and $t$, which can appear as indices of 
explicit quark masses $\chi$, as well as among the indices of the 
residue functions $R$ or the predefined combinations thereof. These 
sea-quark summation indices should be interpreted as follows: If an 
index $t$ is present once or several times, then there will always be 
an occurrence of the $s$ index as well, and the entire term is then to 
be summed over all pairs of different sea quark indices. If the index 
$s$ is present but $t$ is not, then the entire term is to be summed 
over all sea quark indices. Elementary examples are
\begin{eqnarray}
\chi_s &=& \sum_{i\,=\,4,5,6} \chi_i, \nonumber \\
\chi_{st} &=& \hspace{-1.2cm} \sum_{\hspace{1cm}
(ij)\,=\,(45),(46),(56)} \hspace{-1.22cm} \chi_{ij}.
\end{eqnarray}
Thus all the terms in the end result, where the dependence on the 
sea-quark masses is expressed in terms of the indices $s$ and $t$, 
explicitly have the (required) symmetry under interchange of sea-quark 
masses. It should be noted that the summation is over all three 
sea-quark flavors irrespective of the value of $d_{\mathrm{sea}}$.

Further symmetries exist which can also be used to advantage in the 
compactification of the NNLO results. Firstly, the valence quark sector 
has a symmetry under the interchange of 
valence quark masses. This symmetry has been implemented by 
introduction of the summation indices $p$ and $q$. In this paper, they 
are necessary only for the expressions with $d_{\mathrm{val}}=2$, and 
then always occur for the valence quark masses $\chi_1$ and $\chi_3$. 
If the index $q$ is present, there will always be an index $p$ and the 
resulting sum is over the pairs $(p,q) = (1,3)$ and $(p,q) = (3,1)$. If 
only $p$ is present, the sum is over the indices $1$ and~$3$. As an 
example, consider
\begin{eqnarray}
\bar A(\chi_p)\,R^p_{q\eta}\,\chi_p &=&
\bar A(\chi_1)\,R^1_{3\eta}\,\chi_1 \:+\: [1\leftrightarrow 3],
\end{eqnarray}
which demonstrates that any contribution written in terms of the 
$(p,q)$ notation is symmetric (as required) under the interchange of 
the valence quark masses $\chi_1$ and $\chi_3$. Secondly, for 
$d_{\mathrm{sea}}=3$ the sea-quark sector exhibits an additional 
symmetry, under the interchange of the lowest order neutral meson 
masses $\chi_\pi$ and $\chi_\eta$. This symmetry has been implemented 
by the indices $m$ and $n$. If the index $m$ is present, there will 
always be an index $n$ and the corresponding sum is over the pairs 
$(m,n) = (\pi,\eta)$ and $(m,n) = (\eta,\pi)$. For example
\begin{eqnarray}
\bar A(\chi_m)\,R^m_{n11}\,\chi_m &=&
\bar A(\chi_\eta)\,R^{\eta}_{\pi 11}\,\chi_\eta \:+\: 
[\eta\leftrightarrow \pi].
\end{eqnarray}
Similar to the earlier cases, if only the index $n$ is present, 
then the term is to be summed over the $\chi_\pi$ and $\chi_\eta$ 
masses.

The summation techniques described above already incorporate 
most of the recurring combinations of terms in the end results, which 
has the added benefit of avoiding the introduction of an 
unreasonable amount of specialized notation. However, certain 
combinations of quark masses $\chi$ and propagator residues $R$ appear 
throughout the results and have therefore been given dedicated names. 
As an example, combinations of the type
\begin{eqnarray}
\bar\chi_g &=& \frac{1}{3}\sum_{i\,=\,4,5,6} \chi_i^g, \nonumber \\
\bar\chi^{a}_{bg} &=& \frac{1}{3}\sum_{i\,=\,4,5,6}
R^a_{bii}\,\chi_i^g,
\label{average}
\end{eqnarray}
where the index $g$ indicates the power of the sea-quark masses 
averaged, appear throughout the expressions. The corresponding ones for 
$d_{\mathrm{sea}}=2$ can be obtained by setting $R^a_{bii} \to 
R^a_{bi}$. Again, it should be kept in mind that the summation always 
runs over all three quark flavors, regardless of the value of 
$d_{\mathrm{sea}}$. The calculation of the pseudoscalar meson decay 
constant of Ref.~\cite{BL1} made use of many more complicated averages 
of the above type, with products of up to three residues. However, most 
of the quantities $\bar\chi$ which involve the residues $R$ can be 
reexpressed in terms of simpler quantities, such that e.g. 
$\bar\chi^{\pi}_{\eta 0} = -1$, $\bar\chi^{\pi}_{\eta 1} = -\chi_\pi$
and $\bar\chi^{\eta}_{\pi 1} = -\chi_\eta$. The use of the $\bar\chi$ 
notation has therefore been discontinued, and the only quantities of 
the type given in Eq.~(\ref{average}) that are used in this paper 
are $\bar\chi_1$ and $\bar\chi_2$. The other named combinations 
consist of sums of products of quark masses and propagator residues. 
For $d_{\mathrm{sea}}=3$, these are
\begin{eqnarray}
R^v_{ijkl} &=& R^i_{jkk} + R^i_{jll} - 2 R^i_{jkl}\,,
\nonumber \\
R^u_{ijkl} &=& R^i_{\pi\eta k} R^j_{\pi\eta k} 
- R^i_{\pi\eta k} R^j_{\pi\eta l} \nonumber \\
 &-& R^i_{\pi\eta l} R^j_{\pi\eta k} 
+ R^i_{\pi\eta l} R^j_{\pi\eta l}, \nonumber \\
R^{um}_{ijkl} &=& R^i_{mm'k} R^m_{m'jk} 
- R^i_{mm'k} R^m_{m'jl} \nonumber \\
 &-& R^i_{mm'l} R^m_{m'jk} 
+ R^i_{mm'l} R^m_{m'jl}, \nonumber \\
R^{umn}_{ijkl} &=& R^m_{m'ik} R^n_{n'jk} 
- R^m_{m'ik} R^n_{n'jl} \nonumber \\
 &-& R^m_{m'il} R^n_{n'jk} 
+ R^m_{m'il} R^n_{n'jl}, \nonumber \\
R^w_{ijkl} &=& R^i_{\pi\eta k} R^j_{\pi\eta k} \chi_k 
+ 2 R^i_{\pi\eta k} R^j_{\pi\eta l} \chi_{kl} \nonumber \\
 &+& 2 R^i_{\pi\eta l} R^j_{\pi\eta k} \chi_{kl} 
+ R^i_{\pi\eta l} R^j_{\pi\eta l} \chi_l, \nonumber \\
R^{wm}_{ijkl} &=& R^i_{mm'k} R^m_{m'jk} \chi_k 
+ 2 R^i_{mm'k} R^m_{m'jl} \chi_{kl} \nonumber \\
 &+& 2 R^i_{mm'l} R^m_{m'jk} \chi_{kl} 
+ R^i_{mm'l} R^m_{m'jl} \chi_l, \nonumber \\
R^{wmn}_{ijkl} &=& R^m_{m'ik} R^n_{n'jk} \chi_k 
+ 2 R^m_{m'ik} R^n_{n'jl} \chi_{kl} \nonumber \\
 &+& 2 R^m_{m'il} R^n_{n'jk} \chi_{kl} 
+ R^m_{m'il} R^n_{n'jl} \chi_l,
\label{firstcombi}
\end{eqnarray}
and the analogous ones for $d_{\mathrm{sea}}=2$ are
\begin{eqnarray}
R^v_{ijk} &=& R^i_{jj} + R^i_{kk} - 2 R^i_{jk}\,,
\nonumber \\
R^u_{ijkl} &=& R^i_{\eta k} R^j_{\eta k} 
- R^i_{\eta k} R^j_{\eta l} \nonumber \\
 &-& R^i_{\eta l} R^j_{\eta k} 
+ R^i_{\eta l} R^j_{\eta l}, \nonumber \\
R^{um}_{ijkl} &=& R^i_{mk} R^m_{jk} 
- R^i_{mk} R^m_{jl} \nonumber \\
 &-& R^i_{ml} R^m_{jk} 
+ R^i_{ml} R^m_{jl}, \nonumber \\
R^w_{ijkl} &=& R^i_{\eta k} R^j_{\eta k} \chi_k 
+ 2 R^i_{\eta k} R^j_{\eta l} \chi_{kl} \nonumber \\
 &+& 2 R^i_{\eta l} R^j_{\eta k} \chi_{kl} 
+ R^i_{\eta l} R^j_{\eta l} \chi_l, \nonumber \\
R^{wm}_{ijkl} &=& R^i_{mk} R^m_{jk} \chi_k 
+ 2 R^i_{mk} R^m_{jl} \chi_{kl} \nonumber \\
 &+& 2 R^i_{ml} R^m_{jk} \chi_{kl} 
+ R^i_{ml} R^m_{jl} \chi_l.
\label{lastcombi}
\end{eqnarray}
For the $R^u$ and $R^w$ type terms, the indices $m,n$ denote either 
$\pi$ or $\eta$ (independently of each other). If $m$ denotes a $\pi$ 
then the corresponding index $m'$ denotes an $\eta$ and vice
versa. Similarly an $n$ denoting a $\pi$ implies that the $n'$ denotes an
$\eta$. As an example, consider
\begin{eqnarray}
R^{u\pi\eta}_{ijkl} &=& R^\pi_{\eta ik} R^\eta_{\pi jk} 
- R^\pi_{\eta ik} R^\eta_{\pi jl} \nonumber \\
 &-& R^\pi_{\eta il} R^\eta_{\pi jk} + R^\pi_{\eta il} R^\eta_{\pi jl},
\end{eqnarray}
which should suffice to explain the usage of the indices $m,n$ and 
$m',n'$. Note that this particular combination is symmetric under the 
interchange of $\pi$ and $\eta$ only if $i=j$, which means that this 
particular $R^u$ can appear in the final result only in combinations
that are symmetric in the indices $i$ and $j$.

For the case of $d_{\mathrm{sea}}=2$, the indices $m',n'$ no longer 
appear, since $\chi_\pi=\chi_4$ and has therefore canceled out from 
the expressions. In fact, the index $m$ in Eq.~(\ref{lastcombi}) always 
represents an $\eta$. However, Eq.~(\ref{lastcombi}) has been written 
in a slightly more formal way to illustrate the similarity with the 
expressions for $d_{\mathrm{sea}}=3$. It is actually possible to obtain 
the expressions for $d_{\mathrm{sea}}=2$ from the corresponding ones 
for $d_{\mathrm{sea}}=3$ in Eq.~(\ref{firstcombi}) by removing all 
occurrences of the indices $\pi$ and $m'$ there. It should be noted 
that the letters $u,v,w$ in $R^u,R^v$ and $R^w$ are not indices, but 
rather indicate different types of residue combinations. The 
indices $m,n$ used in the above formulas have nothing to do with 
the summation indices $m,n$ defined earlier, which are encountered in 
the analytical expressions in the next sections. In 
Eqs.~(\ref{firstcombi}) and~(\ref{lastcombi}) they only indicate an 
occurrence of either a $\pi$ or an $\eta$.

The implementation of the flavor permutation symmetries described in 
this section is a complicated task, as the direct output of a NNLO 
calculation of masses or decay constants using the propagators defined 
in Eqs.~(\ref{npropij}) and~(\ref{npropij2}) produces highly redundant 
expressions. Especially for the higher values of $d_{\mathrm{sea}}$ and 
$d_{\mathrm{val}}$, the direct output sometimes consists of tens of 
thousands of terms. Thus the output has to be cleaned up using the many 
algebraic relations between the residues $R$, before any attempt at 
implementing the flavor permutation symmetries can be made. This 
process involves the factorization of up to $\sim 200$ different 
expressions of varying length and complexity. As an example for 
$d_{\mathrm{val}} = 2$, the parts of the direct output proportional to 
$\bar A(\chi_\eta;\varepsilon)$ and $\bar A(\chi_1;\varepsilon)$ alone 
are hundreds of terms long, while in factorized and simplified form, 
they contain only~$\sim 5$ terms. After considerable trial and error in 
\texttt{Maple}, the simplification of the results was successfully 
carried through. In addition, the final simplified results have been 
checked algebraically against the original (long) ones.

\section{Analytical results for the masses}
\label{masses}

\subsection{Masses at NNLO}

The corrections to the mass of a pseudoscalar meson are obtained by 
consideration of the self-energy contributions to the propagator of the 
interacting field theory. That propagator is defined in 
terms of the Fourier transform of the two-point Green's function,
\begin{equation}
\label{propagator}
i\Delta(p)= \int d^4x\,e^{ip\cdot x} \langle \Omega \vert 
T[\Phi(x)_{ji}\Phi(0)_{ij}]
\vert \Omega \rangle,
\end{equation}
where $\Phi_{ij}=q_i \bar q_j$ denotes any of the off-diagonal mesons in 
the valence sector of PQ$\chi$PT, and $\Omega$ denotes the vacuum of 
the interacting theory. To lowest order, $i\Delta(p) = G_{ij}^c(p)$ of 
Eq.~(\ref{propc}). When written in terms of the fields in the 
Lagrangian, the resulting self-energy diagrams can be summed as a 
geometric series~\cite{Peskin}, giving
\begin{equation}
\label{PhysMass}
i\Delta(p)=\frac{i}{p^2-M_0^2-\Sigma(p^2,\chi_i)},
\end{equation}
where $M_0^2$ denotes the lowest order mass of the meson which is being 
considered, and $\chi_i$ in $\Sigma$ denotes the dependence of the 
self-energy on all the lowest order meson masses. The quantity 
$\Sigma(p^2,\chi_i)$ takes into account the contributions from the 
one-particle-irreducible~(1PI) diagrams. The physical masses, 
which include the interaction, of the off-diagonal mesons in 
PQ$\chi$PT are defined by the position of the pole in 
Eq.~(\ref{PhysMass}),
\begin{equation}
M_{\mathrm{phys}}^2=M_0^2+\Sigma(M_{\mathrm{phys}}^2,\chi_i),
\end{equation}
whereas the propagators of the neutral or diagonal mesons in PQ$\chi$PT 
have double poles even after resummation.

\begin{figure*}[t]
\begin{center}
\includegraphics[width=\textwidth]{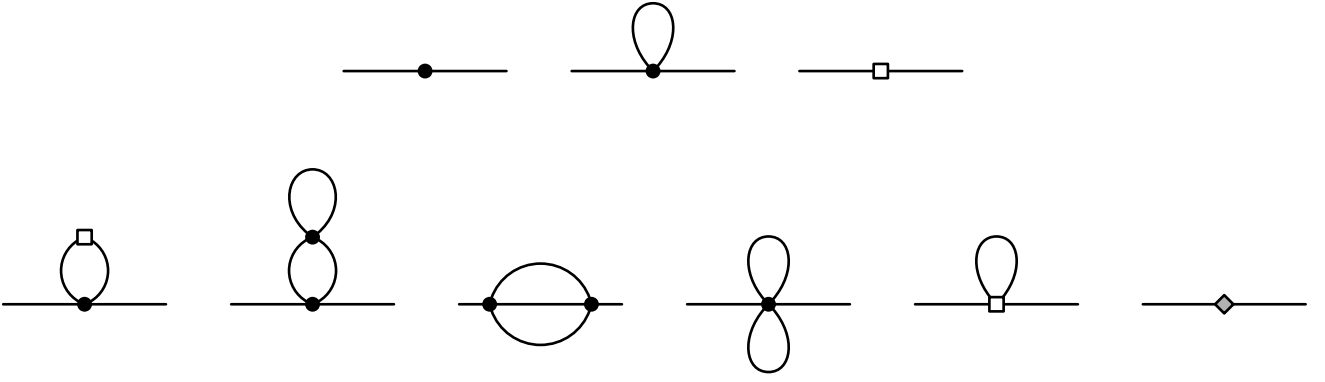}
\caption{Feynman diagrams up to ${\mathcal O}(p^6)$ or two loops, 
for the self-energy $\Sigma(M_{\mathrm{phys}}^2,\chi_i)$. Filled 
circles denote vertices of the ${\mathcal L}_2$ Lagrangian, whereas 
open squares and diamonds denote vertices of the ${\mathcal L}_4$ and 
${\mathcal L}_6$ Lagrangians, respectively. In the top row, the first 
diagram from the left is of ${\mathcal O}(p^2)$, whereas the other 
two diagrams are of ${\mathcal O}(p^4)$. The diagrams of ${\mathcal 
O}(p^6)$, which give the NNLO correction to the meson mass, are shown 
in the bottom row. Of those, the third diagram from the left is called 
the "sunset" diagram in the text.}
\label{massfig}
\end{center}
\end{figure*}

The expression for the self-energy $\Sigma$ can be written 
as a string of terms which denote the 1PI diagrams of progressively 
higher order,
\begin{eqnarray}
\Sigma(M_{\mathrm{phys}}^2,\chi_i) &=& 
\Sigma_4(M_{\mathrm{phys}}^2,\chi_i) \nonumber \\
&+&\Sigma_6(M_{\mathrm{phys}}^2,\chi_i) \:+\: \mathcal{O}(p^8),
\end{eqnarray}
where $\Sigma_4$ contains the self-energy diagrams of 
$\mathcal{O}(p^4)$, and $\Sigma_6$ those of $\mathcal{O}(p^6)$. For the 
present NNLO calculation, the form of $M_{\mathrm{phys}}^2$ has to 
be determined up to $\mathcal{O}(p^6)$. It is sufficient, for this 
purpose, to use the lowest order mass instead of $M_{\mathrm{phys}}^2$ 
in $\Sigma_6$ since the diagrams in that term are already of 
$\mathcal{O}(p^6)$. However, in the case of $\Sigma_4$ the argument 
$M_{\mathrm{phys}}^2$ has to be expanded in a Taylor series around 
$M_0^2$, since the diagrams in $\Sigma_4$ are of $\mathcal{O}(p^4)$. 
If $M_{\mathrm{phys}}^2$ is formally written as
\begin{eqnarray}
M_{\mathrm{phys}}^2 &=& M_0^2 + M_4^2 + M_6^2 \:+\: \mathcal{O}(p^8), 
\end{eqnarray}
then an expansion of $\Sigma_4$ up to $\mathcal{O}(p^6)$ 
gives
\begin{eqnarray}
\Sigma_4(M_{\mathrm{phys}}^2,\chi_i) &=& \Sigma_4(M_0^2,\chi_i) 
\:+\: M_4^2 \left.\frac{\partial \Sigma_4(p^2,\chi_i)}{\partial p^2}
\right\vert_{M_0^2} \hspace{-.5cm} \nonumber \\
&+& \mathcal{O}(p^8),
\end{eqnarray}
where $M_4^2$ represents the NLO correction to the lowest order 
meson mass, which is given by $\Sigma_4(M_0^2,\chi_i)$. The final 
formula for the pseudoscalar meson mass to $\mathcal{O}(p^6)$ is thus
\begin{eqnarray}
M_{\mathrm{phys}}^2 &=& M_0^2 \,+\,\Sigma_4(M_0^2,\chi_i) \nonumber \\ 
&+& \Sigma_4(M_0^2,\chi_i) 
\left.\frac{\partial \Sigma_4(p^2,\chi_i)}{\partial p^2}
\right\vert_{M_0^2} \nonumber \\
&+& \Sigma_6(M_0^2,\chi_i) \:+\:\mathcal{O}(p^8),
\label{masseq}
\end{eqnarray}
where the last two terms are of $\mathcal{O}(p^6)$ and represent the 
NNLO correction to the lowest order mass. The Feynman diagrams that 
contribute to $\Sigma_6(M_0^2,\chi_i)$ are shown in Fig.~\ref{massfig}. 
In the case of the self energies at NNLO, the $\mathcal{O}(p^6)$ vertex 
introduces the LEC:s $K_{17}^r$ through $K_{27}^r$, $K_{39}^r$ 
and $K_{40}^r$. 

The physical mass of a pseudoscalar meson $\Phi_{ij}$ is given to 
NNLO in the form
\begin{equation}
M_{\mathrm{phys}}^2 = \chi_{ij} + \frac{\delta^{(4)\mathrm{vs}}}{F_0^2} 
+ \frac{\delta^{(6)\mathrm{vs}}_{\mathrm{ct}} 
+ \delta^{(6)\mathrm{vs}}_{\mathrm{loops}}}{F_0^4}
\:+\: \mathcal{O}(p^8),
\label{delteq}
\end{equation}
where the LO result $M_0^2 = \chi_{ij}$ has already been inserted, and 
the ${\cal O}(p^4)$ and ${\cal O}(p^6)$ contributions separated. The 
NNLO contribution $\delta^{(6)}$ has been further split into the 
contributions from the chiral loops and from the ${\cal O}(p^6)$ 
counterterms or LEC:s. The superscripts (v) and (s) indicate the values of 
$d_{\mathrm{val}}$ and $d_{\mathrm{sea}}$, respectively. It should be 
noted that we have chosen to give the results to the various orders in 
terms of the lowest order decay constant $F_0$ and in terms of the 
lowest order masses, since these are the fundamental inputs in 
PQ$\chi$PT. The situation is different in standard $\chi$PT, where the 
main objective is comparison with experiment, in which case the 
formulas are most often rewritten in terms of the physical decay 
constants and masses. 

\subsection{Expressions for $d_{\mathrm{val}} = 1$}

In the previously published paper on the NNLO meson masses in 
PQ$\chi$PT~\cite{BDL}, which treated only the most degenerate~(1+1) 
mass case, an overall factor $\chi_1$ was factored out from the 
analytical expressions. However, it has turned out that it is not 
possible to find such an overall factor in a meaningful way for the 
less degenerate mass cases. This is especially evident for 
$d_{\mathrm{val}} = 2$. Thus, for consistency, the factor $\chi_1$
has not been factored out from the expression for the (1+1) mass case
reproduced here either. That expression also uses the new $R$ notation 
as well as a more efficient set of loop integrals, but it is of course 
equivalent to the one presented previously in Ref.~\cite{BDL}.

The NLO result for $d_{\mathrm{val}} = 1$ is rather short, and 
therefore it suffices to give this expression for $d_{\mathrm{sea}} = 
3$ only. The corresponding results for $d_{\mathrm{sea}} = 1,2$ can 
easily be derived from that expression by consideration of the 
appropriate limits, i.e. $\chi_5 \to \chi_4$ for $d_{\mathrm{sea}} = 2$ 
and $\chi_5, \chi_6 \to \chi_4$ for $d_{\mathrm{sea}} = 1$. The 
combined NLO result (loops and counterterms), is
\begin{eqnarray} 
\delta^{(4)13} & = & - \,\:24\,L^r_{4}\,\bar{\chi}_1\,\chi_1
\:-\,8\,L^r_{5}\,\chi_1^2
\:+\,48\,L^r_{6}\,\bar{\chi}_1\,\chi_1 
\nonumber \\ && 
+ \,\:16\,L^r_{8}\,\chi_1^2
\:-\,1/3\,\bar{A}(\chi_m)\,R^m_{n11}\,\chi_1 
\nonumber \\ && 
- \,\:1/3\,\bar{A}(\chi_1)\,R^c_1\,\chi_1 
\nonumber \\ && 
- \,\:1/3\,\bar{B}(\chi_1,\chi_1,0)\,R^d_1\,\chi_1,
\label{M0_NLO_nf3_13} 
\end{eqnarray} 
which is in agreement with Refs.~\cite{BG1,Sharpe1}. The NNLO result 
has been split into the contribution from the loop diagrams and the 
contribution from the ${\cal O}(p^6)$ counterterms. The latter 
consists of the finite part of the rightmost diagram in 
Fig.~\ref{massfig}, and is given for $d_{\mathrm{sea}} = 3$ by
\vspace{.11cm}
\newpage
\begin{eqnarray} 
\delta^{(6)13}_{\mathrm{ct}} & = & - \,\:32\,K^r_{17}\,\chi_1^3 
\:-\,96\,K^r_{18}\,\bar{\chi}_1\,\chi_1^2
\:-\,16\,K^r_{19}\,\chi_1^3 
\nonumber \\ && 
- \,\:48\,K^r_{20}\,\bar{\chi}_1\,\chi_1^2
\:-\,48\,K^r_{21}\,\bar{\chi}_2\,\chi_1 
\nonumber \\ && 
- \,\:144\,K^r_{22}\,\bar{\chi}_1^2\,\chi_1
\:-\,16\,K^r_{23}\,\chi_1^3
\:+\,48\,K^r_{25}\,\chi_1^3 
\nonumber \\ && 
+ \,\:K^r_{26}\,\left[ 96\,\bar{\chi}_1\,\chi_1^2 
+ 48\,\bar{\chi}_2\,\chi_1 \right]
\:+\,432\,K^r_{27}\,\bar{\chi}_1^2\,\chi_1 
\nonumber \\ && 
+ \,\:32\,K^r_{39}\,\chi_1^3
\:+\,96\,K^r_{40}\,\bar{\chi}_1\,\chi_1^2.
\label{M0tree_NNLO_nf3_13} 
\end{eqnarray} 
As for the NLO result above, the contributions from the ${\cal O}(p^6)$ 
counterterms for $d_{\mathrm{sea}} = 1,2$ can be derived 
straightforwardly by taking the appropriate mass limits of this 
expression.

However, the situation is different for the loop contribution at NNLO, 
since it is much larger and has a rather complicated structure. This 
makes it a difficult task to derive the results for $d_{\mathrm{sea}} = 
1,2$ directly from the $d_{\mathrm{sea}} = 3$ case, and therefore the 
different cases are given separately below. The need for such 
expressions is clear since many Lattice QCD simulations work with 
$d_{\mathrm{sea}} = 2$ rather than $d_{\mathrm{sea}} = 3$. As expected, 
all infinities have canceled for all expressions in the renormalization 
procedure, and the result for $\delta^{(6)11}_{\mathrm{loops}}$ is 
equivalent to the one published in Ref.~\cite{BDL}. The 
\texttt{FORM}~\cite{FORM} output with these expressions can be 
downloaded from the website~\cite{website}. The chiral loop 
contributions to the pseudoscalar meson mass at NNLO in PQ$\chi$PT, for 
$d_{\mathrm{val}} = 1$, are

\begin{widetext}

\begin{eqnarray} 
\delta^{(6)11}_{\mathrm{loops}} & = & \pi_{16}\,L^r_{0}\,\left[ 
3\,\chi_1 \chi_4^2 + 26/3\,\chi_1^2 \chi_4 - \chi_1^3 \right]
\:+\,4\,\pi_{16}\,L^r_{1}\,\chi_1^3
\:+\,\pi_{16}\,L^r_{2}\,\left[ 16\,\chi_1 \chi_4^2 + 2\,\chi_1^3 \right] 
\:+\,\pi_{16}\,L^r_{3}\,\left[ 3/2\,\chi_1 \chi_4^2 \right. 
\nonumber \\ && 
+ \,\:17/3\left.\chi_1^2 \chi_4 - 5/2\,\chi_1^3 \right]
\:+\,\pi_{16}^2\,\left[ 73/64\,\chi_1 \chi_4^2 + 15/32\,\chi_1^2 \chi_4 
- 3/32\,\chi_1^3 \right]
\:+\,384\,L^r_{4}L^r_{5}\,\chi_1^2 \chi_4
\:-\,1152\,L^r_{4}L^r_{6}\,\chi_1 \chi_4^2 
\nonumber \\ && 
- \,\:384\,L^r_{4}L^r_{8}\,\chi_1^2 \chi_4
\:+\,576\,L^{r2}_{4}\,\chi_1 \chi_4^2
\:-\,384\,L^r_{5}L^r_{6}\,\chi_1^2 \chi_4
\:-\,128\,L^r_{5}L^r_{8}\,\chi_1^3
\:+\,64\,L^{r2}_{5}\,\chi_1^3
\:-\,8\,\bar{A}(\chi_1)\,L^r_{0}\,\left[ \chi_1^2 \right. 
\nonumber \\ && 
+ \,\:R^d_1\left.\chi_1 \right]
\:+\,8\,\bar{A}(\chi_1)\,L^r_{1}\,\chi_1^2
\:+\,20\,\bar{A}(\chi_1)\,L^r_{2}\,\chi_1^2
\:-\,8\,\bar{A}(\chi_1)\,L^r_{3}\,\left[ \chi_1^2 + R^d_1\,\chi_1 
\right]
\:+\,16\,\bar{A}(\chi_1)\,L^r_{4}\,\chi_1 \chi_4 
\nonumber \\ && 
+ \,\:\bar{A}(\chi_1)\,L^r_{5}\,\left[ 32/3\,\chi_1^2 + 16/3\,R^d_1 
\,\chi_1 \right]
\:-\,\bar{A}(\chi_1)\,L^r_{6}\,\left[ 16\,\chi_1 \chi_4 - 32\,\chi_1^2 
\right]
\:+\,32\,\bar{A}(\chi_1)\,L^r_{7}\,R^d_1\,\chi_1 
\nonumber \\ && 
- \,\:64/3\,\bar{A}(\chi_1)\,L^r_{8}\,\chi_1^2
\:+\,5/9\,\bar{A}(\chi_1)^2\,\chi_1
\:+\,\bar{A}(\chi_1) \bar{B}(\chi_1,\chi_1,0)\,\left[ 11/9\,\chi_1^2 
+ 1/9\,R^d_1\,\chi_1 \right] 
\nonumber \\ && 
+ \,\:2/9\,\bar{A}(\chi_1) \bar{C}(\chi_1,\chi_1,\chi_1,0)\,R^d_1 
\,\chi_1^2
\:-\,\bar{A}(\chi_1,\varepsilon)\,\pi_{16}\,\left[ 11/12\,\chi_1^2 
- 1/4\,R^d_1\,\chi_1 \right]
\:+\,3\,\bar{A}(\chi_{14})\,\pi_{16}\,\chi_1 \chi_4 
\nonumber \\ && 
+ \,\:24\,\bar{A}(\chi_{14})\,L^r_{0}\,\chi_1 \chi_{14}
\:+\,60\,\bar{A}(\chi_{14})\,L^r_{3}\,\chi_1 \chi_{14}
\:-\,48\,\bar{A}(\chi_{14})\,L^r_{5}\,\chi_1 \chi_{14}
\:+\,96\,\bar{A}(\chi_{14})\,L^r_{8}\,\chi_1 \chi_{14}
\:-\,9/4\,\bar{A}(\chi_{14})^2\,\chi_1 
\nonumber \\ && 
- \,\:2\,\bar{A}(\chi_{14}) \bar{B}(\chi_1,\chi_1,0)\,\chi_1 \chi_4
\:-\,\bar{A}(\chi_{14},\varepsilon)\,\pi_{16}\,\left[ 9/2\,\chi_1 \chi_4 
+ 5/2\,\chi_1^2 \right]
\:+\,128\,\bar{A}(\chi_4)\,L^r_{1}\,\chi_1 \chi_4 
\nonumber \\ && 
+ \,\:32\,\bar{A}(\chi_4)\,L^r_{2}\,\chi_1 \chi_4
\:-\,128\,\bar{A}(\chi_4)\,L^r_{4}\,\chi_1 \chi_4
\:+\,128\,\bar{A}(\chi_4)\,L^r_{6}\,\chi_1 \chi_4
\:+\,8/9\,\bar{A}(\chi_4) \bar{B}(\chi_1,\chi_1,0)\,\chi_1 \chi_4 
\nonumber \\ && 
- \,\:2\,\bar{A}(\chi_4,\varepsilon)\,\pi_{16}\,\chi_1 \chi_4 
\:-\,8\,\bar{B}(\chi_1,\chi_1,0)\,L^r_{0}\,R^d_1\,\chi_1^2
\:-\,8\,\bar{B}(\chi_1,\chi_1,0)\,L^r_{3}\,R^d_1\,\chi_1^2
\:+\,\bar{B}(\chi_1,\chi_1,0)\,L^r_{4}\,\left[ 8\,\chi_1^2 \chi_4 
\right. 
\nonumber \\ && 
+ \,\:24\left.R^d_1\,\chi_1 \chi_4 \right]
\:+\,\bar{B}(\chi_1,\chi_1,0)\,L^r_{5}\,\left[ 8/3\,\chi_1^3 + 16\,R^d_1 
\,\chi_1^2 \right]
\:-\,\bar{B}(\chi_1,\chi_1,0)\,L^r_{6}\,\left[ 16\,\chi_1^2 \chi_4 
+ 32\,R^d_1\,\chi_1 \chi_4 \right] 
\nonumber \\ && 
+ \,\:16\,\bar{B}(\chi_1,\chi_1,0)\,L^r_{7}\,(R^d_1)^2 \chi_1
\:-\,\bar{B}(\chi_1,\chi_1,0)\,L^r_{8}\,\left[ 16/3\,\chi_1^3 
+ 32\,R^d_1\,\chi_1^2 - 16/3\,(R^d_1)^2 \chi_1 \right] 
\nonumber \\ && 
+ \,\:\bar{B}(\chi_1,\chi_1,0)^2\,\left[ 2/9\,R^d_1\,\chi_1^2 
+ 1/18\,(R^d_1)^2 \chi_1 \right]
\:+\,2/9\,\bar{B}(\chi_1,\chi_1,0) 
\bar{C}(\chi_1,\chi_1,\chi_1,0)\,(R^d_1)^2 \chi_1^2 
\nonumber \\ && 
+ \,\:29/36\,\bar{B}(\chi_1,\chi_1,0,\varepsilon)\,\pi_{16}\,R^d_1 
\,\chi_1^2
\:+\,16\,\bar{C}(\chi_1,\chi_1,\chi_1,0)\,L^r_{4}\,R^d_1\,\chi_1^2 
\chi_4
\:+\,16/3\,\bar{C}(\chi_1,\chi_1,\chi_1,0)\,L^r_{5}\,R^d_1\,\chi_1^3
\nonumber \\ && 
- \,\:32\,\bar{C}(\chi_1,\chi_1,\chi_1,0)\,L^r_{6}\,R^d_1\,\chi_1^2 
\chi_4
\:-\,32/3\,\bar{C}(\chi_1,\chi_1,\chi_1,0)\,L^r_{8}\,R^d_1\,\chi_1^3
\:+\,5/9\,H^{F}(1,\chi_1,\chi_1,\chi_1,\chi_1)\,\chi_1^2 
\nonumber \\ && 
+ \,\:H^{F}(1,\chi_1,\chi_{14},\chi_{14},\chi_1)\,\left[ 1/4\,\chi_1 
\chi_4 - \chi_1^2 \right]
\:+\,2\,H^{F}(1,\chi_{14},\chi_{14},\chi_4,\chi_1)\,\chi_1 \chi_4 
\nonumber \\ && 
+ \,\:4/9\,H^{F}(2,\chi_1,\chi_1,\chi_1,\chi_1)\,R^d_1\,\chi_1^2
\:+\,3/4\,H^{F}(2,\chi_1,\chi_{14},\chi_{14},\chi_1)\,R^d_1\,\chi_1^2
\:+\,2/9\,H^{F}(5,\chi_1,\chi_1,\chi_1,\chi_1)\,(R^d_1)^2 \chi_1^2 
\nonumber \\ && 
- \,\:4\,H^{F}_1(3,\chi_{14},\chi_1,\chi_{14},\chi_1)\,R^d_1\,\chi_1^2
\:+\,3/4\,H^{F}_{21}(1,\chi_1,\chi_{14},\chi_{14},\chi_1)\,\chi_1^2 
\:+\,6\,H^{F}_{21}(1,\chi_4,\chi_{14},\chi_{14},\chi_1)\,\chi_1^2 
\nonumber \\ && 
- \,\:3/4\,H^{F}_{21}(2,\chi_1,\chi_{14},\chi_{14},\chi_1)\,R^d_1 
\,\chi_1^2,
\label{M0loop_NNLO_nf3_11} 
\end{eqnarray} 

\begin{eqnarray} 
\delta^{(6)12}_{\mathrm{loops}} & = & \pi_{16}\,L^r_{0}\,\left[ - 8/9\, 
\chi_\eta \chi_1 \chi_4 - \chi_1^3 + 26/3\,\bar{\chi}_1\,\chi_1^2 
+ 35/9\,\bar{\chi}_2\,\chi_1 \right]
\:+\,4\,\pi_{16}\,L^r_{1}\,\chi_1^3
\:+\,\pi_{16}\,L^r_{2}\,\left[ 22/3\,\chi_\eta \chi_1 \chi_4 
+ 2\,\chi_1^3 \right. 
\nonumber \\ && 
+ \,\:26/3\left.\bar{\chi}_2\,\chi_1 \right] 
\:+\,\pi_{16}\,L^r_{3}\,\left[ - 8/9\,\chi_\eta \chi_1 \chi_4 
- 5/2\,\chi_1^3 + 17/3\,\bar{\chi}_1\,\chi_1^2 + 43/18\,\bar{\chi}_2 
\,\chi_1 \right]
\:+\,\pi_{16}^2\,\left[ 15/32\,\chi_\eta \chi_1 \chi_4 
\right. 
\nonumber \\ && 
- \,\:3/32\left.\chi_1^3 + 15/32\,\bar{\chi}_1\,\chi_1^2 
+ 43/64\,\bar{\chi}_2\,\chi_1 \right]
\:+\,384\,L^r_{4}L^r_{5}\,\bar{\chi}_1\,\chi_1^2
\:-\,1152\,L^r_{4}L^r_{6}\,\bar{\chi}_1^2\,\chi_1
\:-\,384\,L^r_{4}L^r_{8}\,\bar{\chi}_1\,\chi_1^2 
\nonumber \\ && 
+ \,\:576\,L^{r2}_{4}\,\bar{\chi}_1^2\,\chi_1
\:-\,384\,L^r_{5}L^r_{6}\,\bar{\chi}_1\,\chi_1^2
\:-\,128\,L^r_{5}L^r_{8}\,\chi_1^3
\:+\,64\,L^{r2}_{5}\,\chi_1^3
\:-\,8\,\bar{A}(\chi_\eta)\,L^r_{0}\,R^\eta_{11}\,\chi_\eta \chi_1 
\nonumber \\ && 
+ \,\:16\,\bar{A}(\chi_\eta)\,L^r_{1}\,\chi_\eta \chi_1
\:+\,4\,\bar{A}(\chi_\eta)\,L^r_{2}\,\chi_\eta \chi_1
\:-\,8\,\bar{A}(\chi_\eta)\,L^r_{3}\,R^\eta_{11}\,\chi_\eta \chi_1
\:-\,16\,\bar{A}(\chi_\eta)\,L^r_{4}\,\left[ \chi_\eta \chi_1 
- \bar{\chi}_1 R^\eta_{11}\,\chi_1 \right] 
\nonumber \\ && 
+ \,\:16/3\,\bar{A}(\chi_\eta)\,L^r_{5}\,\left[ 
R^\eta_{11}\,\chi_\eta \chi_1 + R^\eta_{11}\,\chi_1^2 \right]
\:+\,16\,\bar{A}(\chi_\eta)\,L^r_{6}\,\left[ \chi_\eta \chi_1 
- \bar{\chi}_1 R^\eta_{11}\,\chi_1 \right]
\:+\,32\,\bar{A}(\chi_\eta)\,L^r_{7}\,R^z_{\eta 461}\,\chi_1
\nonumber \\ && 
- \,\:64/3\,\bar{A}(\chi_\eta)\,L^r_{8}\,R^\eta_{11}\,\chi_1^2
\:+\,1/18\,\bar{A}(\chi_\eta)^2\,(R^\eta_{11})^2 \chi_1
\:+\,1/9\,\bar{A}(\chi_\eta) \bar{A}(\chi_1)\,R^\eta_{11} R^c_1 
\,\chi_1
\:-\,1/3\,\bar{A}(\chi_\eta) \bar{A}(\chi_{46})\,R^\eta_{11} 
\,\chi_1 
\nonumber \\ && 
- \,\:4/9\,\bar{A}(\chi_\eta) \bar{A}(\chi_{1s})\,R^\eta_{1s} 
R^z_{1s\eta}\,\chi_1
\:+\,\bar{A}(\chi_\eta) \bar{B}(\chi_\eta,\chi_\eta,0)\,\left[ 
1/9\,\bar{\chi}_1 R^\eta_{11}\,\chi_1 - 5/18\,R^\eta_{11}\,\chi_\eta 
\chi_1 \right] 
\nonumber \\ && 
+ \,\:2/9\,\bar{A}(\chi_\eta) \bar{B}(\chi_1,\chi_\eta,0)\,\left[  
R^\eta_{11} R^1_{\eta\eta}\,\chi_1^2 - R^\eta_{14} R^1_{\eta\eta} 
\,\chi_\eta \chi_1 \right]
\:+\,\bar{A}(\chi_\eta) \bar{B}(\chi_1,\chi_1,0)\,\left[ 
2/9\,R^\eta_{11} R^c_1\,\chi_1^2 + 1/9\,R^\eta_{11} R^d_1\,\chi_1
\right.
\nonumber \\ &&
- \,\:1/27\left.R^\eta_{ss} (R^1_{s\eta})^2 \chi_1 \chi_s \right] 
\:+\,2/9\,\bar{A}(\chi_\eta) 
\bar{C}(\chi_1,\chi_1,\chi_1,0)\,R^\eta_{11} R^d_1\,\chi_1^2
\:+\,\bar{A}(\chi_\eta,\varepsilon)\,\pi_{16}\,\left[ 5/9\,R^\eta_{11} 
\,\chi_1^2 - 1/4\,R^c_1\,\chi_\eta \chi_1 \right] 
\nonumber \\ && 
- \,\:8\,\bar{A}(\chi_1)\,L^r_{0}\,\left[ R^c_1\,\chi_1^2 + R^d_1\, 
\chi_1 \right]
\:+\,8\,\bar{A}(\chi_1)\,L^r_{1}\,\chi_1^2
\:+\,20\,\bar{A}(\chi_1)\,L^r_{2}\,\chi_1^2
\:-\,8\,\bar{A}(\chi_1)\,L^r_{3}\,\left[ R^c_1\,\chi_1^2 
+ R^d_1\,\chi_1 \right] 
\nonumber \\ && 
+ \,\:16\,\bar{A}(\chi_1)\,L^r_{4}\,\bar{\chi}_1 R^c_1\,\chi_1
\:+\,\bar{A}(\chi_1)\,L^r_{5}\,\left[ 32/3\,R^c_1\,\chi_1^2 
+ 16/3\,R^d_1\,\chi_1 \right]
\:+\,\bar{A}(\chi_1)\,L^r_{6}\,\left[ 32\,\chi_1^2 - 16\,\bar{\chi}_1 
R^c_1\,\chi_1 \right] 
\nonumber \\ && 
+ \,\:32\,\bar{A}(\chi_1)\,L^r_{7}\,R^d_1\,\chi_1
\:-\,64/3\,\bar{A}(\chi_1)\,L^r_{8}\,R^c_1 \chi_1^2
\:+\,\bar{A}(\chi_1)^2\,\left[ 1/2\,\chi_1 + 1/18\,(R^c_1)^2 \chi_1 
\right] 
\nonumber \\ && 
+ \,\:1/3\,\bar{A}(\chi_1) \bar{A}(\chi_{46})\,R^\eta_{11}\,\chi_1 
\:+\,4/9\,\bar{A}(\chi_1) \bar{A}(\chi_{1s})\,R^\eta_{1s} R^z_{1s\eta} 
\,\chi_1
\:+\,2/9\,\bar{A}(\chi_1) \bar{B}(\chi_1,\chi_\eta,0)\,R^\eta_{11} R^c_1 
\,\chi_1^2 
\nonumber \\ && 
+ \,\:\bar{A}(\chi_1) \bar{B}(\chi_1,\chi_1,0)\,\left[ \chi_1^2 
+ 2/9\,(R^c_1)^2 \chi_1^2 + 1/9\,R^c_1 R^d_1\,\chi_1 \right]
\:+\,2/9\,\bar{A}(\chi_1) \bar{C}(\chi_1,\chi_1,\chi_1,0)\,R^c_1 
R^d_1\,\chi_1^2 
\nonumber \\ && 
- \,\:\bar{A}(\chi_1,\varepsilon)\,\pi_{16}\,\left[ 31/18\,\chi_1^2 
- 29/36\,R^c_1\,\chi_1^2 - 1/4\,R^d_1\,\chi_1 \right]
\:-\,\bar{A}(\chi_{14})^2\,\chi_1
\:-\,\bar{A}(\chi_{14}) \bar{A}(\chi_{16})\,\chi_1
\:-\,1/4\,\bar{A}(\chi_{16})^2\,\chi_1 
\nonumber \\ && 
+ \,\:48\,\bar{A}(\chi_4)\,L^r_{1}\,\chi_1 \chi_4
\:+\,12\,\bar{A}(\chi_4)\,L^r_{2}\,\chi_1 \chi_4
\:-\,48\,\bar{A}(\chi_4)\,L^r_{4}\,\chi_1 \chi_4
\:+\,48\,\bar{A}(\chi_4)\,L^r_{6}\,\chi_1 \chi_4 
\nonumber \\ &&
- \,\:1/6\,\bar{A}(\chi_4) 
\bar{B}(\chi_\eta,\chi_\eta,0)\,R^\eta_{11}\,\chi_1 \chi_4
\:+\,2/3\,\bar{A}(\chi_4) \bar{B}(\chi_1,\chi_\eta,0)\,R^\eta_{14} 
R^1_{4\eta}\,\chi_1 \chi_4 
\nonumber \\ && 
+ \,\:1/3\,\bar{A}(\chi_4) \bar{B}(\chi_1,\chi_1,0)\,(R^1_{4\eta})^2 
\chi_1 \chi_4
\:-\,3/4\,\bar{A}(\chi_4,\varepsilon)\,\pi_{16}\,\chi_1 \chi_4 
\:+\,64\,\bar{A}(\chi_{46})\,L^r_{1}\,\chi_1 \chi_{46}
\:+\,16\,\bar{A}(\chi_{46})\,L^r_{2}\,\chi_1 \chi_{46} 
\nonumber \\ && 
- \,\:64\,\bar{A}(\chi_{46})\,L^r_{4}\,\chi_1 \chi_{46}
\:+\,64\,\bar{A}(\chi_{46})\,L^r_{6}\,\chi_1 \chi_{46}
\:+\,4/9\,\bar{A}(\chi_{46}) \bar{B}(\chi_\eta,\chi_\eta,0)\,R^\eta_{11} 
\,\chi_1 \chi_{46} 
\nonumber \\ && 
- \,\:4/27\,\bar{A}(\chi_{46}) \bar{B}(\chi_1,\chi_\eta,0)\,\left[
 R^{u\eta}_{1146}\,\chi_1^2 - R^{w\eta}_{1146}\,\chi_1 \right] 
\:+\,4/9\,\bar{A}(\chi_{46}) 
\bar{B}(\chi_1,\chi_1,0)\,R^1_{\eta\eta}\,\chi_1 \chi_{46} 
\nonumber \\ && 
- \,\:\bar{A}(\chi_{46},\varepsilon)\,\pi_{16}\,\chi_1 \chi_{46} 
\:+\,\bar{A}(\chi_{1s})\,\pi_{16}\,\left[ 1/4\,\chi_1 \chi_s 
+ 3/4\,\bar{\chi}_1\,\chi_1 \right]
\:+\,8\,\bar{A}(\chi_{1s})\,L^r_{0}\,\chi_1 \chi_{1s}
\:+\,20\,\bar{A}(\chi_{1s})\,L^r_{3}\,\chi_1 \chi_{1s} 
\nonumber \\ && 
- \,\:16\,\bar{A}(\chi_{1s})\,L^r_{5}\,\chi_1 \chi_{1s}
\:+\,32\,\bar{A}(\chi_{1s})\,L^r_{8}\,\chi_1 \chi_{1s}
\:-\,\bar{A}(\chi_{1s}) \bar{B}(\chi_1,\chi_\eta,0)\,\left[ 
4/9\,R^\eta_{1s}\,\chi_1 \chi_{1s} + 2/9\,R^\eta_{1s}\,\chi_1^2 \right] 
\nonumber \\ && 
- \,\:2/3\,\bar{A}(\chi_{1s}) \bar{B}(\chi_1,\chi_1,0)\,R^1_{s\eta} 
\,\chi_1 \chi_s
\:-\,\bar{A}(\chi_{1s},\varepsilon)\,\pi_{16}\,\left[ 3/4\,\chi_1 \chi_s 
+ 5/6\,\chi_1^2 + 3/4\,\bar{\chi}_1\,\chi_1 \right] 
\nonumber \\ && 
+ \,\:8\,\bar{B}(\chi_\eta,\chi_\eta,0)\,L^r_{4}\,\bar{\chi}_1 
R^\eta_{11}\,\chi_\eta \chi_1
\:+\,8/3\,\bar{B}(\chi_\eta,\chi_\eta,0)\,L^r_{5}\,R^\eta_{11} 
\,\chi_\eta^2 \chi_1
\:-\,16\,\bar{B}(\chi_\eta,\chi_\eta,0)\,L^r_{6}\,\bar{\chi}_1 
R^\eta_{11}\,\chi_\eta \chi_1 
\nonumber \\ && 
+ \,\:16\,\bar{B}(\chi_\eta,\chi_\eta,0)\,L^r_{7}\,(R^z_{\eta 
461})^2 \chi_1
\:-\,16/3\,\bar{B}(\chi_\eta,\chi_\eta,0)\,L^r_{8}\,\left[ 
R^\eta_{11}\,\chi_\eta^2 \chi_1 - (R^z_{\eta 461})^2 \chi_1 \right] 
\nonumber \\ && 
+ \,\:32\,\bar{B}(\chi_1,\chi_\eta,0)\,L^r_{7}\,R^z_{\eta 461} 
R^d_1\,\chi_1
\:+\,32/3\,\bar{B}(\chi_1,\chi_\eta,0)\,L^r_{8}\,R^z_{\eta 461} 
R^d_1\,\chi_1 
\nonumber \\ && 
+ \,\:2/9\,\bar{B}(\chi_1,\chi_\eta,0) \bar{B}(\chi_1,\chi_1,0) 
R^\eta_{11} R^d_1\,\chi_1^2
\:-\,8\,\bar{B}(\chi_1,\chi_1,0)\,L^r_{0}\,R^d_1\,\chi_1^2 
\:-\,8\,\bar{B}(\chi_1,\chi_1,0)\,L^r_{3}\,R^d_1\,\chi_1^2 
\nonumber \\ && 
+ \,\:\bar{B}(\chi_1,\chi_1,0)\,L^r_{4}\,\left[ 8\,\bar{\chi}_1 R^c_1 
\,\chi_1^2 + 24\,\bar{\chi}_1 R^d_1\,\chi_1 \right]
\:+\,\bar{B}(\chi_1,\chi_1,0)\,L^r_{5}\,\left[ 8/3\,R^c_1\,\chi_1^3 
+ 16\,R^d_1\,\chi_1^2 \right] 
\nonumber \\ && 
- \,\:\bar{B}(\chi_1,\chi_1,0)\,L^r_{6}\,\left[ 16\,\bar{\chi}_1 
R^c_1\,\chi_1^2 + 32\,\bar{\chi}_1 R^d_1\,\chi_1 \right]
\:+\,16\,\bar{B}(\chi_1,\chi_1,0)\,L^r_{7}\,(R^d_1)^2 \chi_1
\:-\,\bar{B}(\chi_1,\chi_1,0)\,L^r_{8}\,\left[ 16/3\,R^c_1\,\chi_1^3 
\right. 
\nonumber \\ && 
+ \,\:32\left.R^d_1\,\chi_1^2 - 16/3\,(R^d_1)^2 \chi_1 \right] 
\:+\,\bar{B}(\chi_1,\chi_1,0)^2\,\left[ 2/9\,R^c_1 R^d_1\,\chi_1^2 
+ 1/18\,(R^d_1)^2 \chi_1 \right] 
\nonumber \\ && 
+ \,\:2/9\,\bar{B}(\chi_1,\chi_1,0) \bar{C}(\chi_1,\chi_1,\chi_1,0)   
\,(R^d_1)^2 \chi_1^2
\:+\,29/36\,\bar{B}(\chi_1,\chi_1,0,\varepsilon)\,\pi_{16}\,R^d_1 
\,\chi_1^2 
\nonumber \\ && 
+ \,\:16\,\bar{C}(\chi_1,\chi_1,\chi_1,0)\,L^r_{4}\,\bar{\chi}_1 
R^d_1\,\chi_1^2
\:+\,16/3\,\bar{C}(\chi_1,\chi_1,\chi_1,0)\,L^r_{5}\,R^d_1\,\chi_1^3
\:-\,32\,\bar{C}(\chi_1,\chi_1,\chi_1,0)\,L^r_{6}\,\bar{\chi}_1 
R^d_1\,\chi_1^2 
\nonumber \\ && 
- \,\:32/3\,\bar{C}(\chi_1,\chi_1,\chi_1,0)\,L^r_{8}\,R^d_1 
\,\chi_1^3
\:+\,2/9\,H^{F}(1,\chi_\eta,\chi_\eta,\chi_1,\chi_1)\,(R^\eta_{11})^2 
\chi_1^2
\:+\,4/9\,H^{F}(1,\chi_\eta,\chi_1,\chi_1,\chi_1)\,R^\eta_{11} 
R^c_1\,\chi_1^2 
\nonumber \\ && 
- \,\:H^{F}(1,\chi_\eta,\chi_{1s},\chi_{1s},\chi_1)\,\left[ 
1/3\,R^\eta_{11}\,\chi_1^2 + 1/12\,R^v_{\eta 1s}\,\chi_\eta \chi_1 
\right]
\:+\,H^{F}(1,\chi_1,\chi_1,\chi_1,\chi_1)\,\left[ 1/3\,\chi_1^2 
+ 2/9\,(R^c_1)^2 \chi_1^2 \right] 
\nonumber \\ && 
- \,\:H^{F}(1,\chi_1,\chi_{1s},\chi_{1s},\chi_1)\,\left[ 
1/2\,R^1_{s\eta}\,\chi_1^2 - 1/4\,R^c_1\,\chi_1^2 + 1/12\,R^d_1\,\chi_1 
\right]
\:+\,3/4\,H^{F}(1,\chi_{14},\chi_{14},\chi_4,\chi_1)\,\chi_1 
\chi_4 
\nonumber \\ && 
+ \,\:H^{F}(1,\chi_{14},\chi_{16},\chi_{46},\chi_1)\,\chi_1 \chi_{46} 
\:+\,4/9\,H^{F}(2,\chi_1,\chi_\eta,\chi_1,\chi_1)\,R^\eta_{11} 
R^d_1\,\chi_1^2
\:+\,4/9\,H^{F}(2,\chi_1,\chi_1,\chi_1,\chi_1)\,R^c_1 R^d_1 
\,\chi_1^2 
\nonumber \\ && 
+ \,\:1/4\,H^{F}(2,\chi_1,\chi_{1s},\chi_{1s},\chi_1)\,R^d_1\,\chi_1^2 
\:+\,2/9\,H^{F}(5,\chi_1,\chi_1,\chi_1,\chi_1)\,(R^d_1)^2\,\chi_1^2
\nonumber \\
&& - \,\:2/3\,H^{F}_1(1,\chi_\eta,\chi_{1s},\chi_{1s},\chi_1) 
R^\eta_{1s} R^z_{1s\eta}\,\chi_1^2
\:-\,4/3\,H^{F}_1(1,\chi_{1s},\chi_{1s},\chi_1,\chi_1)\,R^\eta_{1s} 
R^z_{1s\eta}\,\chi_1^2 
\nonumber \\ && 
- \,\:4/3\,H^{F}_1(3,\chi_{1s},\chi_1,\chi_{1s},\chi_1)\,R^d_1 
\,\chi_1^2
\:-\,1/4\,H^{F}_{21}(1,\chi_\eta,\chi_{1s},\chi_{1s},\chi_1)\,R^v_{\eta 
1s}\,\chi_1^2 
\nonumber \\ && 
+ \,\:H^{F}_{21}(1,\chi_1,\chi_{1s},\chi_{1s},\chi_1)\,\left[ 1/2\, 
R^1_{s\eta}\,\chi_1^2 - 1/4\,R^c_1\,\chi_1^2 \right]
\:+\,9/4\,H^{F}_{21}(1,\chi_4,\chi_{14},\chi_{14},\chi_1)\,\chi_1^2
\nonumber \\ && 
+ \,\:3\,H^{F}_{21}(1,\chi_{46},\chi_{14},\chi_{16},\chi_1)\,\chi_1^2 
\:-\,1/4\,H^{F}_{21}(2,\chi_1,\chi_{1s},\chi_{1s},\chi_1)\,R^d_1 
\,\chi_1^2,
\label{M0loop_NNLO_nf3_12} 
\end{eqnarray} 

\begin{eqnarray} 
\delta^{(6)13}_{\mathrm{loops}} & = & \pi_{16}\,L^r_{0}\,\left[ - 8/9\, 
\chi_\pi \chi_\eta \chi_1 - \chi_1^3 + 26/3\,\bar{\chi}_1\,\chi_1^2 
+ 35/9\,\bar{\chi}_2\,\chi_1 \right]
\:+\,4\,\pi_{16}\,L^r_{1}\,\chi_1^3
\:+\,\pi_{16}\,L^r_{2}\,\left[ 22/3\,\chi_\pi \chi_\eta \chi_1 
+ 2\,\chi_1^3 \right. 
\nonumber \\ && 
+ \,\:26/3\left.\bar{\chi}_2\,\chi_1 \right] 
\:+\,\pi_{16}\,L^r_{3}\,\left[ - 8/9\,\chi_\pi \chi_\eta \chi_1 
- 5/2\,\chi_1^3 + 17/3\,\bar{\chi}_1\,\chi_1^2 + 43/18\,\bar{\chi}_2 
\,\chi_1 \right]
\:+\,\pi_{16}^2\,\left[ 15/32\,\chi_\pi \chi_\eta \chi_1 \right. 
\nonumber \\ && 
- \,\:3/32\left.\chi_1^3 + 15/32\,\bar{\chi}_1\,\chi_1^2 
+ 43/64\,\bar{\chi}_2\,\chi_1 \right]
\:+\,384\,L^r_{4}L^r_{5}\,\bar{\chi}_1\,\chi_1^2
\:-\,1152\,L^r_{4}L^r_{6}\,\bar{\chi}_1^2 \chi_1
\:-\,384\,L^r_{4}L^r_{8}\,\bar{\chi}_1\,\chi_1^2 
\nonumber \\ && 
+ \,\:576\,L^{r2}_{4}\,\bar{\chi}_1^2 \chi_1
\:-\,384\,L^r_{5}L^r_{6}\,\bar{\chi}_1\,\chi_1^2
\:-\,128\,L^r_{5}L^r_{8}\,\chi_1^3
\:+\,64\,L^{r2}_{5}\,\chi_1^3
\:-\,8\,\bar{A}(\chi_m)\,L^r_{0}\,R^m_{n11}\,\chi_m \chi_1 
\nonumber \\ && 
+ \,\:16\,\bar{A}(\chi_m)\,L^r_{1}\,\chi_m \chi_1
\:+\,4\,\bar{A}(\chi_m)\,L^r_{2}\,\chi_m \chi_1 
\:-\,8\,\bar{A}(\chi_m)\,L^r_{3}\,R^m_{n11}\,\chi_m \chi_1 
\:-\,16\,\bar{A}(\chi_m)\,L^r_{4}\,\left[ \chi_m \chi_1
- \bar{\chi}_1 R^m_{n11}\,\chi_1 \right] 
\nonumber \\ &&
+ \,\:16/3\,\bar{A}(\chi_m)\,L^r_{5}\,\left[ R^m_{n11}\,\chi_m \chi_1 
+ R^m_{n11}\,\chi_1^2 \right]
\:+\,16\,\bar{A}(\chi_m)\,L^r_{6}\,\left[ \chi_m \chi_1 
- \bar{\chi}_1 R^m_{n11}\,\chi_1 \right]
\:+\,32\,\bar{A}(\chi_m)\,L^r_{7}\,R^z_{m456n1}\,\chi_1
\nonumber \\ &&
- \,\:64/3\,\bar{A}(\chi_m)\,L^r_{8}\,R^m_{n11}\,\chi_1^2
\:+\,1/18\,\bar{A}(\chi_m)^2\,(R^m_{n11})^2 \chi_1  
\:+\,1/9\,\bar{A}(\chi_m) \bar{A}(\chi_1)\,R^m_{n11} R^c_1\,\chi_1 
\nonumber \\ &&
- \,\:4/9\,\bar{A}(\chi_m) \bar{A}(\chi_{1s})\,R^m_{n1s} 
R^z_{1sm}\,\chi_1
\:+\,\bar{A}(\chi_m) \bar{A}(\chi_{st})\,\left[1/27\,R^m_{n11} 
R^v_{mnst}\,\chi_1 
+ 2/27\,R^m_{n1s} R^n_{m1s} R^z_{tssmn}\,\chi_1 \right]
\nonumber \\ &&
- \,\:1/27\,\bar{A}(\chi_m) \bar{B}(\chi_m,\chi_m,0)\,
(R^m_{n1s})^2 R^m_{nss}\,\chi_1 \chi_s
\:-\,2/27\,\bar{A}(\chi_m) \bar{B}(\chi_m,\chi_n,0)\,
R^m_{n1s} R^m_{nss} R^n_{m1s}\,\chi_1 \chi_s 
\nonumber \\ &&
+ \,\:\bar{A}(\chi_m) \bar{B}(\chi_m,\chi_1,0)\,\left[ 2/9\,
(R^m_{n11})^2 \chi_1^2
- 2/27\,R^m_{n1s} R^m_{nss} R^1_{smn}
\,\chi_1 \chi_s \right]
\nonumber \\ &&
- \,\:1/27\,\bar{A}(\chi_m) \bar{B}(\chi_n,\chi_n,0)\,
R^m_{nss} (R^n_{m1s})^2 \chi_1 \chi_s 
\:-\,\bar{A}(\chi_m) \bar{B}(\chi_n,\chi_1,0)\,\left[ 2/27\,
R^m_{nss} R^n_{m1s} R^1_{smn}\,\chi_1 \chi_s \right.
\nonumber \\ &&
- \,\:2/9\left.R^\pi_{\eta 11} R^\eta_{\pi 11}\,\chi_1^2 \right]
\:+\,\bar{A}(\chi_m) \bar{B}(\chi_1,\chi_1,0)\,\left[ 2/9\,
R^m_{n11} R^c_1\,\chi_1^2 + 1/9\,R^m_{n11} R^d_1\,\chi_1
- 1/27\,R^m_{nss} (R^1_{smn})^2 \chi_1 \chi_s \right] 
\nonumber \\ &&
+ \,\:2/9\,\bar{A}(\chi_m) \bar{C}(\chi_1,\chi_1,\chi_1,0)\,R^m_{n11} 
R^d_1\,\chi_1^2  
\:+\,\bar{A}(\chi_m,\varepsilon)\,\pi_{16}\,\left[ 
- 1/4\,\chi_m \chi_1 
+ 1/4\,R^m_{n11}\,\chi_m \chi_1
+ 5/9\,R^m_{n11}\,\chi_1^2 \right]
\nonumber \\ &&
+ \,\:1/9\,\bar{A}(\chi_\pi) \bar{A}(\chi_\eta)\,R^\pi_{\eta 11} 
R^\eta_{\pi 11}\,\chi_1
\:-\,8\,\bar{A}(\chi_1)\,L^r_{0}\,\left[ R^c_1\,\chi_1^2 + R^d_1\,\chi_1 
\right] 
\:+\,8\,\bar{A}(\chi_1)\,L^r_{1}\,\chi_1^2
\nonumber \\ &&
+ \,\:20\,\bar{A}(\chi_1)\,L^r_{2}\,\chi_1^2
\:-\,8\,\bar{A}(\chi_1)\,L^r_{3}\,\left[ R^c_1\,\chi_1^2 + R^d_1\,\chi_1 
\right]
\:+\,16\,\bar{A}(\chi_1)\,L^r_{4}\,\bar{\chi}_1 R^c_1\,\chi_1 
\nonumber \\ && 
+ \,\:\bar{A}(\chi_1)\,L^r_{5}\,\left[ 32/3\,R^c_1\,\chi_1^2 
+ 16/3\,R^d_1\,\chi_1 \right]
\:+\,\bar{A}(\chi_1)\,L^r_{6}\,\left[ 32\,\chi_1^2 - 16\,\bar{\chi}_1 
R^c_1\,\chi_1 \right]
\:+\,32\,\bar{A}(\chi_1)\,L^r_{7}\,R^d_1\,\chi_1 
\nonumber \\ && 
- \,\:64/3\,\bar{A}(\chi_1)\,L^r_{8}\,R^c_1\,\chi_1^2
\:+\,\bar{A}(\chi_1)^2\,\left[ 1/2\,\chi_1 + 1/18\,(R^c_1)^2 \chi_1 
\right]
\:+\,4/9\,\bar{A}(\chi_1) \bar{A}(\chi_{1s})\,R^m_{n1s} 
R^z_{1sm}\,\chi_1 
\nonumber \\ && 
- \,\:1/27\,\bar{A}(\chi_1) \bar{A}(\chi_{st})\,(R^1_{s\pi\eta})^2 
R^z_{tss11}\,\chi_1
\:+\,2/9\,\bar{A}(\chi_1) \bar{B}(\chi_m,\chi_1,0)\,R^m_{n11} 
R^c_1\,\chi_1^2
\:+\,\bar{A}(\chi_1) \bar{B}(\chi_1,\chi_1,0)\,\left[ \chi_1^2 \right. 
\nonumber \\ && 
+ \,\:2/9\left.(R^c_1)^2 \chi_1^2 + 1/9\,R^c_1 R^d_1\,\chi_1 \right] 
\:+\,2/9\,\bar{A}(\chi_1) \bar{C}(\chi_1,\chi_1,\chi_1,0)\,R^c_1 
R^d_1\,\chi_1^2
\:+\,\bar{A}(\chi_1,\varepsilon)\,\pi_{16}\,\left[ - 31/18\,\chi_1^2 
\right. 
\nonumber \\ && 
+ \,\:29/36\left.R^c_1\,\chi_1^2 + 1/4\,R^d_1\,\chi_1 \right] 
\:+\,\bar{A}(\chi_{1s})\,\pi_{16}\,\left[ 1/4\,\chi_1 \chi_s 
+ 3/4\,\bar{\chi}_1\,\chi_1 \right]
\:+\,8\,\bar{A}(\chi_{1s})\,L^r_{0}\,\chi_1 \chi_{1s} 
\nonumber \\ && 
+ \,\:20\,\bar{A}(\chi_{1s})\,L^r_{3}\,\chi_1 \chi_{1s}
\:-\,16\,\bar{A}(\chi_{1s})\,L^r_{5}\,\chi_1 \chi_{1s}
\:+\,32\,\bar{A}(\chi_{1s})\,L^r_{8}\,\chi_1 \chi_{1s}
\:-\,1/4\,\bar{A}(\chi_{1s})^2\,\chi_1 
\nonumber \\ && 
- \,\:1/2\,\bar{A}(\chi_{1s}) \bar{A}(\chi_{1t})\,\chi_1
\:-\,\bar{A}(\chi_{1s}) \bar{B}(\chi_m,\chi_1,0)\,\left[ 4/9\, 
R^m_{n1s}\,\chi_1 \chi_{1s} + 2/9\,R^m_{n1s}\,\chi_1^2 \right] 
\nonumber \\ && 
- \,\:2/3\,\bar{A}(\chi_{1s}) \bar{B}(\chi_1,\chi_1,0)\,R^1_{s\pi\eta} 
\,\chi_1 \chi_s
\:-\,\bar{A}(\chi_{1s},\varepsilon)\,\pi_{16}\,\left[ 3/4\,\chi_1 \chi_s 
+ 5/6\,\chi_1^2 + 3/4\,\bar{\chi}_1\,\chi_1 \right] 
\:+\,32\,\bar{A}(\chi_{st})\,L^r_{1}\,\chi_1 \chi_{st} 
\nonumber \\ && 
+ \,\:8\,\bar{A}(\chi_{st})\,L^r_{2}\,\chi_1 \chi_{st}
\:-\,32\,\bar{A}(\chi_{st})\,L^r_{4}\,\chi_1 \chi_{st}
\:+\,32\,\bar{A}(\chi_{st})\,L^r_{6}\,\chi_1 \chi_{st}
\:+\,2/9\,\bar{A}(\chi_{st}) \bar{B}(\chi_m,\chi_m,0)\,R^m_{n11} 
R^m_{nst}\,\chi_1 \chi_{st} 
\nonumber \\ && 
- \,\:2/27\,\bar{A}(\chi_{st}) \bar{B}(\chi_m,\chi_1,0)\,\left[
R^{um}_{11st}\,\chi_1^2 - R^{wm}_{11st}\,\chi_1 \right]
\:+\,2/27\,\bar{A}(\chi_{st}) \bar{B}(\chi_\pi,\chi_\eta,0)
\,R^{w\pi\eta}_{11st}\,\chi_1
\nonumber \\ &&
- \,\:2/27\,\bar{A}(\chi_{st}) \bar{B}(\chi_\pi,\chi_\eta,0,k)\,
R^\pi_{\eta 1s} R^\eta_{\pi 1s} R^z_{tss\pi\eta}\,\chi_1
\:+\,2/9\,\bar{A}(\chi_{st}) \bar{B}(\chi_1,\chi_1,0)\,R^1_{s\eta\eta} 
R^1_{t\pi\pi}\,\chi_1 \chi_{st} 
\nonumber \\ && 
- \,\:1/2\,\bar{A}(\chi_{st},\varepsilon)\,\pi_{16}\,\chi_1 \chi_{st} 
\:+\,8\,\bar{B}(\chi_m,\chi_m,0)\,L^r_{4}\,\bar{\chi}_1 
R^m_{n11}\,\chi_m \chi_1
\:+\,8/3\,\bar{B}(\chi_m,\chi_m,0)\,L^r_{5}\,R^m_{n11}\,\chi_m^2 
\chi_1 
\nonumber \\ && 
- \,\:16\,\bar{B}(\chi_m,\chi_m,0)\,L^r_{6}\,\bar{\chi}_1 R^m_{n11} 
\,\chi_m \chi_1
\:+\,16\,\bar{B}(\chi_m,\chi_m,0)\,L^r_{7}\,(R^z_{m456n1})^2 \chi_1 
\nonumber \\ && 
- \,\:16/3\,\bar{B}(\chi_m,\chi_m,0)\,L^r_{8}\,\left[ R^m_{n11} 
\,\chi_m^2 \chi_1 - (R^z_{m456n1})^2 \chi_1 \right]
\:+\,32\,\bar{B}(\chi_m,\chi_1,0)\,L^r_{7}\,R^z_{m456n1} R^d_1 
\,\chi_1 
\nonumber \\ && 
+ \,\:32/3\,\bar{B}(\chi_m,\chi_1,0)\,L^r_{8}\,R^z_{m456n1} 
R^d_1\,\chi_1
\:+\,2/9\,\bar{B}(\chi_m,\chi_1,0) \bar{B}(\chi_1,\chi_1,0)\,R^m_{n11} 
R^d_1\,\chi_1^2 
\nonumber \\ && 
+ \,\:32\,\bar{B}(\chi_\pi,\chi_\eta,0)\,L^r_{7}\,R^z_{\pi456\eta 
1} R^z_{\eta456\pi 1}\,\chi_1
\:+\,32/3\,\bar{B}(\chi_\pi,\chi_\eta,0)\,L^r_{8}\,R^z_{\pi456\eta 
1} R^z_{\eta456\pi 1}\,\chi_1
\:-\,8\,\bar{B}(\chi_1,\chi_1,0)\,L^r_{0}\,R^d_1\,\chi_1^2 
\nonumber \\ && 
- \,\:8\,\bar{B}(\chi_1,\chi_1,0)\,L^r_{3}\,R^d_1\,\chi_1^2
\:+\,\bar{B}(\chi_1,\chi_1,0)\,L^r_{4}\,\left[ 8\,\bar{\chi}_1 R^c_1 
\,\chi_1^2 + 24\,\bar{\chi}_1 R^d_1\,\chi_1 \right]
\:+\,\bar{B}(\chi_1,\chi_1,0)\,L^r_{5}\,\left[ 8/3\,R^c_1\,\chi_1^3 
\right. 
\nonumber \\ && 
+ \,\:16\left.R^d_1\,\chi_1^2 \right]
\:-\,\bar{B}(\chi_1,\chi_1,0)\,L^r_{6}\,\left[ 16\,\bar{\chi}_1 R^c_1 
\,\chi_1^2 + 32\,\bar{\chi}_1 R^d_1\,\chi_1 \right]
\:+\,16\,\bar{B}(\chi_1,\chi_1,0)\,L^r_{7}\,(R^d_1)^2 \chi_1 
\nonumber \\ && 
- \,\:\bar{B}(\chi_1,\chi_1,0)\,L^r_{8}\,\left[ 16/3\,R^c_1\,\chi_1^3 
+ 32\,R^d_1\,\chi_1^2 - 16/3\,(R^d_1)^2 \chi_1 \right]
\:+\,\bar{B}(\chi_1,\chi_1,0)^2\,\left[ 2/9\,R^c_1 R^d_1\,\chi_1^2 
+ 1/18\,(R^d_1)^2 \chi_1 \right] 
\nonumber \\ && 
+ \,\:2/9\,\bar{B}(\chi_1,\chi_1,0) 
\bar{C}(\chi_1,\chi_1,\chi_1,0)\,(R^d_1)^2 \chi_1^2
\:+\,29/36\,\bar{B}(\chi_1,\chi_1,0,\varepsilon)\,\pi_{16}\,R^d_1 
\,\chi_1^2 
\nonumber \\ && 
+ \,\:16\,\bar{C}(\chi_1,\chi_1,\chi_1,0)\,L^r_{4}\,\bar{\chi}_1 
R^d_1\,\chi_1^2
\:+\,16/3\,\bar{C}(\chi_1,\chi_1,\chi_1,0)\,L^r_{5}\,R^d_1\,\chi_1^3
\:-\,32\,\bar{C}(\chi_1,\chi_1,\chi_1,0)\,L^r_{6}\,\bar{\chi}_1 
R^d_1\,\chi_1^2 
\nonumber \\ && 
- \,\:32/3\,\bar{C}(\chi_1,\chi_1,\chi_1,0)\,L^r_{8}\,R^d_1\,\chi_1^3
\:+\,2/9\,H^{F}(1,\chi_m,\chi_m,\chi_1,\chi_1)\,(R^m_{n11})^2\,\chi_1^2 
\nonumber \\ && 
+ \,\:4/9\,H^{F}(1,\chi_m,\chi_1,\chi_1,\chi_1)\,R^m_{n11} 
R^c_1\,\chi_1^2
\:-\,H^{F}(1,\chi_m,\chi_{1s},\chi_{1s},\chi_1)\,\left[ 1/3\,R^m_{n11} 
\,\chi_1^2 + 1/12\,R^v_{mn1s}\,\chi_m \chi_1 \right] 
\nonumber \\ && 
+ \,\:4/9\,H^{F}(1,\chi_\pi,\chi_\eta,\chi_1,\chi_1)\,R^\pi_{\eta 11} 
R^\eta_{\pi 11}\,\chi_1^2
\:+\,H^{F}(1,\chi_1,\chi_1,\chi_1,\chi_1)\,\left[ 1/3\,\chi_1^2 
+ 2/9\,(R^c_1)^2 \chi_1^2 \right] 
\nonumber \\ && 
- \,\:H^{F}(1,\chi_1,\chi_{1s},\chi_{1s},\chi_1)\,\left[ 
1/2\,R^1_{s\pi\eta}\,\chi_1^2 - 1/4\,R^c_1\,\chi_1^2 
+ 1/12\,R^d_1\,\chi_1 \right]
\:+\,1/2\,H^{F}(1,\chi_{1s},\chi_{1t},\chi_{st},\chi_1)\,\chi_1 
\chi_{st}
\nonumber \\ &&
+ \,\:4/9\,H^{F}(2,\chi_1,\chi_m,\chi_1,\chi_1)\,R^m_{n11} R^d_1 
\,\chi_1^2 
\:+\,4/9\,H^{F}(2,\chi_1,\chi_1,\chi_1,\chi_1)\,R^c_1 R^d_1 
\,\chi_1^2 
\nonumber \\ &&
+ \,\:1/4\,H^{F}(2,\chi_1,\chi_{1s},\chi_{1s},\chi_1)\,R^d_1 
\,\chi_1^2  
\:+\,2/9\,H^{F}(5,\chi_1,\chi_1,\chi_1,\chi_1)\,(R^d_1)^2 \chi_1^2 
\nonumber \\ &&
- \,\:2/3\,H^{F}_1(1,\chi_m,\chi_{1s},\chi_{1s},\chi_1)\,R^m_{n1s} 
R^z_{1sm}\,\chi_1^2  
\:-\,4/3\,H^{F}_1(1,\chi_{1s},\chi_{1s},\chi_1,\chi_1)\,R^m_{n1s} 
R^z_{1sm}\,\chi_1^2 
\nonumber \\ &&
- \,\:4/3\,H^{F}_1(3,\chi_{1s},\chi_1,\chi_{1s},\chi_1)\,R^d_1 
\,\chi_1^2 
\:-\,1/4\,H^{F}_{21}(1,\chi_m,\chi_{1s},\chi_{1s},\chi_1)\,R^v_{mn1s} 
\,\chi_1^2
\nonumber \\ &&
+ \,\:H^{F}_{21}(1,\chi_1,\chi_{1s},\chi_{1s},\chi_1)\,\left[ 1/2\, 
R^1_{s\pi\eta}\,\chi_1^2 - 1/4\,R^c_1\,\chi_1^2 \right] 
\:+\,3/2\,H^{F}_{21}(1,\chi_{st},\chi_{1s},\chi_{1t},\chi_1)\,\chi_1^2 
\nonumber \\ && 
- \,\:1/4\,H^{F}_{21}(2,\chi_1,\chi_{1s},\chi_{1s},\chi_1)\,R^d_1 
\,\chi_1^2.
\label{M0loop_NNLO_nf3_13} 
\end{eqnarray} 

\end{widetext}

\subsection{Results for $d_{\mathrm{val}} = 2$}

In general, the results for $d_{\mathrm{val}} = 2$ are similar in form 
to those for $d_{\mathrm{val}} = 1$. However, they are about twice as 
long, since the number of independent combinations of quark masses that 
can appear in the propagator residues and loop integrals is much 
larger. As already noted for $d_{\mathrm{val}} = 1$, the expressions 
for $d_{\mathrm{val}} = 2$ also become longer with increasing values of 
$d_{\mathrm{sea}}$. The NLO expression for $d_{\mathrm{val}} = 2$ and 
$d_{\mathrm{sea}} = 3$, which agrees with the result of 
Refs.~\cite{BG1,Sharpe1}, is
\begin{eqnarray} 
\delta^{(4)23} & = & -\,\:24\,L^r_{4}\,\bar{\chi}_1\,\chi_{13}
\:-\,8\,L^r_{5}\,\chi_{13}^2
\:+\,48\,L^r_{6}\,\bar{\chi}_1\,\chi_{13} 
\nonumber \\ && 
+ \,\:16\,L^r_{8}\,\chi_{13}^2
\:-\,1/3\,\bar{A}(\chi_p)\,R^p_{q\pi\eta}\,\chi_{13} 
\nonumber \\ && 
- \,\:1/3\,\bar{A}(\chi_m)\,R^m_{n13}\,\chi_{13},
\label{M0_NLO_nf3_23} 
\end{eqnarray} 
while the contribution from the ${\cal O}(p^6)$ counterterms for 
$d_{\mathrm{val}} = 2$ and $d_{\mathrm{sea}} = 3$ is given by
\newpage
\begin{eqnarray} 
\delta^{(6)23}_{\mathrm{ct}} & = & -\,\:32\,K^r_{17}\,\chi_{13}^3
\:-\,96\,K^r_{18}\,\bar{\chi}_1\,\chi_{13}^2
\:-\,8\,K^r_{19}\,\chi_p^2 \chi_{13} 
\nonumber \\ && 
- \,\:48\,K^r_{20}\,\bar{\chi}_1\,\chi_{13}^2
\:-\,48\,K^r_{21}\,\bar{\chi}_2\,\chi_{13} 
\nonumber \\ && 
- \,\:144\,K^r_{22}\,\bar{\chi}_1^2 \chi_{13}
\:-\,16\,K^r_{23}\,\chi_1 \chi_{13} \chi_3 
\nonumber \\ && 
+ \,\:24\,K^r_{25}\,\chi_p^2 \chi_{13}
\:+\,K^r_{26}\,\left[ 96\,\bar{\chi}_1\,\chi_{13}^2 \right. 
\nonumber \\ && 
+ \,\:48\left.\bar{\chi}_2\,\chi_{13} \right]
\:+\,432\,K^r_{27}\,\bar{\chi}_1^2 \chi_{13}
\:+\,32\,K^r_{39}\,\chi_{13}^3 
\nonumber \\ && 
+ \,\:96\,K^r_{40}\,\bar{\chi}_1\,\chi_{13}^2.
\label{M0tree_NNLO_nf3_23} 
\end{eqnarray} 
Similarly to the case of $d_{\mathrm{val}} = 1$, the expressions for 
other values of $d_{\mathrm{sea}}$ can be straightforwardly obtained 
by taking the appropriate quark mass limits in the above expressions. 
Finally, the chiral loop contributions for $d_{\mathrm{val}} = 2$ at 
NNLO are, for $d_{\mathrm{sea}}=1$, $d_{\mathrm{sea}}=2$, and 
$d_{\mathrm{sea}}=3$, respectively,

\begin{widetext}

\begin{eqnarray} 
\delta^{(6)21}_{\mathrm{loops}} & = & \pi_{16}\,L^r_{0}\,\left[ \chi_1 
\chi_{13} \chi_3 + 3\,\chi_{13} \chi_4^2 + 26/3\,\chi_{13}^2 \chi_4 
- 2\,\chi_{13}^3 \right]
\:+\,4\,\pi_{16}\,L^r_{1}\,\chi_{13}^3
\:+\,\pi_{16}\,L^r_{2}\,\left[ 16\,\chi_{13} \chi_4^2 
+ 2\,\chi_{13}^3 \right] 
\nonumber \\ && 
+ \,\:\pi_{16}\,L^r_{3}\,\left[ 7/6\,\chi_1 \chi_{13} \chi_3 
+ 3/2\,\chi_{13} \chi_4^2 + 17/3\,\chi_{13}^2 \chi_4 
- 11/3\,\chi_{13}^3 \right]
\:+\,\pi_{16}^2\,\left[ 59/192\,\chi_1 \chi_{13} \chi_3 
+ 73/64\,\chi_{13} \chi_4^2 \right. 
\nonumber \\ && 
+ \,\:15/32\left.\chi_{13}^2 \chi_4 
- 77/192\,\chi_{13}^3 \right]
\:+\,384\,L^r_{4}L^r_{5}\,\chi_{13}^2 \chi_4
\:-\,1152\,L^r_{4}L^r_{6}\,\chi_{13} \chi_4^2
\:-\,384\,L^r_{4}L^r_{8}\,\chi_{13}^2 \chi_4
\:+\,576\,L^{r2}_{4}\,\chi_{13} \chi_4^2 
\nonumber \\ && 
- \,\:384\,L^r_{5}L^r_{6}\,\chi_{13}^2 \chi_4
\:-\,128\,L^r_{5}L^r_{8}\,\chi_{13}^3
\:+\,64\,L^{r2}_{5}\,\chi_{13}^3
\:-\,\bar{A}(\chi_p)\,\pi_{16}\,\left[ 1/24\,\chi_q \chi_{13} 
+ 1/8\,\chi_{13} \chi_4 - 1/4\,R^p_q\,\chi_{13} \chi_4 \right. 
\nonumber \\ && 
- \,\:1/12\left.R^p_q\,\chi_{13}^2 \right]
\:-\,\bar{A}(\chi_p)\,L^r_{0}\,\left[ 4/3\,\chi_p \chi_{13} + 
16/3\,R^p_q\,\chi_p \chi_{13} + 4/3\,R^d_p\,\chi_{13} \right]
\:-\,\bar{A}(\chi_p)\,L^r_{3}\,\left[ 10/3\,\chi_p \chi_{13} \right. 
\nonumber \\ && 
+ \,\:4/3\left.R^p_q\,\chi_p \chi_{13} + 10/3\,R^d_p\,\chi_{13} \right] 
\:+\,16\,\bar{A}(\chi_p)\,L^r_{4}\,R^p_q\,\chi_{13} \chi_4
\:+\,\bar{A}(\chi_p)\,L^r_{5}\,\left[ 8/3\,\chi_p \chi_{13} 
+ 16/3\,R^p_q\,\chi_p \chi_{13} \right] 
\nonumber \\ && 
- \,\:16\,\bar{A}(\chi_p)\,L^r_{6}\,R^p_q\,\chi_{13} \chi_4
\:+\,\bar{A}(\chi_p)\,L^r_{7}\,\left[ 16/3\,R^d_p\,\chi_{p4} 
+ 32/3\,R^d_p\,\chi_{13} \right]
\:-\,\bar{A}(\chi_p)\,L^r_{8}\,\left[ 16/3\,\chi_p \chi_{13} 
+ 32/9\,R^p_q\,\chi_1 \chi_3 \right. 
\nonumber \\ && 
+ \,\:64/9\left.R^p_q\,\chi_{13}^2 
+ 8/9\,(R^d_p)^2 \right]
\:+\,\bar{A}(\chi_p)^2\,\left[ 5/36\,\chi_{13} + 1/18\,(R^p_q)^2 
\chi_{13} - 1/54\,R^q_p\,\chi_p \right]
\:+\,\bar{A}(\chi_p) \bar{A}(\chi_{p4})\,\left[ 1/24\,\chi_p 
\right. 
\nonumber \\ && 
- \,\:1/8\left.\chi_q + 1/24\,R^p_q\,\chi_p + 19/72\,R^p_q\,\chi_q 
\right]
\:+\,\bar{A}(\chi_p) \bar{A}(\chi_{q4})\,\left[ 1/4\,\chi_{13} 
+ 1/6\,R^p_q\,\chi_q - 1/24\,R^q_p\,\chi_p + 1/72\,R^q_p\,\chi_q \right] 
\nonumber \\ && 
+ \,\:\bar{A}(\chi_p) \bar{A}(\chi_{13})\,\left[ 1/18\,\chi_p 
- 1/36\,\chi_{13} \right]
\:-\,4/27\,\bar{A}(\chi_p) \bar{A}(\chi_4)\,\chi_4 
\:+\,\bar{A}(\chi_p) \bar{B}(\chi_p,\chi_p,0)\,\left[ 11/54\,R^p_q 
\,\chi_p \chi_{13} \right. 
\nonumber \\ && 
- \,\:1/27\left.R^p_q\,\chi_p^2 - 1/18\,R^p_q\,\chi_{13}^2 
+ 1/54\,R^p_q R^d_p\,\chi_p \right]
\:-\,\bar{A}(\chi_p) \bar{B}(\chi_q,\chi_q,0)\,\left[ 1/54\,R^p_q R^d_q 
\,\chi_q + 1/36\,R^d_q\,\chi_{13} \right] 
\nonumber \\ && 
+ \,\:2/27\,\bar{A}(\chi_p) \bar{B}(\chi_1,\chi_3,0)\,R^q_p\,\chi_p 
\chi_{13}
\:+\,1/27\,\bar{A}(\chi_p) \bar{B}(\chi_1,\chi_3,0,k)\,R^q_p\,\chi_p
\:+\,\bar{A}(\chi_p,\varepsilon)\,\pi_{16}\,\left[ 13/72\,\chi_p 
\chi_{13} \right. 
\nonumber \\ && 
+ \,\:5/72\left.\chi_q \chi_{13} + 5/36\,\chi_{13} \chi_4 
+ 5/18\,R^p_q\,\chi_p \chi_{13} - 1/4\,R^p_q\,\chi_{13} \chi_4 \right] 
\:-\,\bar{A}(\chi_{p4})\,\pi_{16}\,\left[ 3/4\,\chi_p \chi_{13} 
- 3/2\,\chi_{13} \chi_4 \right. 
\nonumber \\ && 
- \,\:3/4\left.\chi_{13}^2 \right] 
\:+\,12\,\bar{A}(\chi_{p4})\,L^r_{0}\,\chi_{p4} \chi_{13}
\:+\,30\,\bar{A}(\chi_{p4})\,L^r_{3}\,\chi_{p4} \chi_{13}
\:-\,24\,\bar{A}(\chi_{p4})\,L^r_{5}\,\chi_{p4} \chi_{13}
\:+\,48\,\bar{A}(\chi_{p4})\,L^r_{8}\,\chi_{p4} \chi_{13} 
\nonumber \\ && 
+ \,\:1/4\,\bar{A}(\chi_{p4}) 
\bar{B}(\chi_q,\chi_q,0)\,R^d_q\,\chi_{13}
\:-\,\bar{A}(\chi_{p4}) \bar{B}(\chi_1,\chi_3,0)\,\left[ 4/9\,\chi_p 
\chi_{13} + 2/9\,\chi_{13} \chi_4 - 1/9\,R^p_q\,\chi_p^2 \right] 
\nonumber \\ && 
- \,\:\bar{A}(\chi_{p4}) \bar{B}(\chi_1,\chi_3,0,k)\,\left[ 2/9 
\,\chi_{p4} + 2/9\,\chi_{13} - 1/9\,R^q_p\,\chi_q \right]
\:-\,\bar{A}(\chi_{p4},\varepsilon)\,\pi_{16}\,\left[ 9/4\,\chi_{13} 
\chi_4 + 5/4\,\chi_{13}^2 \right] 
\nonumber \\ && 
- \,\:\bar{A}(\chi_1) \bar{A}(\chi_3)\,\left[ 1/36\,\chi_{13} 
+ 1/54\,\chi_4 - 1/9\,R^1_3 R^3_1\,\chi_{13} \right]
\:+\,8\,\bar{A}(\chi_{13})\,L^r_{1}\,\chi_{13}^2
\:+\,20\,\bar{A}(\chi_{13})\,L^r_{2}\,\chi_{13}^2 
\nonumber \\ && 
+ \,\:32\,\bar{A}(\chi_{13})\,L^r_{6}\,\chi_{13}^2
\:-\,1/4\,\bar{A}(\chi_{13})^2\,\chi_{13}
\:+\,\bar{A}(\chi_{13}) \bar{B}(\chi_1,\chi_3,0)\,\left[ 1/9\,\chi_1 
\chi_3 + 4/9\,\chi_{13}^2 \right] 
\nonumber \\ && 
+ \,\:4/9\,\bar{A}(\chi_{13}) \bar{B}(\chi_1,\chi_3,0,k)\,\chi_{13}
\:-\,31/18\,\bar{A}(\chi_{13},\varepsilon)\,\pi_{16}\,\chi_{13}^2
\:-\,9/4\,\bar{A}(\chi_{14}) \bar{A}(\chi_{34})\,\chi_{13}
\:+\,128\,\bar{A}(\chi_4)\,L^r_{1}\,\chi_{13} \chi_4 
\nonumber \\ && 
+ \,\:32\,\bar{A}(\chi_4)\,L^r_{2}\,\chi_{13} \chi_4
\:-\,128\,\bar{A}(\chi_4)\,L^r_{4}\,\chi_{13} \chi_4
\:+\,128\,\bar{A}(\chi_4)\,L^r_{6}\,\chi_{13} \chi_4
\:+\,16/27\,\bar{A}(\chi_4) \bar{B}(\chi_1,\chi_3,0)\,\chi_{13} \chi_4 
\nonumber \\ && 
+ \,\:8/27\,\bar{A}(\chi_4) \bar{B}(\chi_1,\chi_3,0,k)\,\chi_4
\:-\,2\,\bar{A}(\chi_4,\varepsilon)\,\pi_{16}\,\chi_{13} \chi_4 
\:-\,\bar{B}(\chi_p,\chi_p,0)\,\pi_{16}\,\left[ 1/24\,R^d_p\,\chi_q 
\chi_{13} + 1/8\,R^d_p\,\chi_{13} \chi_4 \right] 
\nonumber \\ && 
- \,\:4/3\,\bar{B}(\chi_p,\chi_p,0)\,L^r_{0}\,R^d_p\,\chi_p \chi_{13}
\:-\,10/3\,\bar{B}(\chi_p,\chi_p,0)\,L^r_{3}\,R^d_p\,\chi_p \chi_{13}
\:+\,8\,\bar{B}(\chi_p,\chi_p,0)\,L^r_{4}\,R^p_q\,\chi_p \chi_{13} 
\chi_4 
\nonumber \\ && 
+ \,\:\bar{B}(\chi_p,\chi_p,0)\,L^r_{5}\,\left[ 8/3\,R^p_q\,\chi_p^3 
+ 4/3\,R^d_p\,\chi_1 \chi_3 \right]
\:-\,16\,\bar{B}(\chi_p,\chi_p,0)\,L^r_{6}\,R^p_q\,\chi_p 
\chi_{13} \chi_4 
\nonumber \\ && 
- \,\:\bar{B}(\chi_p,\chi_p,0)\,L^r_{8}\,\left[ 16/3\,R^p_q\,\chi_p^3 
+ 8/3\,R^d_p\,\chi_1 \chi_3 \right]
\:+\,\bar{B}(\chi_p,\chi_p,0)^2\,\left[ 1/8\,R^p_q R^d_p\,\chi_p 
\chi_{13} - 1/72\,R^p_q R^d_p\,\chi_q \chi_{13} \right] 
\nonumber \\ && 
+ \,\:2/27\,\bar{B}(\chi_p,\chi_p,0) \bar{B}(\chi_1,\chi_3,0)\,R^q_p 
R^d_p\,\chi_p \chi_{13}
\:+\,1/27\,\bar{B}(\chi_p,\chi_p,0) \bar{B}(\chi_1,\chi_3,0,k)\,R^q_p 
R^d_p\,\chi_p 
\nonumber \\ && 
+ \,\:\bar{B}(\chi_p,\chi_p,0,\varepsilon)\,\pi_{16}\,\left[ 1/4\, 
R^d_p\,\chi_{p4} \chi_{13} + 5/36\,R^d_p\,\chi_{13}^2 \right]
\:-\,1/36\,\bar{B}(\chi_1,\chi_1,0) \bar{B}(\chi_3,\chi_3,0)\,R^d_1 
R^d_3\,\chi_{13} 
\nonumber \\ && 
+ \,\:32/3\,\bar{B}(\chi_1,\chi_3,0)\,L^r_{7}\,R^d_1 R^d_3\,\chi_{13}
\:+\,32/9\,\bar{B}(\chi_1,\chi_3,0)\,L^r_{8}\,R^d_1 R^d_3\,\chi_{13}
\:+\,16/3\,\bar{B}(\chi_1,\chi_3,0,k)\,L^r_{7}\,R^d_1 R^d_3 
\nonumber \\ && 
+ \,\:16/9\,\bar{B}(\chi_1,\chi_3,0,k)\,L^r_{8}\,R^d_1 R^d_3
\:-\,H^{F}(1,\chi_p,\chi_p,\chi_{13},\chi_{13})\,\left[ 1/4\,\chi_p 
\chi_{13} + 5/12\,\chi_{13}^2 - 1/6\,R^p_q\,\chi_{13}^2 \right. 
\nonumber \\ && 
+ \,\:1/18\left.(R^p_q)^2 \chi_{13}^2 \right]
\:+\,H^{F}(1,\chi_p,\chi_{14},\chi_{34},\chi_{13})\,\left[ 1/4\, 
\chi_{13}^2 + 3/4\,R^q_p\,\chi_p \chi_{13} + 1/4\,R^q_p\,\chi_q 
\chi_{13} - 1/4\,R^q_p\,\chi_{13} \chi_4 \right] 
\nonumber \\ && 
- \,\:H^{F}(1,\chi_1,\chi_{13},\chi_3,\chi_{13})\,\left[ 1/36\, 
\chi_{13}^2 - 1/3\,R^1_3 R^3_1\,\chi_{13}^2 \right]
\:+\,1/4\,H^{F}(1,\chi_{13},\chi_{13},\chi_{13},\chi_{13})\,\chi_{13}^2 
\nonumber \\ && 
+ \,\:2\,H^{F}(1,\chi_{14},\chi_{34},\chi_4,\chi_{13})\,\chi_{13} 
\chi_4
\:-\,H^{F}(2,\chi_p,\chi_p,\chi_{13},\chi_{13})\,\left[ 1/18\,R^p_q 
R^d_p\,\chi_{13}^2 - 5/36\,R^d_p\,\chi_{13}^2 \right] 
\nonumber \\ && 
- \,\:H^{F}(2,\chi_p,\chi_{13},\chi_q,\chi_{13})\,\left[ 1/18\,R^q_p 
R^d_p\,\chi_{13}^2 - 1/36\,R^d_p\,\chi_{13}^2 \right]
\:-\,1/4\,H^{F}(2,\chi_p,\chi_{14},\chi_{34},\chi_{13})\,R^d_p\,\chi_{p4} 
\chi_{13} 
\nonumber \\ && 
+ \,\:5/72\,H^{F}(5,\chi_p,\chi_p,\chi_{13},\chi_{13})\,(R^d_p)^2 
\,\chi_{13}^2
\:+\,1/36\,H^{F}(5,\chi_1,\chi_3,\chi_{13},\chi_{13})\,R^d_1 
R^d_3\,\chi_{13}^2 
\nonumber \\ && 
+ \,\:H^{F}_1(1,\chi_p,\chi_p,\chi_{13},\chi_{13})\,\left[ 20/9\, 
\chi_{13}^2 - 4/9\,R^p_q R^q_p\,\chi_{13}^2 \right]
\:-\,2\,H^{F}_1(1,\chi_p,\chi_{14},\chi_{34},\chi_{13})\,R^q_p\,\chi_{13}^2 
\nonumber \\ && 
- \,\:2\,H^{F}_1(1,\chi_{p4},\chi_{q4},\chi_p,\chi_{13})\,\chi_{13}^2 
\:+\,H^{F}_1(1,\chi_{13},\chi_1,\chi_3,\chi_{13})\,\left[ 
2/9\,\chi_{13}^2 - 4/9\,R^1_3 R^3_1\,\chi_{13}^2 \right] 
\nonumber \\ && 
+ \,\:2/9\,H^{F}_1(3,\chi_{13},\chi_p,\chi_p,\chi_{13})\,\left[ 
R^p_q R^d_p\,\chi_{13}^2 - R^d_p\,\chi_{13}^2 \right]
\:+\,2/9\,H^{F}_1(3,\chi_{13},\chi_p,\chi_q,\chi_{13})\,R^q_p 
R^d_p\,\chi_{13}^2 
\nonumber \\ && 
- \,\:1/9\,H^{F}_1(7,\chi_{13},\chi_p,\chi_p,\chi_{13})\,(R^d_p)^2 
\chi_{13}^2 
\:-\,3/4\,H^{F}_{21}(1,\chi_p,\chi_p,\chi_{13},\chi_{13})\,\chi_{13}^2
\nonumber \\ && 
+ \,\:3/4\,H^{F}_{21}(1,\chi_p,\chi_{14},\chi_{34},\chi_{13})\,R^q_p 
\,\chi_{13}^2
\:+\,3/4\,H^{F}_{21}(1,\chi_{p4},\chi_{q4},\chi_p,\chi_{13})\,R^p_q 
\,\chi_{13}^2 
\nonumber \\ && 
- \,\:3/4\,H^{F}_{21}(1,\chi_{p4},\chi_{q4},\chi_q,\chi_{13})\,\left[  
\chi_{13}^2 - R^q_p\,\chi_{13}^2 \right]
\:+\,H^{F}_{21}(1,\chi_{13},\chi_p,\chi_p,\chi_{13})\,\left[ 5/12\, 
\chi_{13}^2 - 1/6\,R^p_q R^q_p\,\chi_{13}^2 \right] 
\nonumber \\ && 
- \,\:H^{F}_{21}(1,\chi_{13},\chi_1,\chi_3,\chi_{13})\,\left[ 1/12\, 
\chi_{13}^2 - 1/3\,R^1_3 R^3_1\,\chi_{13}^2 \right]
\:+\,3/4\,H^{F}_{21}(1,\chi_{13},\chi_{13},\chi_{13},\chi_{13})\,\chi_{13}^2 
\nonumber \\ && 
+ \,\:6\,H^{F}_{21}(1,\chi_4,\chi_{14},\chi_{34},\chi_{13})\,\chi_{13}^2
\:-\,3/4\,H^{F}_{21}(3,\chi_{p4},\chi_p,\chi_{q4},\chi_{13})\,R^d_p 
\,\chi_{13}^2 
\nonumber \\ && 
- \,\:H^{F}_{21}(3,\chi_{13},\chi_p,\chi_p,\chi_{13})\,\left[ 
1/6\,R^p_q R^d_p\,\chi_{13}^2 - 1/12\,R^d_p\,\chi_{13}^2 \right] 
\:-\,H^{F}_{21}(3,\chi_{13},\chi_p,\chi_q,\chi_{13})\,\left[ 1/6\,R^q_p 
R^d_p\,\chi_{13}^2 \right. 
\nonumber \\ && 
- \,\:1/12\left.R^d_p\,\chi_{13}^2 \right]
\:+\,1/24\,H^{F}_{21}(7,\chi_{13},\chi_p,\chi_p,\chi_{13})\,(R^d_p)^2 
\chi_{13}^2
\:+\,1/12\,H^{F}_{21}(7,\chi_{13},\chi_1,\chi_3,\chi_{13})\,R^d_1 R^d_3 
\,\chi_{13}^2,
\label{M0loop_NNLO_nf3_21} 
\end{eqnarray} 

 

\end{widetext}

\section{Analytical Results for the Decay Constants}
\label{decayconstant}
\label{decays}

In the previous section, explicit analytical results for the masses of 
charged pseudoscalar mesons to NNLO in PQ$\chi$PT were presented for 
all possible degrees of degeneracy in the quark masses. The decay 
constants of the pseudoscalar mesons are similarly known, and have for 
the most part been published earlier in Ref.~\cite{BL1}. In that 
reference, analytical results were presented for all cases except for 
the most difficult one with $d_{\mathrm{val}}=2$ 
and $d_{\mathrm{sea}}=3$. However, the more advanced simplification 
methods used for the meson masses in this paper has also made it 
possible to bring that expression down to a manageable size. This is not 
surprising, since the results for the decay constants are in general 
slightly less complex than the ones for the meson masses. In order to 
obtain a complete set of NNLO calculations, the expression for the 
decay constant with $d_{\mathrm{val}}=2$ and $d_{\mathrm{sea}}=3$ is 
presented here.

\begin{figure*}[t]
\begin{center}
\includegraphics[width=\textwidth]{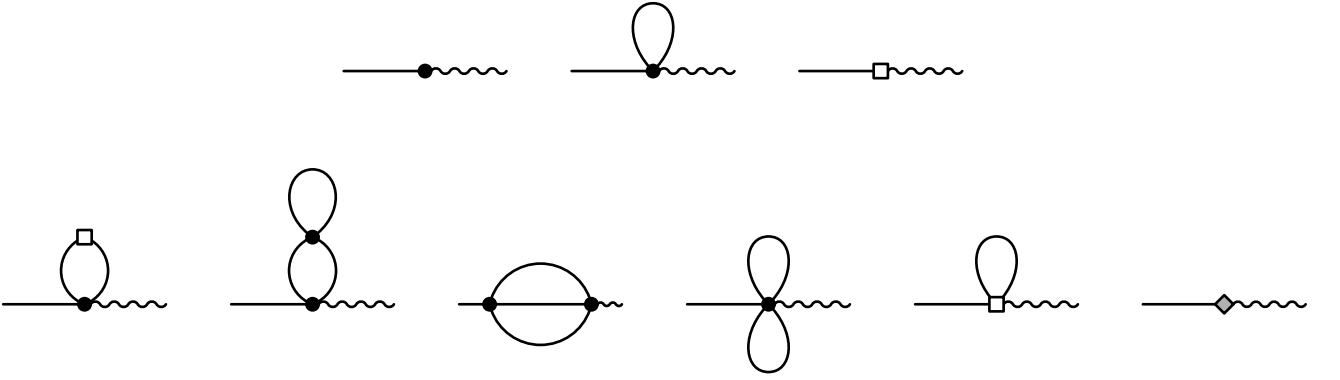}
\caption{Feynman diagrams up to ${\mathcal O}(p^6)$ or two loops, for 
the matrix element $F(M_{\mathrm{phys}}^2,\chi_i)$ of the axial current 
operator $A_a^\mu(0)$. Filled circles denote vertices of the ${\mathcal 
L}_2$ Lagrangian, whereas open squares and diamonds denote vertices of 
the ${\mathcal L}_4$ and ${\mathcal L}_6$ Lagrangians, respectively. 
In the top row, the first diagram from the left is of ${\mathcal 
O}(p^2)$, whereas the other two diagrams are of ${\mathcal 
O}(p^4)$. The diagrams of ${\mathcal O}(p^6)$, which give the NNLO 
correction to the decay constant, are shown in the bottom row.}
\label{decfig}
\end{center}
\end{figure*}

The decay constants $F_a$ of the pseudoscalar mesons are obtained
from the definition
\begin{equation}
\langle 0| A_a^\mu(0) |\phi_a(p)\rangle = i\sqrt{2}\,p^\mu\,F_a,
\label{decdef}
\end{equation}
in terms of the axial current operator $A_a^\mu(0)$. In the following 
developments, the flavor index $a$ has been suppressed for simplicity.
The Feynman diagrams that contribute to the axial current operator at 
NNLO, or ${\mathcal O}(p^6)$, are those shown in Fig.~\ref{decfig}. 
Diagrams of ${\mathcal O}(p^2)$ and 
${\mathcal O}(p^4)$ also contribute to Eq.~(\ref{decdef}) via the
renormalization of the pseudoscalar meson wave function $\phi_a(p)$. 
From Eq.~(\ref{decdef}), the expression for the decay constant of a 
pseudoscalar meson to ${\mathcal O}(p^6)$ is given by
\begin{eqnarray}
\frac{F_{\mathrm{phys}}}{\sqrt{Z}} &=& 
F_0 + F_4(M_{\mathrm{phys}}^2,\chi_i) \nonumber \\
&+& F_6(M_{\mathrm{phys}}^2,\chi_i) \:+\: {\mathcal O}(p^8),
\label{deceq}
\end{eqnarray}
where $\sqrt{Z}$ is the wave function renormalization factor. Here the 
subscripts of the matrix elements $F$ indicate the chiral order, and 
should not be confused with the flavor index $a$ in Eq.~(\ref{decdef}). 
Thus $F_4$ and $F_6$ denote the matrix elements of the axial current 
operator at NLO and NNLO, respectively. In the above equation, the 
lowest order contribution $F_2$ has already been identified with $F_0$. 
The wave function renormalization is given in terms of the self-energy 
diagrams by
\begin{eqnarray}
Z^{-1} &\equiv&
1 - \left.\frac{\partial\Sigma(p^2,\chi_i)}
{\partial p^2}\right|_{M_{\mathrm{phys}}^2}
\end{eqnarray}
which becomes, when expanded such that all contributions up to 
${\mathcal O}(p^6)$ are taken into account,
\begin{eqnarray}
\sqrt{Z} &=& 1 \:+\: \frac{\Sigma'}{2} 
\:+\: \frac{3}{8}{\Sigma'}^2 \:+\: \cdots \\
\Sigma' &=& 
\left.\frac{\partial\Sigma_4(p^2,\chi_i)}
{\partial p^2}\right|_{M_{\mathrm{phys}}^2} 
\hspace{-.2cm} + \:
\left.\frac{\partial\Sigma_6(p^2,\chi_i)}
{\partial p^2}\right|_{M_{\mathrm{phys}}^2}
\end{eqnarray}
and is thus seen to contain self-energy contributions of 
${\mathcal O}(p^4)$ as well as ${\mathcal O}(p^6)$. It should be noted 
that the two terms in the expression for $\Sigma'$ are of chiral order 
$p^2$ and $p^4$, respectively. That the self-energy of ${\mathcal 
O}(p^8)$ does not contribute at NNLO is evident from Eq.~(\ref{deceq}), 
since the matrix element of the axial current operator itself is of 
${\mathcal O}(p^2)$ or higher. By combining all the above results and 
expressing everything, analogously to Eq.~(\ref{masseq}), in terms of 
the lowest order meson mass $M_0$, the final result for the decay 
constant at NNLO is obtained as
\begin{eqnarray}
F_{\mathrm{phys}} &=& F_0 \:+\: \overbrace{
F_4(\chi_i) \:+\: F_0 \left.\frac{\partial\Sigma_4(p^2,\chi_i)}
{2\,\partial p^2}\right|_{M_0^2}}^
{{\mathcal O}(p^4)\:\mathrm{contribution}} \\
&+& F_0\:\frac{3}{8} \!\left(
\left.\frac{\partial\Sigma_4(p^2,\chi_i)}{\partial p^2}\right|_{M_0^2}
\right)^{\!2} + F_0 \left.\frac{\partial\Sigma_6(p^2,\chi_i)}
{2\,\partial p^2}\right|_{M_0^2} \nonumber \\
&+& F_4(\chi_i) \left.\frac{\partial\Sigma_4(p^2,\chi_i)}
{2\,\partial p^2}\right|_{M_0^2}
\!+\, F_6(M_0^2,\chi_i) \,+\, {\mathcal O}(p^8), \nonumber 
\end{eqnarray}
where all noncontributing terms have been discarded. In principle, the 
NLO result indicated by braces in the above equation depends on 
$M_{\mathrm{phys}}^2$ and has to be expanded, analogously to the 
calculation of the meson masses, around $M_0^2$ up to ${\mathcal 
O}(p^6)$. However, the extra terms generated by such an expansion can 
be eliminated using the identities 
\begin{equation}
\frac{\partial F_4(p^2,\chi_i)}{\partial p^2} = 0, \quad
\frac{\partial^2 \Sigma_4(p^2,\chi_i)}{\partial (p^2)^2} = 0. 
\end{equation}
The former one follows from the fact that the matrix element 
$F_4$ does not depend on $p^2$, and the latter one is valid since the 
highest power of the momentum $p$ that appears in the self-energy 
$\Sigma_4$ is $p^2$. The NNLO decay constants of the charged 
pseudoscalar mesons so obtained depend on the ${\mathcal O}(p^6)$ LEC:s 
$K_{19}^r$ through $K_{23}^r$. 

The physical decay constant of a charged pseudoscalar meson $\Phi_{ij}$ 
is given to NNLO in the form
\begin{equation}
F_{\mathrm{phys}} = F_0 \left[ 1 + \frac{f^{(4)\mathrm{vs}}}{F_0^2} 
+ \frac{f^{(6)\mathrm{vs}}_{\mathrm{ct}} 
+ f^{(6)\mathrm{vs}}_{\mathrm{loops}}}{F_0^4}
+ \mathcal{O}(p^8) \right],
\label{delteqF}
\end{equation}
where the ${\cal O}(p^4)$ and ${\cal O}(p^6)$ contributions have been
separated and denoted by $f$ rather than $\delta$ in order to minimize 
the potential for confusion with the rather similar expressions for the 
pseudoscalar meson masses. Again, the NNLO contribution $f^{(6)}$ has 
been further split into the contributions from the chiral loops and 
from the ${\cal O}(p^6)$ counterterms. As for the meson masses, the 
superscripts (v) and (s) indicate the values of $d_{\mathrm{val}}$ and 
$d_{\mathrm{sea}}$, respectively. Up to NLO, the pseudoscalar meson 
decay constant for $d_{\mathrm{val}}=2$ and $d_{\mathrm{sea}}=3$ is 
given by
\begin{eqnarray} 
f^{(4)23} & = &
12\,L^r_{4}\,\bar{\chi}_1
\:+\,4\,L^r_{5}\,\chi_{13} 
\nonumber \\ &&
+ \,\:\bar{A}(\chi_p)\,\left[ 1/6\,R^p_{q\pi\eta} - 1/12\,R^c_p \right]
\nonumber \\ && 
+ \,\:1/4\,\bar{A}(\chi_{ps})
\:-\,1/12\,\bar{A}(\chi_m)\,R^v_{mn13} 
\nonumber \\ &&
- \,\:1/12\,\bar{B}(\chi_p,\chi_p,0)\,R^d_p,
\label{F0_NLO_nf3_23} 
\end{eqnarray} 
and agrees as expected with Refs.~\cite{BG1,Sharpe1}. The LEC:s at 
${\cal O}(p^6)$ give the following contribution to the pseudoscalar 
meson decay constant for $d_{\mathrm{val}}=2$ and 
$d_{\mathrm{sea}}=3$, 
\begin{eqnarray} 
f^{(6)23}_{\mathrm{ct}} & = &
4\,K^r_{19}\,\chi_p^2
\:+\,24\,K^r_{20}\,\bar{\chi}_1\,\chi_{13}
\:+\,24\,K^r_{21}\,\bar{\chi}_2 
\nonumber \\ &&
+ \,\:72\,K^r_{22}\,\bar{\chi}_1^2
\:+\,8\,K^r_{23}\,\chi_1 \chi_3.
\label{F0tree_NNLO_nf3_23} 
\end{eqnarray} 
The expression for the chiral loop contribution is rather long, but it 
can be brought to a manageable form with the new notation introduced 
earlier in this paper. The result for $d_{\mathrm{val}}=2$ and 
$d_{\mathrm{sea}}=3$ is 

\begin{widetext}
 

\begin{figure*}[h]
\begin{center}
\includegraphics{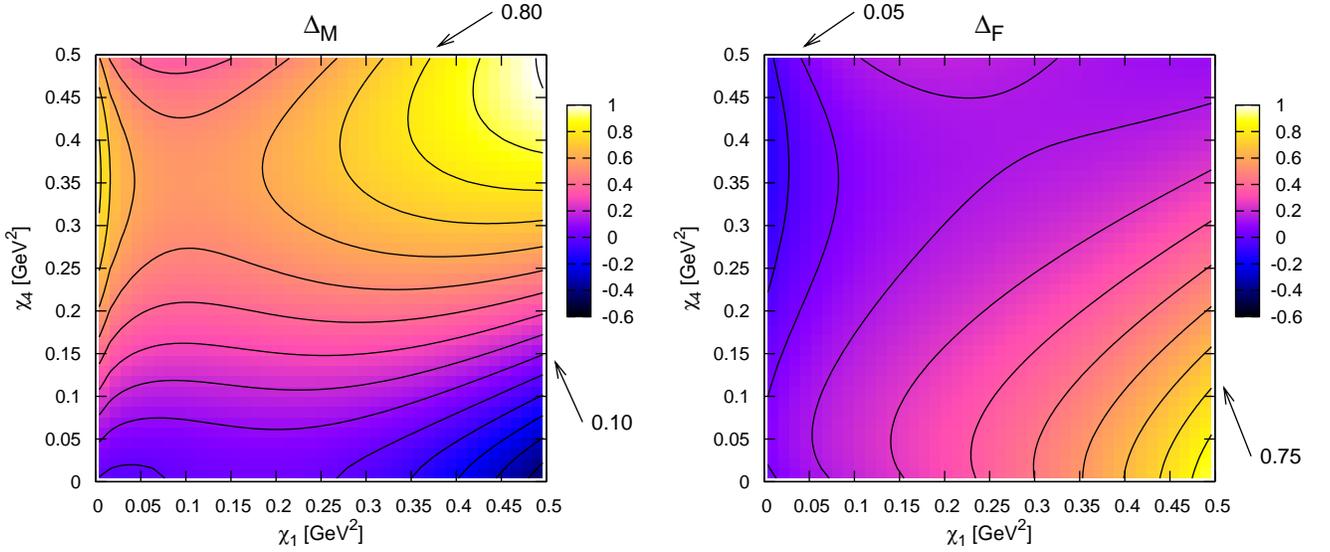}
\vspace{-1cm}
\caption{The relative shifts of the charged pseudoscalar meson mass 
$\Delta_M$ and decay constant $\Delta_F$ to NNLO for $d_{\mathrm{val}} 
= 1$ and $d_{\mathrm{sea}} = 1$, as a function of the valence and 
sea-quark masses $\chi_1$ and $\chi_4$. The quantity plotted represents 
the sum of the NLO and NNLO shifts, and the difference between two 
successive contour lines in the plots is~$0.10$. The values chosen for 
the LEC:s correspond to "fit~10" as discussed in the text.}
\label{figmass113d}
\end{center}
\end{figure*}   

\end{widetext}

\section{Numerical Results and Discussion}
\label{discussion}

\subsection{Checks on the Calculation}
\label{checks}

\begin{figure*}[t]
\begin{center}
\includegraphics{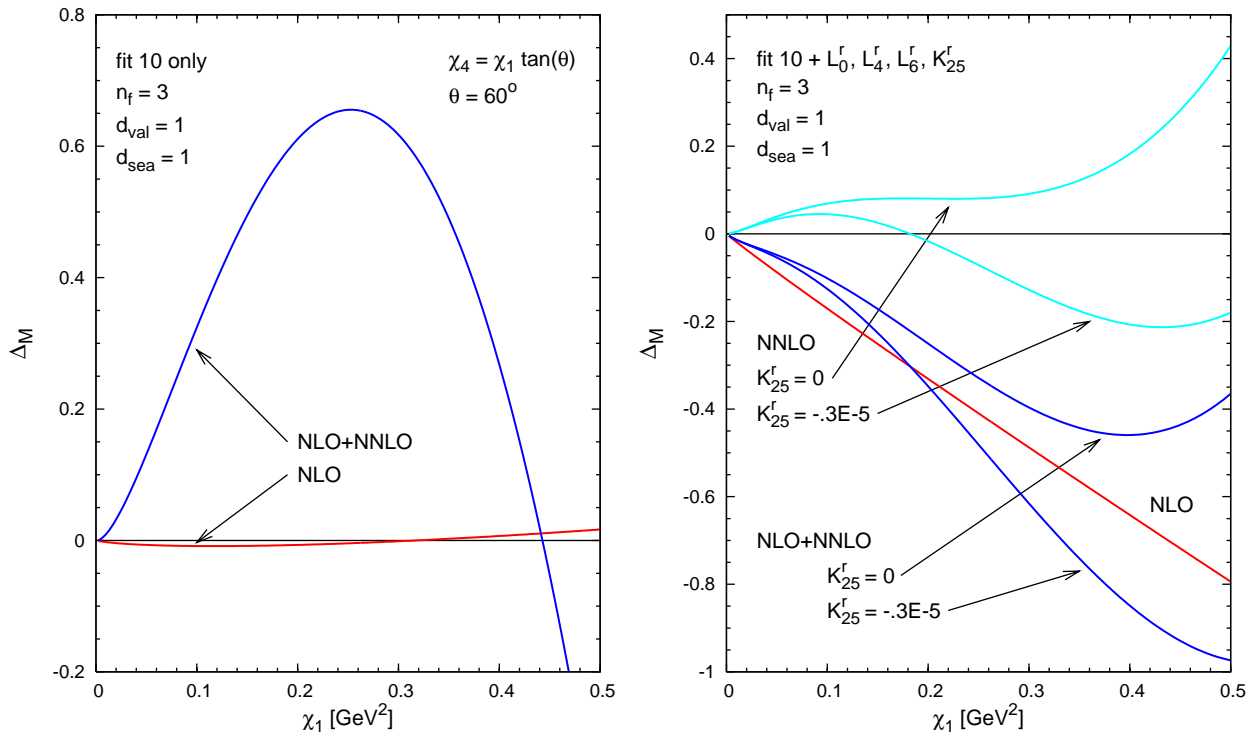}
\caption{Comparison of the NLO, NNLO and total NLO+NNLO shifts of the 
charged meson mass for $d_{\mathrm{val}} = 1$ and $d_{\mathrm{sea}} = 
1$. The left-hand plot shows the NLO and NLO+NNLO results for the set 
of LEC:s labeled 'fit~10', which has been used for all the other 
figures in this paper. The right-hand plot shows the result when 
"fit~10" is augmented by $L_0^r = -0.2 \times 10^{-2}$, 
$L_4^r = 0.1 \times 10^{-3}$, and $L_6^r = -0.1 \times 10^{-3}$. The 
effects of introducing a nonzero value for the $\mathcal{O}(p^6)$ LEC
$K_{25}^r$ is also demonstrated for the NNLO and NLO+NNLO results.}
\label{figconvergence}
\end{center}
\end{figure*}   

It is mandatory, in such a lengthy and complex calculation, that all 
possible measures be taken to ensure the correctness of the analytical 
as well as the numerical results. First of all, the divergence 
structure and the cancellation of nonlocal divergencies has been found 
to behave as expected. However, finiteness alone is not a very rigorous 
check on the end results, as several of the two-loop sunset integrals 
are convergent because of the appearance of double poles in the 
propagators. A much more rigorous check is provided by the fact that 
most parts of the calculation have been performed independently by each 
one of the authors, such that the computer programs written to handle 
the symbolic manipulations (mainly using \texttt{FORM}) have been 
developed independently. At the end of the calculation, and at several 
intermediate steps during the process, the output from each program 
has been compared and cross-checked, to make sure that perfect 
agreement was found. These checks are highly nontrivial, since the form 
of the (equivalent) outputs is by no means unique. Firstly, the exact 
appearance of the result depends on the order in which the various 
symmetries inherent in the Feynman diagrams have been implemented.
Secondly, the various loop integrals and propagator residues $R$ 
satisfy a large number of nontrivial relations, yielding further 
possibilities to rewrite the result. In the very end, the output from 
one computer program was simplified and compressed (by more than an 
order of magnitude) to bring it into a publishable size. Again, that 
expression was thoroughly checked to determine whether the compactified 
analytical expressions were still equivalent to the other two versions.

The computer programs for the numerical treatment of the analytical 
expressions have also been produced independently of each other. 
Thus the risk for accidental cross-contamination has been minimized as
each code was produced in a different programming language 
(\texttt{f77}, \texttt{f90} and \texttt{C++}). All three programs have 
been found to agree with each other, and the permutation symmetries 
under exchange of quark masses have been verified up to the numerical 
precision. As a final check, the general mass cases have been found to 
converge numerically, as expected, towards the less complex ones when 
pairs of quark masses become degenerate.

\subsection{Presentation of Numerical Results}
\label{numeric}

In view of the complexity of the analytical NNLO results in PQ$\chi$PT, 
a graphical presentation of the results is clearly called for. In 
general, the corrections to the pseudoscalar meson masses and decay 
constants are real-valued functions of all the sea and valence quark 
masses. Since, for the nondegenerate cases, this involves the graphical 
representation of a function which depends on up to five different 
quark masses, an exhaustive plot is obviously difficult to produce. 
Except for the most degenerate~(1+1) mass case, the choice has 
therefore been to present everything as a function of the valence quark mass 
$\chi_1$ only, by parameterizing the dependence on all other quark 
masses in terms of $\chi_1$. Furthermore, due to the size and 
complexity of the analytical expressions, their practical usefulness 
depends highly on the availability of software which can produce 
numerical output from these expressions in a reliable manner. In view 
of this, the software which was used for the figures in this paper will 
be made available in the near future from the website~\cite{website}.

\begin{figure}[t]
\begin{center}
\includegraphics{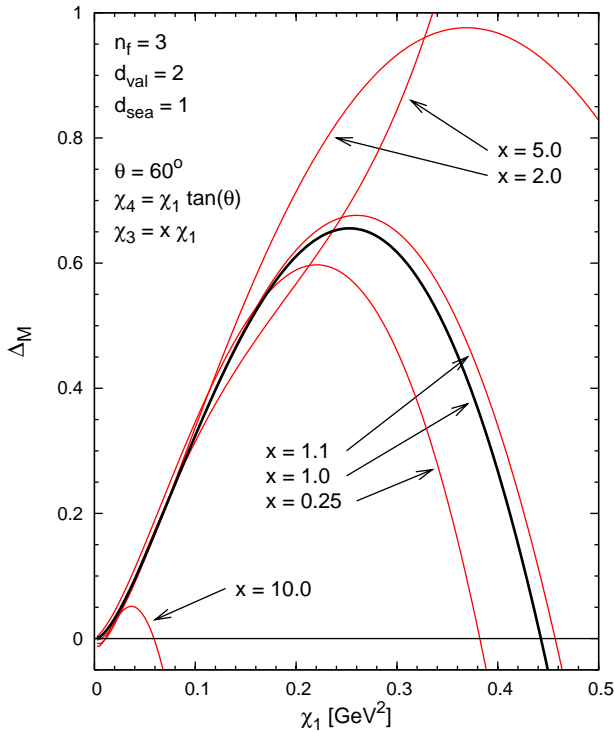}
\caption{The combined NLO and NNLO shifts $\Delta_M$ of the charged 
pseudoscalar meson mass, plotted for $d_{\mathrm{val}} = 2$ and 
$d_{\mathrm{sea}} = 1$, for $\theta=60^\circ$ in the $\chi_1 - \chi_4$ 
plane, and for the proportionality factors $x=\{0.25, 1.1, 
2.0, 5.0, 10.0\}$ between the valence quark masses $\chi_3$ and 
$\chi_1$.}
\label{21fig}
\end{center}
\end{figure}

In addition to the quark mass dependence, the NNLO expressions are also 
functions of a number of largely unknown LEC:s. In the long run, these 
LEC:s should of course be determined by a fit of the PQ$\chi$PT 
formulas to Lattice QCD data. At the present time, this is not yet 
possible, although suitable simulation results should become available 
in the near future. Therefore, the present work makes use of the LEC:s 
determined by a fit to experimental data, referred to as "fit~10", 
which is presented in Ref.~\cite{ABT2}. That fit has $F_0=87.7$~MeV and 
a renormalization scale of $\mu=770$ MeV. The NNLO LEC:s $K_i^r$ and 
the NLO LEC:s $L_4^r, L_6^r$ and $L_0^r$ were not determined in 
"fit~10", and they have thus been set to zero for simplicity. However, 
some results for nonzero values of these LEC:s are presented in 
Fig.~\ref{figconvergence}. It should be noted that $L_0^r$ cannot, as 
discussed earlier in this paper, be determined from experimental data, 
since it is a distinguishable quantity only in the PQ theory. Some 
recent results on $L_4^r$ and $L_6^r$ have been obtained in 
Ref.~\cite{piK}, but they have nevertheless been set to zero in most 
of the plots in this paper, since the present numerics are mainly 
intended for illustrative purposes.

In the next subsections, the NNLO meson masses and decay constants are 
presented in terms of the relative shifts $\Delta_M$ and $\Delta_F$ 
respectively, which represent the change in the indicated quantity 
due to the NLO and NNLO contributions. For the pseudoscalar meson mass, 
$\Delta_M$ is defined by
\begin{equation}
\label{P6masshift}
\Delta_M = M_{\mathrm{phys}}^2/\chi_{ij}-1,
\end{equation}
where $\chi_{ij}$ has again been substituted for the lowest order 
result. The calculation of $\Delta_M$ thus involves the numerical 
evaluation of $M_{\mathrm{phys}}^2$ in Eq.~(\ref{delteq}) to the order 
of the plot in question. For clarity, it is useful to recall here the 
definitions 
\begin{eqnarray}
\chi_i &=& 2 B_0 m_{qi}, 
\nonumber \\
\chi_{ij} &=& (\chi_i + \chi_j)/2, 
\label{chidef}
\end{eqnarray}
where $m_{qi}$ denotes the (current) mass of the quark $q_i$. In 
$\chi$PT, this corresponds to the lowest order mass of a meson composed 
of a quark and an antiquark, each of mass $m_{qi}$. As an example, the 
quantity $\chi_1$ plotted on the horizontal axes of 
Figs.~\ref{figmass113d} and~\ref{figconvergence} represents the lowest 
order mass of the valence quark meson $\bar q_1 q_1$, and is given by 
$\chi_1 = 2B_0 m_1$. Where no confusion can arise, the quantities 
$\chi_i$ are sometimes referred to as "quark masses".

\begin{figure*}[t]
\begin{center}
\includegraphics{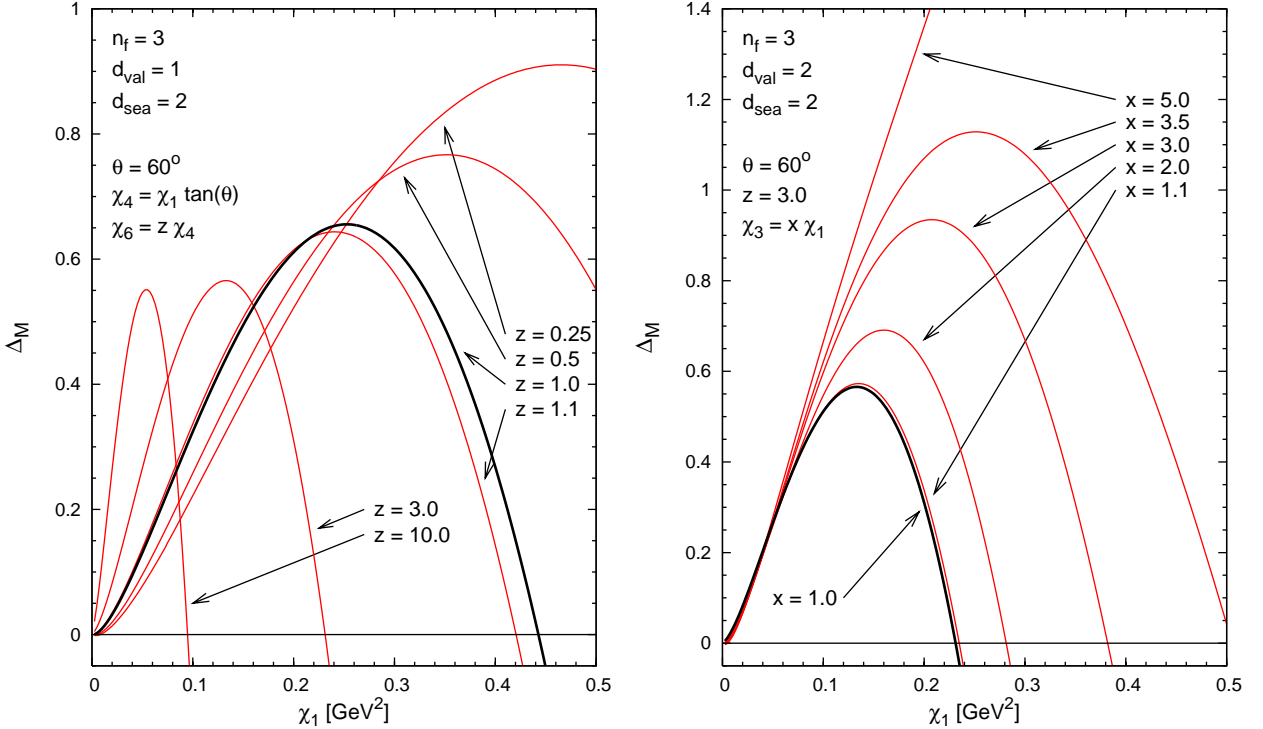}
\caption{The combined NLO and NNLO shifts $\Delta_M$ of the charged 
meson mass plotted for $d_{\mathrm{sea}} = 2$ and $\theta=60^\circ$. 
The left-hand plot shows the results for $d_{\mathrm{val}} = 1$ and a 
ratio between the sea-quark masses $\chi_6$ and $\chi_4$ of 
$z=\{0.25,0.5,1.0,1.1,3.0,10.0\}$. The right-hand plot shows the 
$d_{\mathrm{val}} = 2$ result for $z=3.0$ and a ratio between the 
valence quark masses $\chi_3$ and $\chi_1$ of 
$x=\{1.0,1.1,2.0,3.0,3.5,5.0\}$.}
\label{dsea2fig}
\end{center}
\end{figure*}   

\begin{figure*}[t]
\begin{center}
\includegraphics{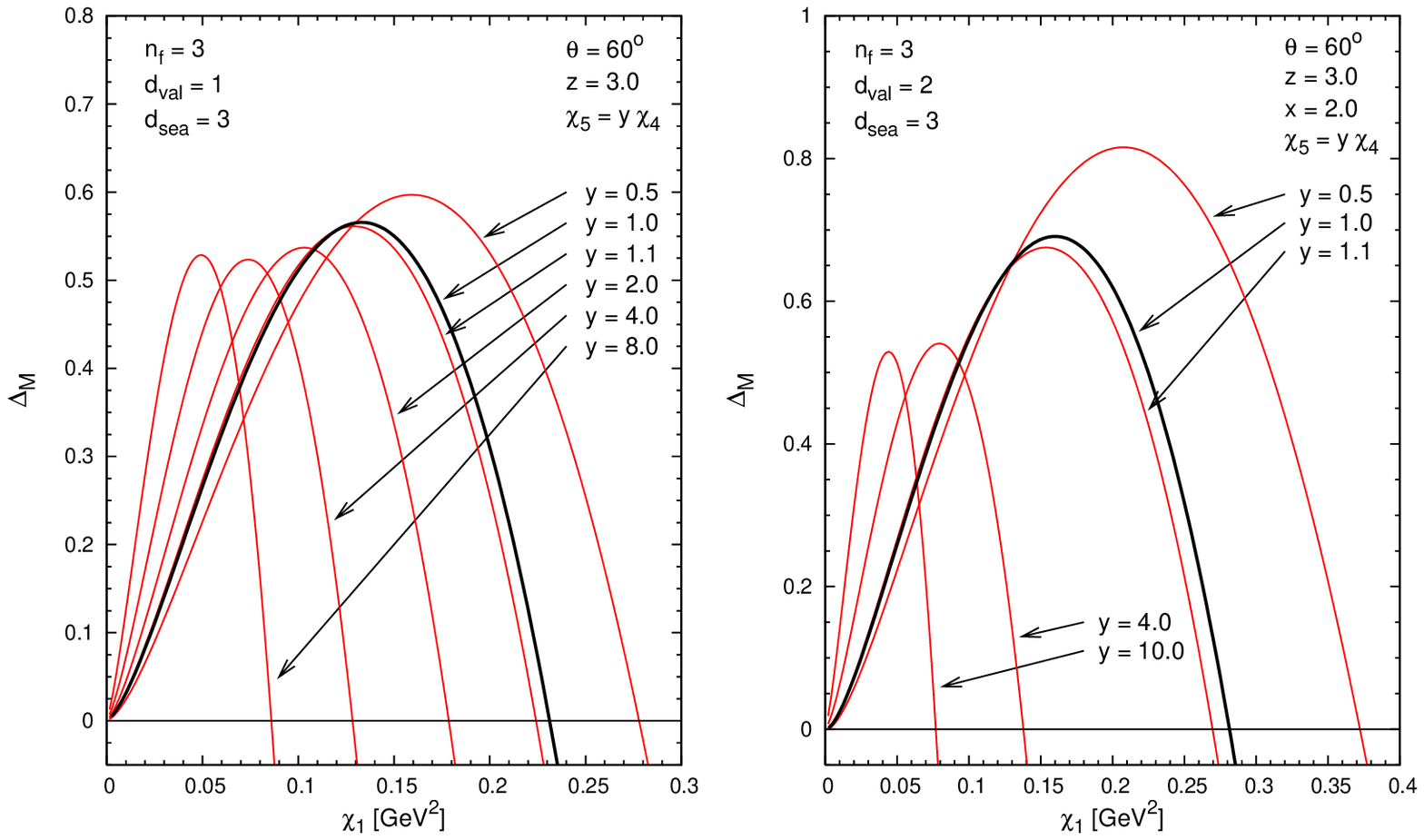}
\caption{The combined NLO and NNLO shifts of the charged meson mass, 
in 
the left-hand plot for $d_{\mathrm{val}} = 1$ and $d_{\mathrm{sea}} = 
3$, with $\theta=60^\circ$, $z=3.0$ and a ratio $\chi_5/\chi_4$ of 
$y=\{0.5,1.0,1.1,2.0,4.0,8.0\}$. The right-hand plot shows the case of 
$d_{\mathrm{val}} = 2$ and $d_{\mathrm{sea}} = 3$, with 
$\theta=60^\circ, z=3.0$, $x=2.0$ and $y=\{0.5,1.0,1.1,4.0,10.0\}$}.
\label{dsea3fig}
\end{center}
\end{figure*}   

The decay constant shift $\Delta_F$ is defined in a 
similar way, such that
\begin{equation}
\label{P6decayshift}
\Delta_F = F_{\mathrm{phys}}/F_0-1,
\end{equation}
where $F_{\mathrm{phys}}$ of Eq.~(\ref{delteqF}) is again evaluated to 
the desired order. Since the decay constant has been treated in detail 
in Ref.~\cite{BL1}, the numerical analysis of this paper will mainly 
focus on the pseudoscalar meson masses. The dependence of the 
results on the sea-quark mass $\chi_4$ and the valence quark mass 
$\chi_3$ is parameterized in terms of an angle $\theta$ in the 
$\chi_1 - \chi_4$ plane and a proportionality factor~$x$, according to 
\begin{eqnarray}
\chi_4 &=& \tan\theta\,\chi_1,
\nonumber\\
\chi_3 &=& x\,\chi_1,
\label{thxfact}
\end{eqnarray}
and the deviations of the remaining sea quark masses from $\chi_4$ are 
similarly given by
\begin{eqnarray}
\chi_5 &=& y\,\chi_4,
\nonumber\\
\chi_6 &=& z\,\chi_4.
\label{yzfact}
\end{eqnarray}
The values of $x,y,z$ and $\theta$ shown in the various plots in this 
paper have been chosen for convenience, and do not carry any particular 
significance. A value $\theta > 45^\circ$ was chosen since the 
sea-quark masses are then heavier than the valence ones, a situation 
which is encountered in realistic Lattice QCD simulations. The values 
for the other quark mass ratios have been picked so as to illustrate a 
large range of possible quark mass combinations, but they have 
otherwise been arbitrarily chosen.
 
The expressions calculated to NNLO for degenerate quark masses should 
be numerically recoverable as limits of the more general ones, which 
serves as a useful consistency check. This has been checked for all 
cases studied, and is explicitly demonstrated for each of the 
proportionality factors $x,y$ and $z$. For each diagram, one of the 
plot curves from a more degenerate mass case (the thicker black curve) 
has been included for comparison. The curve with the new mass ratio,
$x$, $y$ or $z$ equal to~1 shows the more degenerate case and
the curve with the ratio equal to~1.1 shows how the more degenerate 
case is approached.

A naive validity criterion for PQ$\chi$PT is that all masses should
satisfy approximately $\chi_i\le0.3$~GeV$^2$, but this should naturally 
be separately checked for each quantity in question. Note that the 
plots presented in this paper have been extended to include also 
regions where this constraint is not satisfied. A general study of the 
convergence of the PQ$\chi$PT expansion up to NNLO is beyond the scope 
of this paper, as realistic values obtained from Lattice QCD 
simulations should be used for all LEC:s before any meaningful 
statements concerning the relative magnitude of the NNLO corrections 
can be made. The question of convergence then hinges on whether it is 
possible to describe the Lattice QCD data in such a way that the NNLO 
correction is of a reasonable magnitude. However, on a qualitative 
level, the convergence is typically better for the decay constant
than for the mass, and the inclusion of nonzero values for the $L_i^r$ 
tends to improve the convergence. Nevertheless, it should be kept in 
mind that these statements are mainly valid for the LEC:s of "fit~10" 
since a general study of the convergence has not yet been performed. 

It should also be noted that an example which demonstrates the 
effect of variation of the $\mathcal{O}(p^4)$ LEC:s as well as the 
introduction of nonzero values for the $\mathcal{O}(p^6)$ LEC:s on the 
convergence of the chiral expansion (in the most degenerate mass case) 
is shown in Fig.~\ref{figconvergence} and discussed in 
Sect.~\ref{numeric1}.

\subsection{Numerical Results for $d_{\mathrm{sea}} = 1$}
\label{numeric1}

Numerical results for the simplest possible case, where both valence 
quarks are degenerate, i.e. $x=1$, have already appeared in a number of 
earlier works. In particular, the case of degenerate sea-quark 
masses, $y=z=1$, has been considered extensively in Refs.~\cite{BDL} 
and~\cite{BL1} both for the meson mass and the decay constant, as 
well as for different choices of LEC:s. In addition to this, 
Fig.~\ref{figmass113d} shows a two-dimensional plot of the mass shift 
$\Delta_M$ and the decay constant shift $\Delta_F$ over the whole 
parameter space up to $0.5\:\mathrm{GeV}^2$. These include
thus $\Delta_M$ and $\Delta_F$ for all values of $\theta$.

An important question in the context of PQ$\chi$PT calculations is the 
issue of convergence of the chiral expansion as a function of the input 
quark masses. At present, this question cannot be easily answered since 
the behavior of the NNLO expressions given in this paper depends rather 
sensitively on the values of the LEC:s. This point is well illustrated 
by Fig.~\ref{figconvergence}, where the NLO mass shift for 
$\theta=60^\circ$, $d_{\mathrm{val}} = 1$ and $d_{\mathrm{sea}} = 1$ 
has been given explicitly. For the set of LEC:s labeled "fit~10", the 
NLO result is very small and the total result is completely dominated 
by the NNLO contribution. For this set of LEC:s, there is obviously no 
good convergence and the overall corrections are rather large for most 
values of $\chi_1$. Furthermore, the strong curvature exhibited by the 
NNLO result is not typical of the behavior seen in Lattice QCD 
simulations. It should be noted that unquenched $\chi$PT also shows 
similar behavior~\cite{ABT2} for these values of the LEC:s.

This apparently pathological behavior can be completely changed by 
appropriately tuning the values of the LEC:s, as shown in 
Fig.~\ref{figconvergence}. From the results shown in that plot, one can 
conclude that "fit~10" with small changes is actually completely 
compatible with excellent convergence of the PQ$\chi$PT expansion. In 
fact, by choosing $L_0^r = -0.2 \times 10^{-2}$, $L_4^r = 0.1 \times 
10^{-3}$ and $L_6^r = -0.1 \times 10^{-3}$, the dramatic curvature in 
the NNLO contribution is straightened out and the near total 
cancellation in the NLO result is eliminated. The total NLO+NNLO result 
then lies close to the NLO one even for rather large values of $\chi_1$ 
and $\chi_4$. The effects of considering nonzero values of the LEC:s at 
$\mathcal{O}(p^6)$ can be sizeable as well. The changes that result 
from the introduction of a naturally sized value of $K^r_{25}=-0.3 
\times 10^{-5}$ are also shown in Fig.~\ref{figconvergence}. Overall, 
these changes in the values of the LEC:s yield a nicely convergent 
chiral expansion and a rather smooth and featureless final result.

For the case of $d_{\mathrm{val}} = 2$ and $d_{\mathrm{sea}} = 1$, the 
mass shift $\Delta_M$ has been plotted in Fig.~\ref{21fig} as a 
function of the input valence quark mass $\chi_1$ for several values of 
the valence quark mass ratio $x$. Note that the shifts $\Delta$ should 
vanish in the chiral limit, $\chi_i \rightarrow 0$, which is obviously 
satisfied by the plots since the sea-quark masses also vanish for 
$\chi_1 \rightarrow 0$. On the other hand, it is apparent from 
Fig.~\ref{figmass113d} that the shifts $\Delta$ do not vanish if the 
valence quark mass is set to zero for a constant value of the sea-quark 
mass. This effect is due to the quenched chiral logarithms and 
constitutes a generic feature of the PQ theory. Furthermore, 
Fig.~\ref{21fig} shows that the results for the (2+1)~mass case 
converge well toward the (1+1)~mass case (the thick 
black curve) in the limit $x\rightarrow 1$. 

\begin{figure}[t]
\begin{center}
\includegraphics{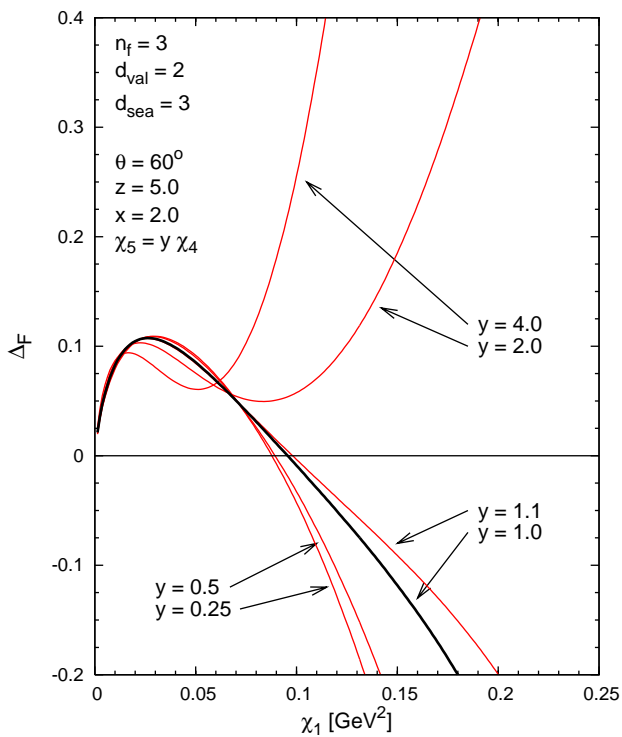}
\caption{The combined relative
NLO and NNLO shifts of the charged meson decay constant, $\Delta_F$,
plotted for $d_{\mathrm{val}} = 2$ and $d_{\mathrm{sea}} = 3$, with 
$\theta=60^\circ, z=5.0$, $x=2.0$ and $y=\{0.25,0.5,1.1,1,2,4\}$. Note 
that further plots for smaller values of $d_{\mathrm{val}}$ 
and $d_{\mathrm{sea}}$ can be found in Ref.~\cite{BL2}.}
\label{figdecay23}
\end{center}
\end{figure}

\subsection{Numerical Results for $d_{\mathrm{sea}} = 2$ and
$d_{\mathrm{sea}} = 3$}
\label{numeric2}

The expressions for $d_{\mathrm{sea}} = 2$ have two sea-quark mass 
parameters, $\chi_4$ and $\chi_6$. In the numerical analysis, the 
latter is given in terms of the former by the ratio $z$ of 
Eq.~(\ref{yzfact}). As for $d_{\mathrm{sea}} = 1$, the sea-quark masses 
are parameterized with respect to the valence quark mass $\chi_1$ by 
the angle $\theta$, as in Eq.~(\ref{thxfact}). Similarly, for 
$d_{\mathrm{val}} = 2$ the relationship between the valence quark 
masses $\chi_1$ and $\chi_3$ is given by the ratio $x$ as defined in 
Eq.~(\ref{thxfact}). For the case of $d_{\mathrm{sea}} = 2$, the 
behavior of $\Delta_M$ as a function of $\chi_1$ is illustrated, for 
different choices of $x$ and $z$, in Fig.~\ref{dsea2fig}. As for the 
previous mass case, all mass shifts vanish in the chiral 
limit for all cases shown in Fig.~\ref{dsea2fig}. In the left-hand plot 
of Fig.~\ref{dsea2fig}, the thick black curve of the mass case (1+1) 
has been plotted for the same parameters as in Fig.~\ref{21fig}, and as 
expected, the NNLO correction again approaches this curve in the limit 
$z \rightarrow 1$. In the right-hand plot of 
Fig.~\ref{dsea2fig}, the thick black curve corresponds to the mass case 
(1+2), with $60^\circ$ and $z=3.0$. There, the convergence to 
the more degenerate quark mass configuration is quite fast, as the plot 
for $x=1.1$ lies very close to that curve.

Finally, the case with $d_{\mathrm{sea}} = 3$ depends on all three 
sea-quark mass parameters $\chi_4, \chi_5$ and $\chi_6$, given in terms 
of the $\theta$ and $y,z$ defined in Eqs.~(\ref{thxfact}) 
and~(\ref{yzfact}). In the left-hand plot of Fig.~\ref{dsea3fig}, 
$\Delta_M$ is shown as a function of $\chi_1$ for $d_{\mathrm{val}} = 
1$ and $d_{\mathrm{sea}} = 3$, and in the right-hand plot the case with 
$d_{\mathrm{val}} = 2$ and $d_{\mathrm{sea}} = 3$ is shown. As for all 
previous cases, the mass shifts vanish in the chiral limit. In both 
plots of Fig.~\ref{dsea3fig}, the thick black curves can be found among 
the lines plotted for $d_{\mathrm{sea}} = 2$ in Fig.~\ref{dsea2fig}. 
Again, the convergence to the respective thick black lines is seen to 
be properly attained when $y \rightarrow 1$. For completeness, a plot 
of the relative NNLO correction to the decay constant in shown in 
Fig.~\ref{figdecay23}. Plots for the more degenerate mass cases can 
be found in Ref.~\cite{BL1}.

\subsection{Determination of LECs from Lattice QCD simulations}
\label{LECdetermination}

As noted in the previous sections, the main motivation for NNLO 
calculations in PQ$\chi$PT is that such expressions can be used to 
analyze the results of PQ Lattice QCD simulations in terms of the LEC:s 
of unquenched QCD. Up to this point, the numerical results presented in 
this paper have illustrated the quark mass dependence of the NNLO 
expressions for a given set of LEC:s. In this section, the attention is 
turned towards finding efficient methods of determining the LEC:s from 
fits of the NNLO expressions to future Lattice QCD data. It should be 
kept in mind that the LEC:s denoted by $L_i^r$ and $K_i^r$ always refer 
to the $L_i^{r(3pq)}$ and the $K_i^{r(3pq)}$ of the PQ theory with 
$n_\mathrm{sea} = 3$. An equivalent discussion for the case of
$n_\mathrm{sea} = 2$ can be found in Ref.~\cite{BL2}.

At NLO, the expressions for the masses and decay constants of the 
pseudoscalar mesons can be separated into a tree-level contribution, 
which is a function of the $L_i^r$, and a remaining part independent of 
the $L_i^r$, which involves the chiral logarithms. In the following 
developments, the average quark mass is denoted by $\bar\chi$, where 
the index denoting the power of the sea quark masses averaged has been 
dropped for simplicity. Note also the use of the product of 
lowest order sea-meson masses $\chi_\pi\chi_\eta$, which was defined in 
Eq.~(\ref{neutral_sea_masses}). For the most general case of 
$d_{\mathrm{val}} = 2$ and $d_{\mathrm{sea}} = 3$, the dependence on 
the $L_i^r$ is then proportional to
\begin{eqnarray}
f^{(4)23}_{\mathrm{ct}} & \sim &
3\bar\chi\,L_4^r \:+\: \chi_{13}\,L_5^r
\label{decayL}
\end{eqnarray}
for the decay constant, and 
\begin{eqnarray}
\delta^{(4)23}_{\mathrm{ct}} & \sim &
3\bar\chi \left(2L_6^r-L_4^r\right) \:+\:
\chi_{13} \left(2L_8^r-L_5^r\right)
\label{massL}
\end{eqnarray}
for the mass~\cite{Sharpe1}, from which an overall factor of 
$\chi_{13}$ has been removed. The expressions for the more degenerate 
quark mass cases can then be obtained by consideration of the 
appropriate limits. For the expressions in Eqs.~(\ref{decayL}) 
and~(\ref{massL}), these limits can be taken straightforwardly, and it 
can be seen at once that only the simplest (1+1) mass case is required 
in order to determine all the NLO LEC:s present in the expressions, 
since one can determine $L_4^r$ and $L_5^r$ by fitting 
Eq.~(\ref{decayL}) to Lattice QCD data, and then use this knowledge to 
determine the remaining two constants from Eq.~(\ref{massL}). The more 
complicated cases with nondegenerate quarks are thus redundant for a 
determination of the LEC:s at NLO.

A similar analysis at NNLO becomes more challenging, not because of the 
analytical complexity of the NNLO expressions, but rather because the 
chiral logarithms are no longer independent of the $L_i^r$. On the 
other hand, the dependence of the NNLO expressions on the $K_i^r$ is 
similar to that of the NLO ones on the $L_i^r$, with the exception that 
the $K_i^r$ parameters at NNLO are slightly more numerous than the  
$L_i^r$ at NLO. In all, the ${\cal O}(p^6)$ expressions depend on a 
total of 12~$K_i^r$ parameters, all of which occur in the expression 
for the meson mass, whereas the expression for the decay constants only 
depends on five of them. For the decay constants, the expression for 
$d_{\mathrm{val}} = 2$ and $d_{\mathrm{sea}} = 3$ can be factorized 
into the form
\begin{eqnarray}
f^{(6)23}_{\mathrm{ct}} & \sim & 2\,\chi_{13}^2\,K_{19}^r
\:-\,\chi_1\chi_3\left(K_{19}^r-K_{23}^r\right)
\nonumber \\ &&
+ \,\:3\,\bar\chi\,\chi_{13}\,K_{20}^r \:+\,9\,\bar\chi^2
\left(K_{21}^r+K_{22}^r\right)
\nonumber \\ &&
- \,\:6\,\chi_\pi\chi_\eta\,K_{21}^r,
\label{decayK}
\end{eqnarray}
up to an overall numerical factor which has been omitted. From this 
expression, it is then possible to determine all 5~LEC:s $K_{19}^r$ 
through $K_{23}^r$. The (1+1) mass case suffices to determine three 
combinations of LEC:s from Eq.~(\ref{decayK}), but in order to separate 
$K_{19}^r$ from $K_{23}^r$ at least one case with $d_{\mathrm{val}} = 
2$ has to be considered, otherwise only the combination 
$K_{19}^r+K_{23}^r$ is accessible, as can be readily seen by setting
$\chi_3 \rightarrow \chi_1$ in Eq.~(\ref{decayK}). Similarly, in 
order to separate $K_{21}^r$ from $K_{22}^r$, at least one case with 
$d_{\mathrm{sea}} = 2$ has to be considered, since only the combination 
$K_{21}^r+3K_{22}^r$ can be distinguished for completely degenerate 
sea-quarks, i.e. $d_{\mathrm{sea}} = 1$. This analysis is identical in 
form to the one for PQ$\chi$PT with $n_{\mathrm{sea}} = 2$ given in 
Ref.~\cite{BL2}. It is actually possible, by making the replacements 
$\chi_{13} \rightarrow \chi_{12}$, $\bar\chi \rightarrow 
2/3\,\chi_{34}$ and $\chi_\pi\chi_\eta \rightarrow \chi_3\chi_4/3$ to 
recover the corresponding expression for $n_{\mathrm{sea}} = 2$ from 
the present one given in Eq.~(\ref{decayK}).

The dependence of the meson masses on the $K_i^r$ is somewhat more 
involved, but the structure in the quark mass combinations is similar 
to that of Eq.~(\ref{decayK}). Explicitly, the dependence of the NNLO 
shift of the pseudoscalar meson mass on the $K_i^r$ is proportional to
\begin{eqnarray}
\delta^{(6)23}_{\mathrm{ct}} & \sim &
- \,\:2\,\chi_{13}^2
\left(K_{17}^r + K_{19}^r - 3K_{25}^r - K_{39}^r\right)
\nonumber \\ &&
+ \,\:\chi_1\chi_3
\left(K_{19}^r - K_{23}^r - 3K_{25}^r\right)
\nonumber \\ &&
- \,\:6\,\bar\chi\,\chi_{13}
\left(K_{18}^r + K_{20}^r/2 - K_{26}^r - K_{40}^r\right)
\nonumber \\ && 
- \,\:9\,\bar\chi^2
\left(K_{21}^r + K_{22}^r - K_{26}^r - 3K_{27}^r\right)
\nonumber \\ && 
+ \,\:6\,\chi_\pi\chi_\eta 
\left(K_{21}^r - K_{26}^r\right),
\label{massK}
\end{eqnarray}
from which a factor of $\chi_{13}$ as well as an overall numerical 
factor has been removed. As for Eq.~(\ref{decayK}), the (1+1) mass 
case here gives access to 3~combinations of LEC:s, plus an additional 
combination for each of the (1+2) and (2+1) mass cases. In all, 
5~additional combinations may be determined from the mass expression, 
even though the dependencies are slightly more entangled. However, it 
is clear that simulations with nondegenerate quark masses are needed in 
order to obtain independent information about as many of the $K_i^r$ as 
possible. On the other hand, when one determines the $L_i^r$ from NNLO 
fits, it may be convenient to work in the (1+1)~mass case only, since 
it is then still possible to distinguish all of the $L_i^r$, while the 
number of distinct combinations of $K_i^r$ parameters is as low as 
possible, which makes the fitting somewhat less complicated. 

It is also useful to compare the three-flavor result in 
Eq.~(\ref{massK}) with the analogous formula in the two-flavor 
treatment of Ref.~\cite{BL2}. As for the ${\mathcal O}(p^6)$ terms from 
the three-flavor decay constant, the terms from the mass expressions 
can be similarly translated into the expression for the two-flavor case 
using the replacements mentioned in connection with Eq.~(\ref{decayK}). 
Any apparent differences in the numerical prefactors are then only due 
to the slightly different notational conventions between the present 
work and that of Ref.~\cite{BL2}. Some additional $L_i^r$ show up 
in the NNLO expressions, namely $L_0^r$ through $L_3^r$ and $L_7^r$. It 
has been noted in Ref.~\cite{Sharpe1} that $L_7^r$ can be determined 
from the properties of the double pole of the PQ neutral meson 
propagators. This quantity has not yet been calculated up to NNLO. The 
other four $L_i^r$ are those which are relevant for meson-meson 
scattering. In the present analysis, their values have been used as 
input, but they could in principle be separated from the $K_i^r$ since 
the terms where they are present depend nonanalytically on the quark 
masses. 

While the number of free parameters in the NNLO expressions which need 
to be fitted to the Lattice QCD data is certainly large, the above
analysis has demonstrated that their number is neither unmanageable nor 
overwhelming. In particular, the possibility of producing Lattice QCD 
data for different combinations of sea and valence quark masses should 
simplify matters considerably.

\section{Conclusions}
\label{conclusions}

In this paper, a complete calculation of the mass of a charged 
pseudoscalar meson to NNLO in PQ$\chi$PT has been performed, and 
explicit analytical formulas of a reasonable length have been given for 
all relevant degrees of degeneracy in the input quark masses. In 
addition, the hitherto missing NNLO result with three nondegenerate 
sea-quarks for the decay constant of a charged pseudoscalar meson has 
been provided here. These expressions could be brought into a 
manageable size, since the residues of the PQ propagators in the 
neutral meson sector satisfy a large number of relations between sums 
of products of ratios of differences of quark masses. As elaborated in 
Sect.~\ref{checks}, the results have passed a variety of nontrivial 
consistency checks, analytical as well as numerical.

In the numerical analysis of the NNLO expressions, various plots for 
different choices of the input quark masses were presented to indicate 
the typical size of the corrections. From these plots, it is certainly 
evident that the total NLO + NNLO corrections may be uncomfortably 
large, unless the input quark masses assume very small values. However, 
it is known that the NNLO corrections in unquenched $\chi$PT are 
rather large as well~\cite{ABT1,ABT2}, which is especially true for the 
case of the meson mass. The fact that the partially quenched as well as 
the unquenched expressions depend sensitively on the values of the 
$L_i^r$ as well as the $K_i^r$ suggests that this problem may be 
entirely due to the omission of the $K_i^r$ from the numerical 
analysis. Thus, once a realistic set of values for the LEC:s is 
available, the convergence of the expansion up to NNLO should be 
dramatically improved. Nevertheless, one may readily conclude that the 
NNLO effects are bound to be nonneglegible for values of the quark 
masses that are presently used in Lattice QCD simulations.

As yet, no fits to available Lattice QCD data 
have been attempted, as such a comparison is beyond the scope of this 
article. It may be noted that an extrapolation of the simulation 
results to zero lattice spacing and to infinite volume is necessary for 
a rigorous comparison with the PQ$\chi$PT expressions, but on the other 
hand it is possible to simply include such effects as extra 
uncertainties in the determined values of the LEC:s.

\section*{Acknowledgments}

The program \texttt{FORM 3.0} has been used extensively in these 
calculations~\cite{FORM}. This work is supported by the European Union 
TMR network, Contract No. HPRN - CT - 2002 - 00311 (EURIDICE) and the 
EU - Research Infrastructure Activity RII3 - CT - 2004 - 506078 
(HadronPhysics). TL acknowledges Martin Savage, \mbox{Aurel} 
\mbox{Bulgac} and Massimiliano Procura for instructive discussions, and 
the Mikael Bj\"ornberg memorial foundation for a travel grant.

\end{document}